\documentclass[twoside,twocolumn,9pt]{article}
\usepackage{extsizes}
\usepackage[super,sort&compress,comma]{natbib} 
\usepackage[version=3]{mhchem}
\usepackage[left=1.5cm, right=1.5cm, top=1.785cm, bottom=2.0cm]{geometry}
\usepackage{balance}
\usepackage{mathptmx}
\usepackage{sectsty}
\usepackage{graphicx} 
\usepackage{lastpage}
\usepackage[format=plain,justification=justified,singlelinecheck=false,font={stretch=1.125,small,sf},labelfont=bf,labelsep=space]{caption}
\usepackage{float}
\usepackage{fancyhdr}
\usepackage{fnpos}
\usepackage[english]{babel}
\addto{\captionsenglish}{%
  
}
\usepackage{array}
\usepackage{droidsans}
\usepackage{charter}
\usepackage[T1]{fontenc}
\usepackage[usenames,dvipsnames]{xcolor}
\usepackage{setspace}
\usepackage[compact]{titlesec}
\usepackage{hyperref}

\usepackage{hyperref}

\usepackage{epstopdf}

\usepackage[normalem]{ulem} 

\usepackage{multicol}

\definecolor{cream}{RGB}{222,217,201}

\begin{document}

\pagestyle{fancy}
\thispagestyle{plain}
\fancypagestyle{plain}{
\renewcommand{\headrulewidth}{0pt}
}

\makeFNbottom
\makeatletter
\renewcommand\LARGE{\@setfontsize\LARGE{15pt}{17}}
\renewcommand\Large{\@setfontsize\Large{12pt}{14}}
\renewcommand\large{\@setfontsize\large{10pt}{12}}
\renewcommand\footnotesize{\@setfontsize\footnotesize{7pt}{10}}
\makeatother

\renewcommand{\thefootnote}{\fnsymbol{footnote}}
\renewcommand\footnoterule{\vspace*{1pt}%
\color{cream}\hrule width 3.5in height 0.4pt \color{black}\vspace*{5pt}} 
\setcounter{secnumdepth}{5}

\makeatletter 
\renewcommand\@biblabel[1]{#1}            
\renewcommand\@makefntext[1]%
{\noindent\makebox[0pt][r]{\@thefnmark\,}#1}
\makeatother 
\renewcommand{\figurename}{\small{Fig.}~}
\sectionfont{\sffamily\Large}
\subsectionfont{\normalsize}
\subsubsectionfont{\bf}
\setstretch{1.125} 
\setlength{\skip\footins}{0.8cm}
\setlength{\footnotesep}{0.25cm}
\setlength{\jot}{10pt}
\titlespacing*{\section}{0pt}{4pt}{4pt}
\titlespacing*{\subsection}{0pt}{15pt}{1pt}

\fancyfoot{}
\fancyfoot[LO,RE]{\vspace{-7.1pt}\includegraphics[height=9pt]{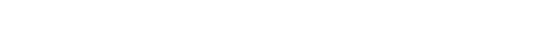}}
\fancyfoot[CO]{\vspace{-7.1pt}\hspace{13.2cm}\includegraphics{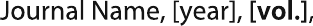}}
\fancyfoot[CE]{\vspace{-7.2pt}\hspace{-14.2cm}\includegraphics{head_foot/RF}}
\fancyfoot[RO]{\footnotesize{\sffamily{1--\pageref{LastPage} ~\textbar  \hspace{2pt}\thepage}}}
\fancyfoot[LE]{\footnotesize{\sffamily{\thepage~\textbar\hspace{3.45cm} 1--\pageref{LastPage}}}}
\fancyhead{}
\renewcommand{\headrulewidth}{0pt} 
\renewcommand{\footrulewidth}{0pt}
\setlength{\arrayrulewidth}{1pt}
\setlength{\columnsep}{6.5mm}
\setlength\bibsep{1pt}

\makeatletter 
\newlength{\figrulesep} 
\setlength{\figrulesep}{0.5\textfloatsep} 

\newcommand{\topfigrule}{\vspace*{-1pt}%
\noindent{\color{cream}\rule[-\figrulesep]{\columnwidth}{1.5pt}} }

\newcommand{\botfigrule}{\vspace*{-2pt}%
\noindent{\color{cream}\rule[\figrulesep]{\columnwidth}{1.5pt}} }

\newcommand{\dblfigrule}{\vspace*{-1pt}%
\noindent{\color{cream}\rule[-\figrulesep]{\textwidth}{1.5pt}} }

\makeatother

\twocolumn[
  \begin{@twocolumnfalse}
{\includegraphics[height=30pt]{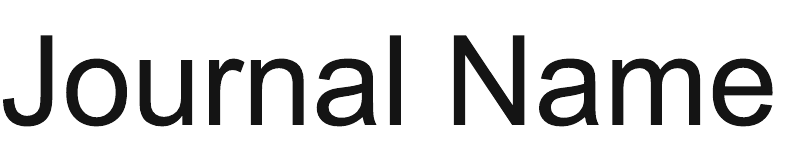}\hfill\raisebox{0pt}[0pt][0pt]{\includegraphics[height=55pt]{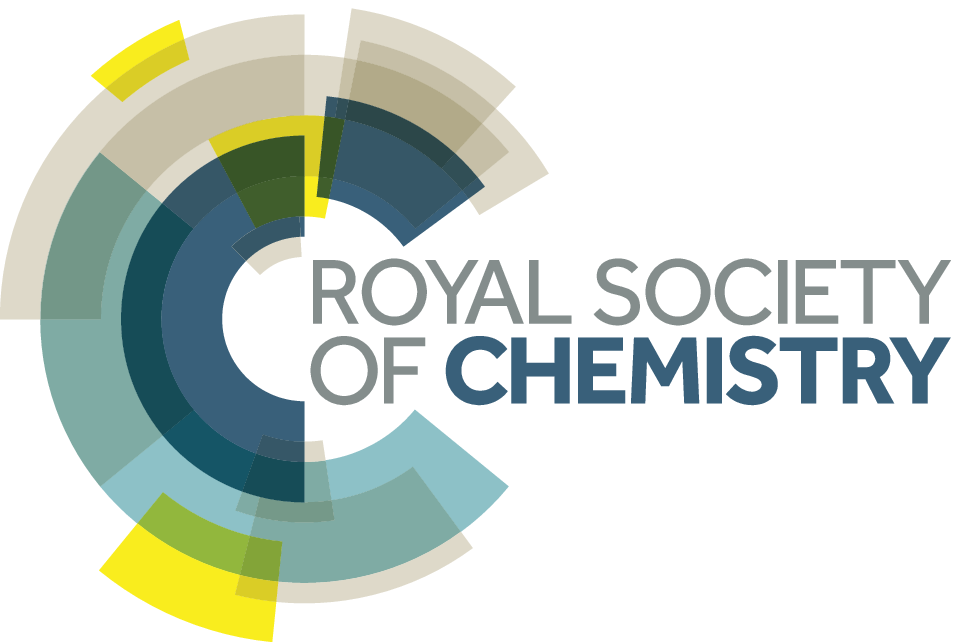}}\\[1ex]
\includegraphics[width=18.5cm]{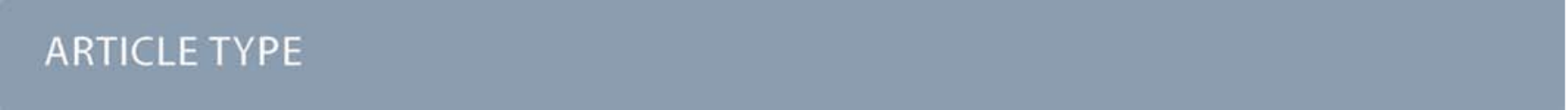}}\par
\vspace{1em}
\sffamily
\begin{tabular}{m{4.5cm} p{13.5cm} }

\includegraphics{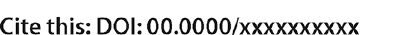} & \noindent\LARGE{\textbf{Lanthanide molecular nanomagnets as probabilistic bits$^\dag$}} \\

\vspace{0.3cm} & \vspace{0.3cm} \\

& \noindent\large{Gerliz M. Gutiérrez-Finol,$\textit{$^{a}$}$, Silvia Giménez-Santamarina,$\textit{$^{a}$}$, Ziqi Hu,$\textit{$^{a}$}$, Lorena E. Rosaleny,$\textit{$^{a}$}$ Salvador Cardona-Serra,$\textit{$^{a}$}$ Alejandro Gaita-Ariño,$^{\ast}$\textit{$^{a}$}}\\

\includegraphics{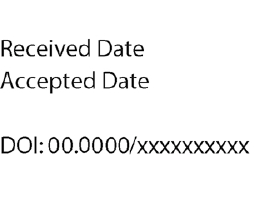} & \noindent\normalsize{Over the decades, the spin dynamics of a large set of lanthanide complexes have been explored. Lanthanide-based molecular nanomagnets are bistable spin systems, generally conceptualized as classical bits, but many lanthanide complexes have also been presented as candidate quantum bits (qubits). Here we offer a third alternative and model them as probabilistic bits (p-bits), where their stochastic behaviour constitutes a computational resource instead of a limitation. \textcolor{black}{Employing a novel} modelling tool for molecular spin p-bits \textcolor{black}{and molecular nanomagnets}, we 
simulate a minimal p-bit network under realistic conditions. Finally, we go back to a \textcolor{black}{recent systematic data gathering/recently published dataset} and screen the best lanthanide complexes for p-bit behaviour, lay out the performance of the different lanthanide ions and chemical families and offer some chemical design considerations.}

\end{tabular}

 \end{@twocolumnfalse} \vspace{0.6cm}

  ]


\renewcommand*\rmdefault{bch}\normalfont\upshape
\rmfamily
\section*{}
\vspace{-1cm}


\footnotetext{\textit{$^{a}$~Instituto de Ciencia Molecular (ICMol), Universitat de Val{\`e}ncia, Paterna, Spain; E-mail: gaita@uv.es }}


\footnotetext{\dag~Electronic Supplementary Information (ESI) available: full technical details of the modelling and extensive screening of the SIMDAVIS dataset. See DOI: 10.1038/s41467-022-35336-9.}



\section{Introduction}

The rising of Artificial Intelligence (AI) has been instrumental for pattern recognition \cite{Fatemehalsadat2022}, reasoning under uncertainty \cite{Robert2016}, control methods \cite{Derbeli2021}, analyzing and classifying big data.\cite{Han2022,Ahmad2022}. Nevertheless, there is a need for scalable and energy-efficient hardware constructed following the same scheme: further progress of AI algorithms depends on the efficiency of its hardware.\cite{Brown2001} In this scenario, 
Neuromorphic Computing promises higher efficiency since it manipulates information with hardware processes that directly mimic the nature of neurons instead of emulating this via software. \cite{Tuma2016,Chanthbouala2012} 

\begin{figure*}
 \centering
 \includegraphics[width=1\textwidth]{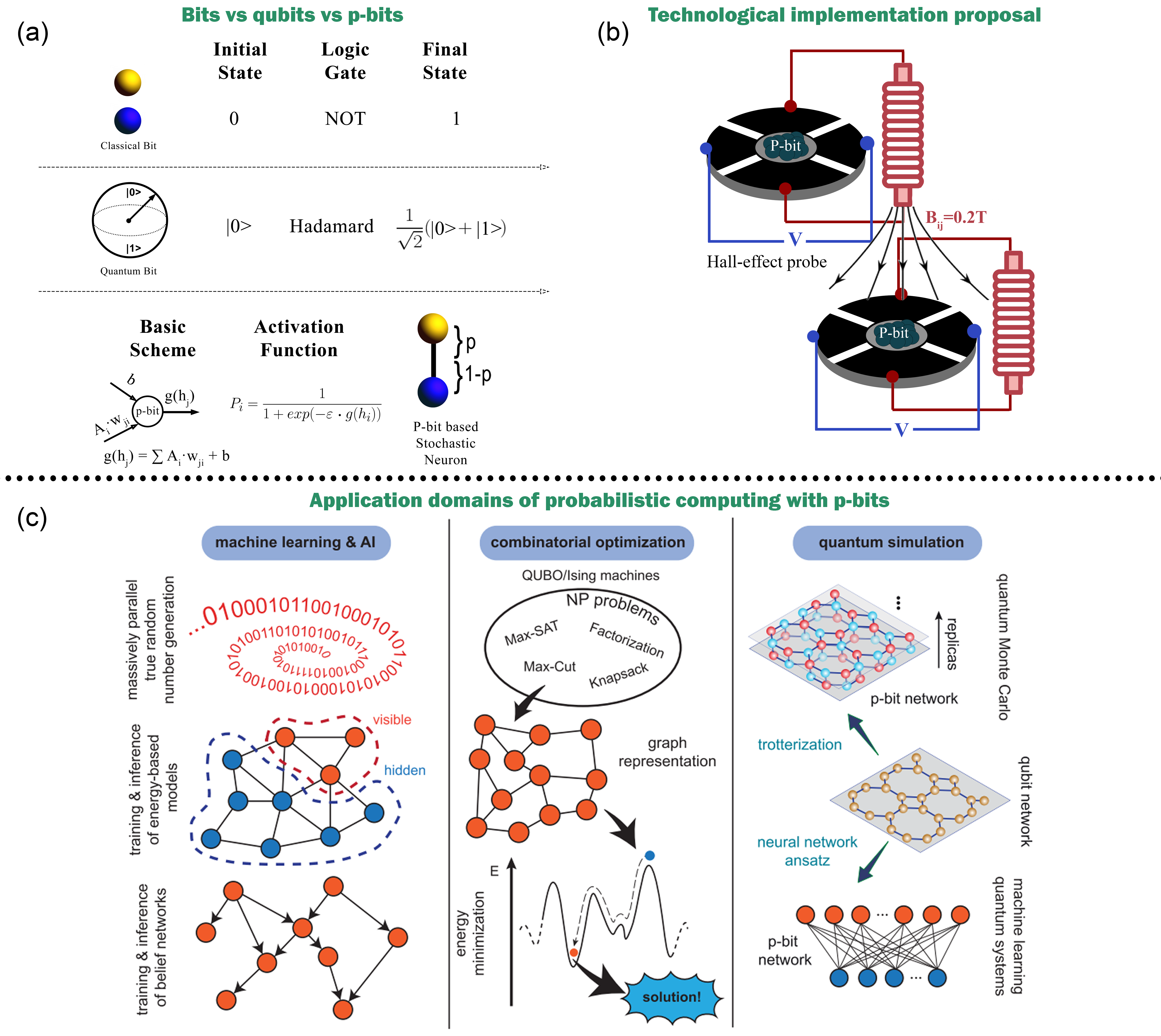}
 \caption{(a) Bits vs qubit vs p-bit based stochastic neurons. For any starting state and logical operation, bits take simple 0 or 1 values, while qubits can take coherent superposition states $\alpha|0>+\beta|1>$. The operation is inherently deterministic, and the same memory position hold the initial and the final states. P-bit based Binary Stochastic Neurons function in layers (as neural networks) rather than by applying logic gates to memory positions. Furthermore, they are inherently stochastic, meaning two identical executions will result in different microscopic values for each neuron. (b) Proposed experimental setup for spin-based p-bits: the state of a p-bit $i$ is read out by a Hall effect probe, and the resulting output voltage is amplified and translated a magnetic field that (after being properly weighted and added to all other signals) acts as input to a p-bit$j$. (c) Application domains of probabilistic computing with p-bits (reused with permission from Chowdhury et al, \cite{Camsari2023} CC BY 4.0).}
 \label{fig:bitpbitqubit}
\end{figure*}

The majority of current computers store and process information using deterministic bits that can take one of two possible values, 0 or 1. On the  other hand, quantum computers are based on quantum-bits (qubits) which can exist in a superposition of |0> and |1> spin states described by a complex wavefunction.\cite{Nielsen2002} Between these two approaches, and sharing some of their qualities, we can find the probabilistic bits (p-bits, see Figure \ref{fig:bitpbitqubit}). \cite{Camsari2023} These information elements can fluctuate stochastically between 0 and 1, but take a well-defined classical value at any given time. 

There are many important advantages for the use of p-bits in computing.\cite{Datta2021} In contrast with qubits, p-bits may operate at room temperatures with current technology, and yet are able to emulate sign-problem–free Hamiltonians (the so-called ``stoquastic'' problems).\cite{Camsari2019} Furthermore, there are no hard physical limitations in terms of strength or distance of connections between p-bits as they can be wired employing known transistor technology. In fact, probabilistic computers are designed as a network of p-bits, autonomously fluctuating between 0 and 1, with probabilities that are controlled through an input constructed from the outputs of other p-bits.\cite{Kaiser2021} Recent landmarks of probabilistic computing include experimental integer factorisation using stochastic magnetic tunnel junctions,\cite{Borders2019} and an experimental spintronic demonstration employing the inherently stochastic behaviour of nanomagnets to implement probabilistic computing.\cite{Grollier2016, Grollier2014}

The nanomagnets employed as p-bits have so far been conventional, solid state materials. This is despite of three decades of research in molecular nanomagnets which have resulted in the theoretical investigation of the switching dynamics of stochastic nanomagnets, describing their relaxation time mechanisms\cite{Shanai2021}; in addition, many studies have been carried on in the area of the experimental characterisation of several hundreds of molecules presenting a measurable magnetic memory at low temperatures,\cite{Duan2022} with some of them being very promising as qubits.\cite{Gaita2019} These systems deserve exploration for their use as nanoscale p-bits. Two challenges hampered the practical application of molecular nanomagnets as bits and qubits: (a) the difficulty to detect and manipulate  a single molecular spin, and (b) the instability of their spin states, since the bit becomes useless as soon as a spin state suffers spontaneous changes and thus, information is lost. Qubits, the quantum version of spin-based memories, are even more challenging, requiring the preservation of quantum coherence and the implementation of coherent quantum operations.\cite{wasielewski2020exploiting,Silvia2020} Notable achievements have been reported in both fields, such as molecular classical bits with (vanishingly thin) magnetic hysteresis up to liquid nitrogen temperature,\cite{Guo2018,gould2022ultrahard} or molecular qubits where one can implement a minimal quantum algorithm.\cite{Godfrin2017} Although the ephemeral nature of spin information continues to be a problem and hinders further progress for these proposed computation technologies,\cite{Gaita2019,atzori2019second} there have been advances regarding the experimental implementation of quantum gates,\cite{aguila2014,ardavan2015} and recent progress in the integration of molecular systems with superconducting circuits.\cite{gimeno2020,carretta2021}

When molecular nanomagnets are considered as p-bits, their stochasticity is a crucial aspect of their nature rather than a defect to be corrected. Then, one can profit from the sophisticated modelling of their spin dynamics.\cite{Aravena2020} Initial studies on molecular nanomagnets assumed that a simple Orbach process was the main relaxation mechanism at high temperatures, thus the focus was set on modelling the effective barrier ($U_\mathrm{eff}$),\cite{Takamatsu2007} with a Quantum Tunneling of the Magnetisation (QTM) model being responsible for spin relaxation at low temperatures.\cite{Liu2012}

As sufficient data were gathered, the key role of other relaxation mechanisms, in particular Raman, was recognized. In parallel, spin-phonon coupling was found to be a key phenomenon, both for Orbach and for Raman processes, and became the focus of attention in the field.\cite{long2017,briganti2021complete,gu2020origins} Recently, the data-science approach also provided some insights and pointers for the design of molecular nanomagnets.\cite{Duan2022} Combining electronic structure calculations with Machine Learning Force Fields is now allowing to unravel the nature of both the Orbach and the Raman relaxation pathways in terms of atomistic processes.\cite{Lunghi2020}

How could one leverage this expertise for the design of modes of computing based on molecular p-bits? Molecular nanomagnets in this context would act as Low Barrier Nanomagnets (LBNs) and be employed as Binary Stochastic Neurons (BSNs, see Figure \ref{fig:bitpbitqubit}).\cite{Hassan2019} BSNs 
would be the Artificial Neural Networks (ANNs)\cite{Sutton2020} equivalents of p-bits: 

the output of a BSN is governed by a combination of weights and biases, with the difference with ANNs being that BSN's activation function also includes a stochastical contribution. 

\textcolor{black}{A reasonable question here is: why molecular nanomagnets rather than any other magnetic nanoparticles? Firstly, with molecules we benefit from a systematization of data, which can allow experimentalists to choose the best system for their particular hardware implementation. Moreover, there is the issue of chemical and crystalline design of molecular nanomagnet-based nanoparticles vs superparamagnetic nanoparticles. Combining the SIMDAVIS dataset with inexpensive tools (such as MAGELLAN)\cite{Chilton2013} one could check for crystals where the easy axis is perpendicular to a crystal plane, to facilitate choosing the direction of the easy axis in the device. But a further, more fundamental, advantage of molecular nanomagnets as LBNs is their reproducible magnetic behavior. Within the top-down approach of ferromagnetic nanoparticles, the magnetic dynamics vary with the size of the nanoparticle. The bottom-up approach of molecular nanomagnets guarantees reproducible magnetic dynamics that only depend on the chosen molecule. This would facilitate schemes that rely on specific operating frequencies, like lock-in amplification.}

And how would the physics of molecular nanomagnets relate to the information processing as p-bits? At any given point in time, the state $m_i$ of the ideal spin p-bit $i$ is given by the signum of:
\begin{equation}
    m_i = \mathrm{sgn} \left( I_{i,[-1,1]}-r_{[-1,1]} \right)
\end{equation}
where $I_{i,[-1,1]}$ is its input signal, normalized between -1 and 1, and $r_{[-1,1]}$ is a random number with a uniform distribution between -1 and +1, meaning that the output is stochastic but biased by the input. It is only unbiased when $I_i=0$ and only deterministic when $I_i=1$ or $I_i=-1$. For a spin p-bit, $I_i$ can be embodied by the polarisation achieved by an external magnetic field $B_i$, i.e. a Zeeman effect that at a given temperature achieves a certain Boltzmann distribution. For the direction of the field corresponding to positive input signals, this is expressed as: 
\begin{equation}
    I_{i,[0,1]} = 2\cdot\frac{e^{-E_{Zeeman}}/kT}{{1+e^{-E_{Zeeman}}}/kT}
\end{equation}
where
\begin{equation}
E_{Zeeman} = g \mu_B B M_J
\end{equation}
where $g$ is Landé's factor, $\mu_B$ is Bohr's magneton, $B$ is the external magnetic field and $M_J$ is the projection of the spin state. Applying $B$ in the opposite direction would let one access the range $I_{i,[-1,0]}$. The same kind of biasing effect can be in principle achieved spintronically, by the flow of a spin-polarized current that acts as an effective magnetic field that couples with the spin p-bit.\cite{Locatelli2014} In any case, the practical limit values of $I_i$, i.e. the saturation magnetisation, will be given by technological limits. In particular, large biases would require either very high effective magnetic fields or very low temperatures, or a combination of both.
In the field of p-bits this behavior is often written, equivalently, as
\begin{equation}
    I_{i,[-1,1]} = \mathrm{tanh}(H_i)
\end{equation}
 
In the general case, for an interconnected network of p-bits, each input $I_i$ results from the states of the rest of the p-bits $m_j$, weighted by their connections $J_{ij}$, plus a local bias $h_i$:
\begin{equation}
    I_i = \sum_j J_{ij}m_j + h_i
\end{equation}

Note that this fundamentally differs from the circuit model of Boolean logic gates that applies to conventional computing based on bits and quantum circuits employing qubits. Instead, it is closer to the way classical neural networks operate, with the difference being the stochastical character of p-bits. In a neural network, and also in a p-bit network, rather than a logical gate acting on a memory unit and altering it, the state of a given memory unit depends on the states of several other memory units with different weights, plus a bias (in the case of p-bits, to the bias one adds a random contribution). The system has to be adapted to solve a specific problem by adjusting the individual biases of each memory unit and the weights of each interaction between every pair of memory units, rather than by selecting an algorithm that applies a sequence of logical gates. 
This proposed use of molecular spin p-bits would add to recent proposals to employ molecular electric switches to emulate synaptic behaviour.\cite{Wang2022}




\section{Methods}

\textcolor{black}{We developed a custom implementation of Markov Chain Monte Carlo, which is based in random transitions for each spin at each time step and uses parameterised magnetization dynamics and Boltzmann distribution for the calculation of the Markov transition probabilities between the two possible spin states. In contrast with many Monte Carlo model implementations, our end goal is not to simulate the thermodynamic properties of a material via averaging of many samples but the dynamical behavior of a full population. Thus, within the many implementations of Markov Chain Monte Carlo methods (including Metropolis Monte Carlo), three crucial features for our goals are (a) we simulate independent particles, meaning we work with $N$ identical and independent Markov chains, (b) instead of accumulating data over many runs with small sizes, we perform a single run with a large size, which will eventually be the actual number of magnetic molecules in our p-bit and (c) each calculation step has an associated natural time in real time units, which is taken into consideration for the calculation of the transition probabilities; this model is empirical and relies on externally determined relaxation parameters.}

Assuming that the dynamics are the same for each individual nanomagnet and that the collective behaviour results from a simple addition of magnetic moment, one can employ the same $\tau$ to describe each individual spin. In particular, we employed the Maclaurin expansion of the exponential function to calculate the spin flip probability at a given time interval, and so allow recovering macroscopic spin dynamics of either a single and an ensemble of molecules. The magnetisation is simulated by the number of uncompensated spins and scaled to the experimental values (see full details in the SI section S1A).

In order to simulate stochastic spin dynamics in presence of a magnetic field, we introduced the Zeeman effect to produce a bias in the spin flip probability. This allows to recover Boltzmann statistics (see full details in SI section S1B). This complication of the model is necessary to be able to reproduce p-bit dynamics in a spin system that is driven by an external field.

All ac (alternate current) simulations were calculated with the same scheme, but under a time-dependent field. In-phase and out-of-phase susceptibilities were fitted employing a generalized Debye model (see full details in the SI section S2). 

The screening of the SIMDAVIS dataset was performed by employing the SIMDAVIS dashboard (see full details in the SI sections S3\textcolor{black}{, S4}).\cite{Duan2022}

All simulations \textcolor{black}{presented herein} were performed on a desktop computer and took, all in all, less than \textcolor{black}{24} h of processor time. The associated software to this work, named STOSS (for STOchastic Spin Simulator),  and the instructions to reproduce all the graphic results are available at \href{https://github.com/gerlizg/STOSS}{https://github.com/gerlizg/STOSS}.

\section{Results and discussion}

\subsection{Lanthanide-based, molecular, isolated spin p-bits at constant field}

\begin{figure*}[th!]
 \centering
 \includegraphics[width=\textwidth]{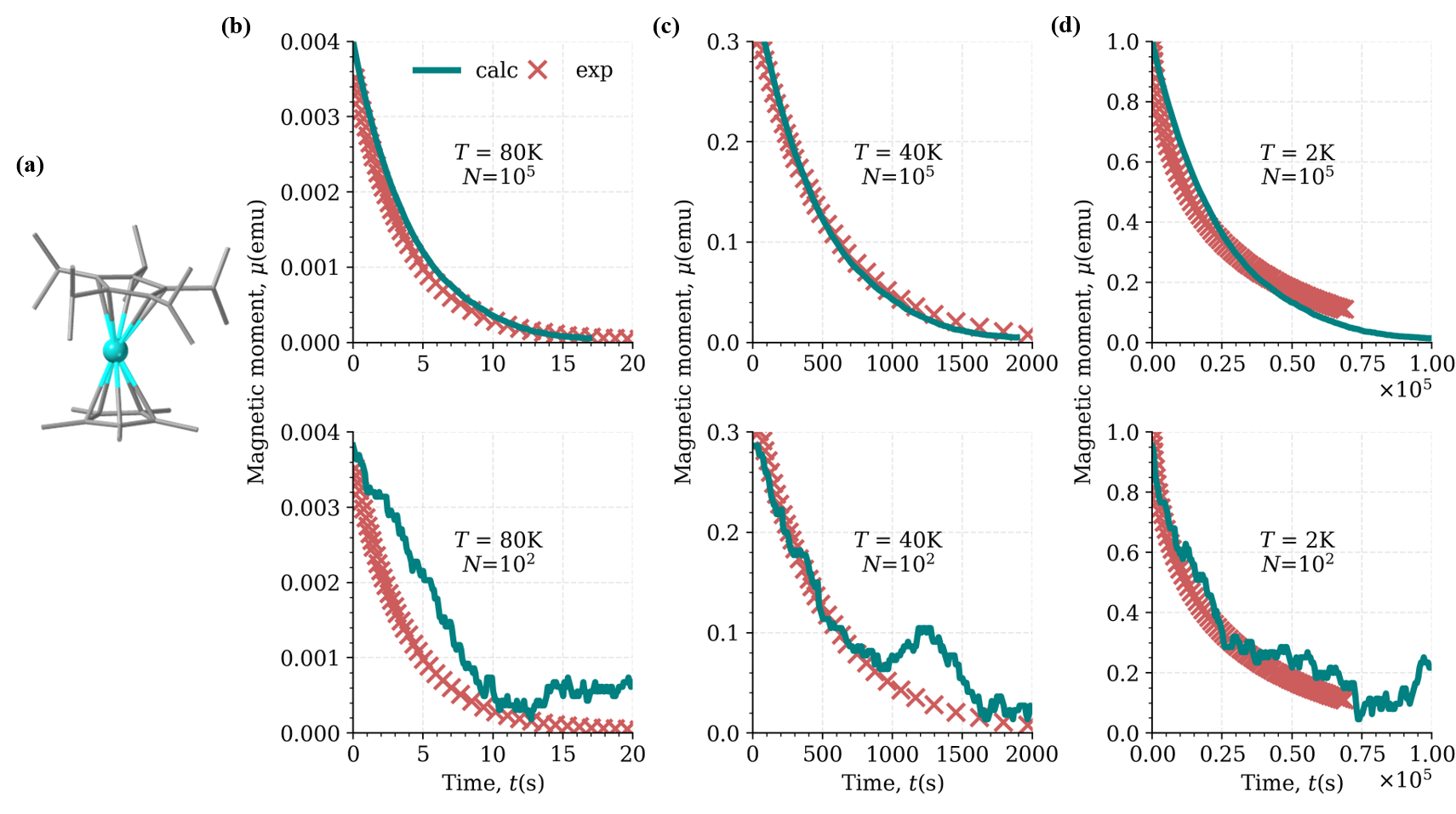}
 \caption{(a) Structure of [(Cp$^\mathrm{iPr5}$)Dy(Cp$^\mathrm{*}$)]$^+$ (Cp$^\mathrm{iPr5}$, penta-iso-propylcyclopentadienyl; Cp$^\mathrm{*}$, pentamethylcyclopentadienyl), (b-d, upper panels) its magnetisation decay vs time, experimental vs (scaled) simulation, at 3 temperatures, from left to right, 2K, 40K, 80K, with the simulations employing $N = 10^5$ spins. Experimental data are as extracted from Guo et al.~\cite{Guo2018} The simulated magnetisation is plotted as difference between spins "up" and "down", starting from a fully polarized population, each evolving as explained in the text, and employing the parameters for Orbach, Raman and QTM mechanisms reported by Guo et al.~\cite{Guo2018} \textcolor{black}{(b-d, lower panels) present the same calculations, in the same conditions, but with a more limited number of spins  $N=10^2$ to evidence the stochastic nature of the model.}}
 \label{fig:casestudyspin}
\end{figure*}

In state-of-the-art p-bit modelling, the spin dynamics caused by spin transfer torque effects on solid state superparamagnetic tunnel junctions are modeled by an Arrhenius/Neel-Brown approach.\cite{Vodenicarevic2018} In the case of molecular spin p-bits, we benefit from the more sophisticated modelling of molecular nanomagnets, in which a commonly employed equation of the relaxation time ($\tau$) reads as follows:
\begin{equation}
    \tau^{-1} = \tau_\mathrm{QTM}^{-1} +CT^n + \tau_0^{-1}\mathrm{exp}(-U_\mathrm{eff}/T)
\label{Orbachetc}
\end{equation}
where the first term presents temperature-independent QTM, and the second and the third terms are thermally-assisted Raman and Orbach processes, respectively.

This relaxation time $\tau$ is employed to describe a collective behaviour. In particular, an ensemble of molecular nanomagnets experiences exponential decay of the magnetisation (see Figure \ref{fig:casestudyspin}):
\begin{equation}
    M(t) = M_\mathrm{eq} + (M_0 - M_\mathrm{eq})\mathrm{exp}[-(t/\tau)^\beta]
\label{Decay}
\end{equation}
where $M_\mathrm{eq}$ and $M_0$ are the equilibrium and initial magnetisations, respectively; $\beta$ is a so-called "stretching" parameter related to the time-dependent decay rate. \textcolor{black}{Here w}e worked with $\beta=1$, meaning independent spins (see details in the SI section S1)\textcolor{black}{, although STOSS can incorporate an averaged internal magnetic field reflecting the effect of dipolar coupling}.



As a basic test for our simulator, we verified that, for a sufficiently large number of spins in the model, an exponential fit of the decay of the ensemble magnetisation recovers the input $\tau$ with the desired accuracy. This served as a basic benchmark of the model and also emphasized a further shortcoming: if the experimental magnetisation relaxes as a stretched exponential, our current model of independent spins cannot reproduce its shape properly. \textcolor{black}{A} simple modification of the \textcolor{black}{code is available where the} spin flip probability is a function of the spin up/down ratio\textcolor{black}{, effectively simulating the average internal magnetic field created by dipolar interaction}. 

As an illustration, we employ our modelling sofware STOSS to simulate the magnetisation decay at three representative temperatures (2 K, 40 K, 80 K) with the parameters obtained experimentally in the case of  [(Cp$^\mathrm{iPr5}$)Dy(Cp$^\mathrm{*}$)]$^+$.\cite{Guo2018} Figure~\ref{fig:casestudyspin} upper panels depicts a simulation of the experimental results based on the evolution of $N=10^5$ spins. While the simulation deviates significantly from the experimental decay curve at 2 K, they coincide well at high temperatures (40 K and 80 K). The deviation at low temperatures is due to the fact that the observed relaxation of magnetisation follows a stretched exponential decay ($0 < \beta < 1$ in Eq. \ref{Decay}), as occurs when spins are interacting. As a result, the value of $\beta$ for 2 K is 0.5527, which differs greatly from the values found for 40 K and 80 K (0.9831 and 0.993, respectively). This variation is due to the time-dependent relaxation rate that likely stems from the redistribution of local dipolar fields when the magnetisation of the molecule is varied. In contrast, single exponential decay is normally adequate to account for the relaxation behaviour at high temperatures, as evidenced by $\beta \approx 1$ at 40 K and 80 K in this case. To prove that this is not a bulk simulation but rather intrinsically microscopic, the evolution of a smaller number of spins $N=100$ at each temperature is shown in Figure~\ref{fig:casestudyspin} bottom panels. It is by the simple addition of a large number of these telegraph noises that the properties of the bulk sample are reproduced.


\subsection{Lanthanide-based, molecular, dynamically driven spin p-bits}

\begin{figure}[htb!]
\centering
  \includegraphics[width=\columnwidth]{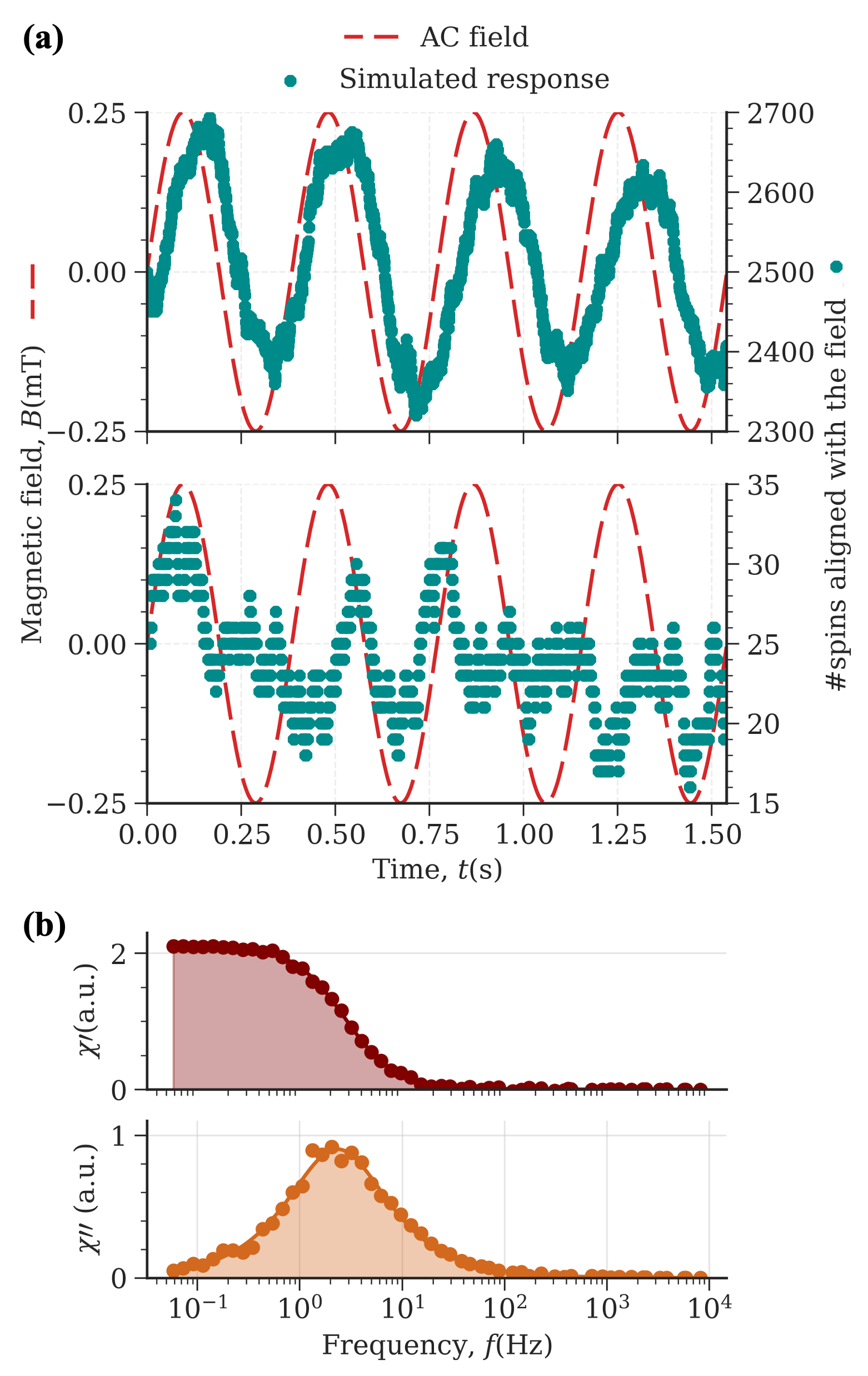}
  \caption {Periodically driven p-bits\textcolor{black}{, when simulated by our Markov Chain Monte Carlo calculations,} behave \textcolor{black}{qualitatively} as spins in an ac susceptometry experiment: (a) calculated time evolution of \textcolor{black}{$N=2.5\times10^4$ (up) or $N=50$ (down) stochastic spins described by the parameters corresponding to [Dy(bath)(tcpb)\textsubscript{3}], at 20~mK and under an external ac field of amplitude $B_\mathrm{max}=0.25$~mT and frequency {$2~\mathrm{Hz}$} that evidences a certain delay 
  in the shape of the simulated magnetisation for $N=2.5\times10^4$, whereas in the same conditions in the simulation of $N=50$ stochastic spins neither the delay nor the periodicity are apparent and (b) simulated $\chi^\prime$, $\chi^{\prime\prime}$, where each value of $\chi^\prime$ ($\chi^{\prime\prime}$) corresponds to the prefactor of the cos(sin) in a weighted sin+cos fit of simulated oscillations for $N=2.5\times10^4$ p-bits described by the parameters corresponding to  [Dy(bath)(tcpb)\textsubscript{3}]}; the simulation was also performed at 20~mK meaning that the spin dynamics is governed by the parameter $\tau_\mathrm{QTM}$. The solid lines indicate the fits using the generalized Debye model.}
  \label{fig:ac}
\end{figure}

After simulating the spontaneous evolution of a single p-bit with constant input signal (e.g. constant magnetic field) and benchmarking the modelling software STOSS against the experimental relaxation of the magnetisation of a molecular nanomagnet, the second step is driving it with a time-dependent signal $I_i=\mathrm{f}(t)$. The natural way to test the behaviour of the model in these conditions is by simulating the out-of-phase susceptibility of a molecular nanomagnet, where $I_i=\mathrm{cos}(t)$ (see Figure \ref{fig:ac}). 

Note that microscopic \textcolor{black}{(spin-by-spin)} simulations of bulk ac behaviour are necessarily limited to low temperatures due to fundamental reasons. Indeed, the population difference between the distinct spin directions for very weak fields that are typical in ac susceptometry $B\leq 0.5~\mathrm{mT}$ is extremely small even at reasonably low temperatures in the order of $T=2-4~\mathrm{K}$, and the population difference gets even smaller at higher temperatures. This is the reason why it can be experimentally challenging to obtain ac signal for molecular nanomagnets that function at high temperatures. Furthermore, while sample sizes of the order of $m=0.1~\mathrm{mg}$ are considered small, they typically contain a number of spins in the order of $N\simeq10^{16}$, which is absolutely out of the bounds of what one can simulate microscopically. For this reason, we performed our simulation at $T=20~\mathrm{mK}$, a temperature in the low limit of the experimentally accessible for researchers studying magnetic molecules. At this temperature, one should be able to access the spin dynamics corresponding to the QTM regime.

\textcolor{black}{As above, the dynamic response is simulated by employing externally obtained parameters. For this simulation we chose [Dy(bath)(tcpb)$_3$], where tcpb = 1-(4-chlorophenyl)-4,4,4-trifluoro-1,3-butanedione and bath = 4,7-diphenyl-1,10-phenanthroline, where $\tau_\mathrm{QTM}=$ 0.067 s, $n=$ 4.90, $C=$  7.80$\times10^{-2}$ s$^{-1}$ K$^{-4.90}$, $\tau_0=$ 2.83$\times10^{-9}$ s, and the effective energy barrier ($U_\mathrm{eff}$) is 116.87 cm$^{-1}$ (167.87 K).}
In a first example simulation, we simulated 4 cycles with a period of an order of magnitude longer than $\tau_\mathrm{QTM}$. Employing $N=10^4$ spins in the simulation, the response is largely in-phase with the magnetic field and yet a certain delay in the signal is already evident (see Figure \ref{fig:ac}(a) up). To offer a more intuitive insight into how this ensemble response signal is obtained, we also represented the same simulation with $N=100$ spins (see Figure \ref{fig:ac}(a) down), where apparently there is just some relatively rapid telegraph noise overlapped with a slower random drift. It is crucial to understand that merely by summing a hundred similarly noisy patterns resulting from microscopic simulations one can reproduce a relatively clean macroscopic behaviour. 

We repeated these calculations both at longer and shorter periods for the oscillation of the magnetic field and fitted the result to a weighted sum of $sin$ and $cos$ functions to extract in-phase and out-of-phase signals (details in SI section S2A). The result, illustrated in Figure \ref{fig:ac}(b), is consistent with the expected behaviour, and the extracted $\tau$ of \textcolor{black}{$5.79(6) \cdot 10^{-2}~\mathrm{s}$} using generalized Debye model is essentially identical to the experimental value.  

In the context of molecular nanomagnets, STOSS should serve as an auxiliary tool in the ongoing effort for a proper interpretation of magnetic relaxation times and magnetic relaxation parameters.\cite{Reta2019} In the context of p-bits, one can see this exercise as a synchronisation with an external periodic drive, but rather than being limited to simulating bulk ac experiments, one can easily think of ways to construct more sophisticated circuit architectures. With an adequate circuitry construction, the output of p-bit $i$ can serve as input for p-bit $j$. \textcolor{black}{To work at temperatures higher than $T=20$~mT, it suffices to rise the applied magnetic field above $B=0.5$~mT.} Precisely this scenario is explored in the next section.


\subsection{Lanthanide-based, molecular spin p-bit network}


As a minimal toy model for a p-bit network, we explored a 2-p-bit architecture \textcolor{black}{where each p-bit is constituted by the collective signal of $10^6$ magnetic molecules, for example forming a thick layer on top of a sensor. We assume a circuit able to a spin excess of $10^3$ spins, as should be possible to do employing nanoscale (100 nm $\times$ 100 nm) Hall probes.\cite{Collomb2019} For simplicity we define as p-bit state 1 the case where there is a spin excess of $10^3$ spins in the "up" direction, and 0 otherwise. Compared with requiring also a spin excess for the 0 state, this means that 0 and 1 states are slightly asymmetrical, but it saves us from having to consider "dead" times where neither an excess of spin "up" nor "down" can be detected. Also for simplicity, we take} identical p-bits in terms of 
$\tau$. 

We allowed a free evolution for the spin dynamics of p-bit $i$ (no input signal, no bias, $B_i=0$). It was also assumed that each \textcolor{black}{p-bit} flip in p-bit $i$ is detected by a sufficiently fast readout and it is used to switch the magnetic field acting on p-bit $j$. 
\textcolor{black}{We chose two systems from the SIMDAVIS dataset which presented distinct advantages and a maximum in the out-of-phase signal around 4 K and 40 K respectively. 
Thus, for the simulations at $T=4$~K we chose [Dy\{$\eta^\mathrm{4}$-C\textsubscript{4}(SiMe\textsubscript{3})\textsubscript{4}\}{$\eta^\mathrm{4}$-C\textsubscript{4}(SiMe\textsubscript{3})\textsubscript{3-k}-(CH\textsubscript{2}SiMe\textsubscript{2}}]$^\mathrm{2-}$ molecules\cite{day2018}, which are air stable and thus in principle easier to process. 
This system presents a $\tau_0=$ 1.83$\times10^{-9}$ s and $U_\mathrm{eff}=$ 464 K. For the simulations at $T=40$~K we chose quadruple decker \{[Dy(obPc)$_2$]Cd[Dy(obPc)$_2$]\} molecules\cite{katoh2012}, where obPC = 2,3,9,10,16,17,23,24-octabutoxyphthalocyaninato. This is a phthalocyaninato-based system that should be easy to deposit on surfaces with a predictable orientation of the easy axis of magnetization.
This system presents a $\tau_0=$ 1.2$\times10^{-7}$ s and $U\mathrm{eff}=$ 30.24 K. At both temperatures we employed an "on" field ($B_j$) of 20$~\mathrm{mT}$. The "off" field was fitted so that the average field cancels over a long experiment.}
In this case, the information flow is unidirectional, but both the scheme and the modelling are scalable to larger networks and to more complex information feedbacks.

\begin{figure*}[h!]
\centering
  \includegraphics[width=.93\textwidth]{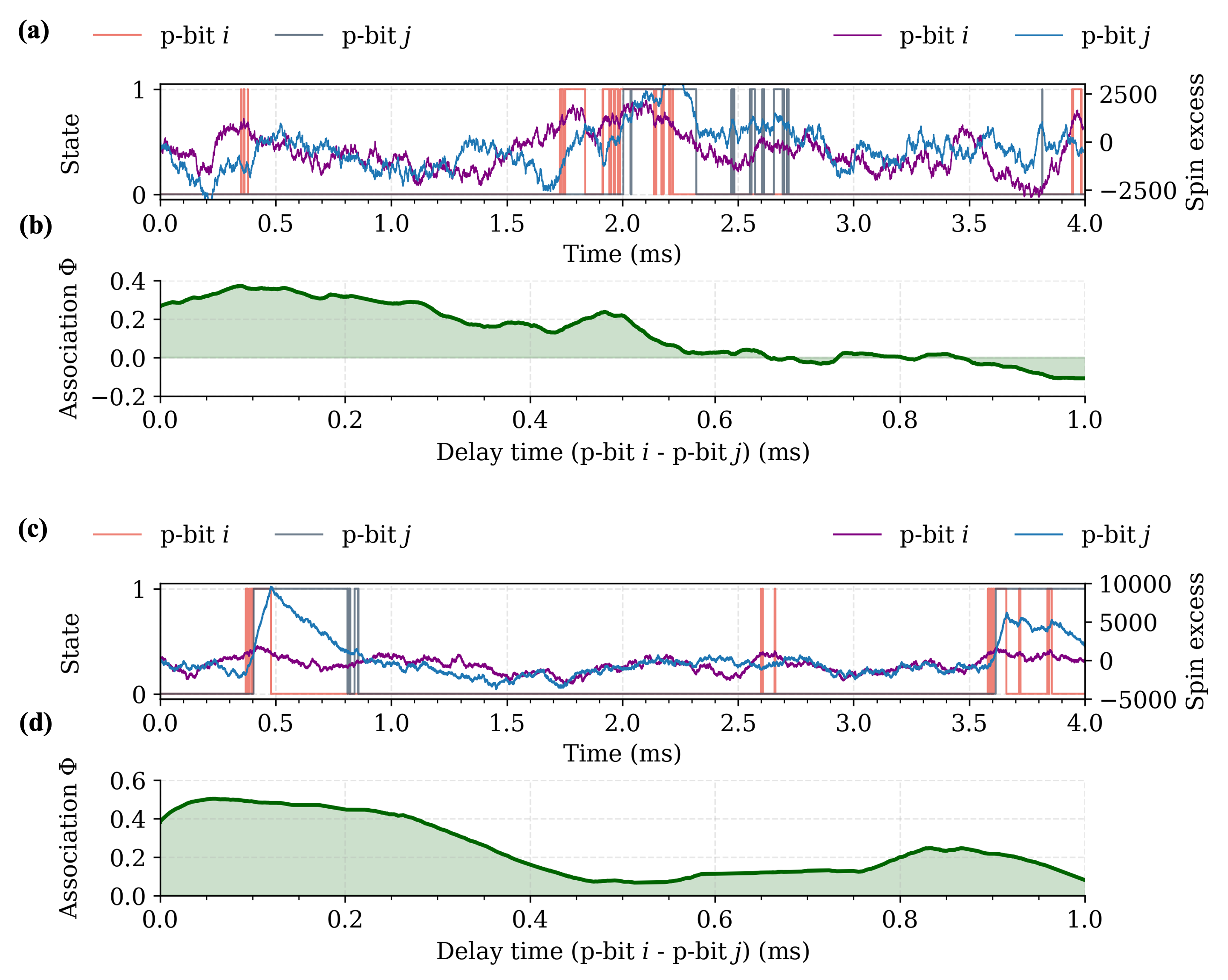}
  \caption{\textcolor{black}{Scheme for a toy network of associated p-bits based on the collective behavior of (a,b) [Dy\{$\eta^\mathrm{4}$-C\textsubscript{4}(SiMe\textsubscript{3})\textsubscript{4}\}{$\eta^\mathrm{4}$-C\textsubscript{4}(SiMe\textsubscript{3})\textsubscript{3-k}-(CH\textsubscript{2}SiMe\textsubscript{2}}]$^\mathrm{2-}$ molecules\cite{day2018} at $T=4~\mathrm{K}$ and  (c,d) quadruple decker \{[Dy(obPc)$_2$]Cd[Dy(obPc)$_2$]\} molecules\cite{katoh2012} at $T=40~\mathrm{K}$, where the state of p-bit $i$ controls a magnetic field $B=0.02~\mathrm{T}$ acting on p-bit $j$. (a,c): State of p-bit $i$ and p-bit $j$ vs time; one can appreciate that after a period where p-bit $i$ takes the value 1 (or 0), p-bit $j$ often follows. (b,d) Association $\phi$ (see equation~\ref{eq:association}) comparing the state of p-bit $i$ at a certain time with state of p-bit $j$ after a certain delay time where the x-axis is the delay time. }}
  \label{fig:pbitpair}
\end{figure*}

To quantify the \textcolor{black}{association between the values of the two pbits}, we consider the four states $(0,0),(0,1),(1,0),(1,1)$ where the first and second index correspond to the state of p-bit $i$ and $j$ respectively, and $N(i,j)$ as the number of such states in a continuous run of the simulation. We define the \textcolor{black}{association $\phi$ between p-bit $i$ and p-bit $j$ as: 
\begin{equation}
    \phi=\frac{N(0,0)\cdot N(1,1)-{N(0,1)\cdot N(1,0)}}{\sqrt{N_{1*}\cdot N_{0*} \cdot N_{*0} \cdot N_{*1}}}
    \label{eq:association}
\end{equation}
with
\begin{equation}
    N_{1*}=N(1,0)+N(1,1)
\end{equation}
\begin{equation}
    N_{0*}=N(0,0)+N(0,1)
\end{equation}
\begin{equation}
    N_{*0}=N(0,0)+N(1,0)
\end{equation}
\begin{equation}
    N_{*1}=N(0,1)+N(1,1) 
\end{equation}
}

By choosing relatively cold temperatures, even moderate fields \textcolor{black}{and a moderate number of magnetic molecules per pbit are} able to achieve a \textcolor{black}{strong association between the p-bits (see Figure \ref{fig:pbitpair}).}

\textcolor{black}{We also calculated the delayed association by considering the state of p-bit $i$ at a certain time $t_0$ and the state of p-bit $j$ at a later time $t_0+d$.}
What one observes \textcolor{black}{both here and also for a single spin as a p-bit in Supplementary} Figure \textcolor{black}{S6} is a base instantaneous correlation, that is \textcolor{black}{initially} enhanced by allowing a \textcolor{black}{short} delay time between the stimulus and the response, to accommodate the intrinsic dynamics, and 
which is then gradually lost for longer delay times, until only noise  \textcolor{black}{around $\phi\simeq0$} is observed, i.e.  \textcolor{black}{mathematically} independent values of p-bits $i$ and $j$. \textcolor{black}{One can also appreciate that the same magnetic field $B=0.02$~T that has an adequate effect at 40 K is excessive at 4 K, with the spin excess of p-bit $j$ overshooting by a wide margin and taking a long time to recover when the state of p-bit $i$ changes.}


How could such a p-bit network be implemented in practice? We revealed a possibility in Figure \ref{fig:bitpbitqubit}(b). Each p-bit would sit on a clover-type Hall effect probe, where a current is passed in one direction and a Hall voltage is measured in the perpendicular direction, so that the sign of the \textcolor{black}{Hall} voltage is \textcolor{black}{controlled by} the sign of the magnetic moment in the p-bit. This \textcolor{black}{output} voltage signal then needs to be amplified and converted into a current. \textcolor{black}{This current, passing through an electromagnet, generates a magnetic field $B$ that acts as input (red arrow) to another p-bit, with a sign that depends on the value of the first p-bit.} A magnetic field $B=0.02~\mathrm{T}$ is realistic if the response time is wanted to be relatively short and able to react to the change in the state of the p-bits. 
Note the crucial effect of the magnetic field here, \textcolor{black}{creating a} Zeeman splitting \textcolor{black}{strong enough to} allow our microscopic simulator to pick up correlations at a \textcolor{black}{temperature of 40 K} instead of requiring an extreme cryogenic \textcolor{black}{temperature of 20 mK} like in the previous section.

%

\textcolor{black}{What we extract from these simulations is that at 4 K  and 40 K even with a moderate number of $M=10^6$ magnetic molecules one can expect a very high association between p-bits communicated by reasonable magnetic fields ($\simeq 0.02$ T) and times ($\simeq 1$ ms). Simulations would be more challenging at 300 K, higher number of molecules and slower nanomagnets would be required, but one can expect these results to carry over to room temperature.}
In the next section 
\textcolor{black}{we discuss} the chemical design of the desired characteristic relaxation times.



\begin{figure*}[th]
\centering
  \includegraphics[width=\textwidth]{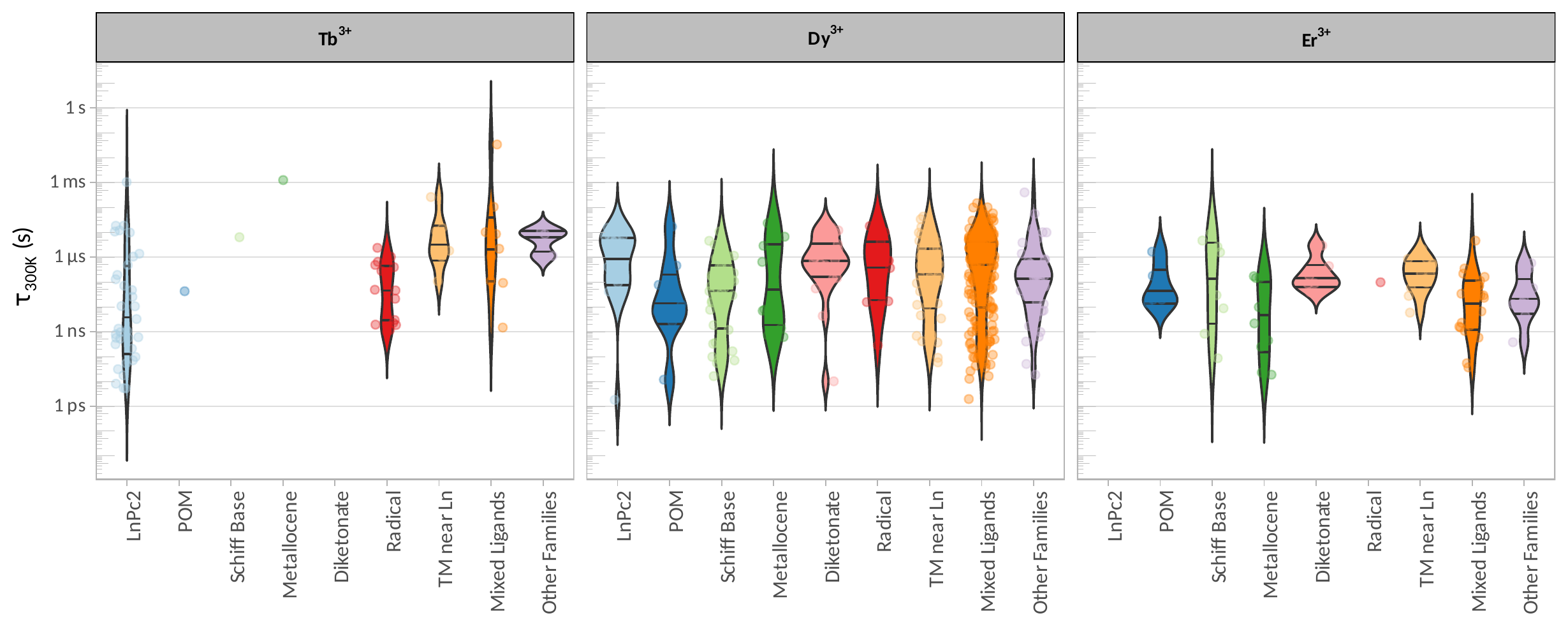}
  \includegraphics[width=\textwidth]{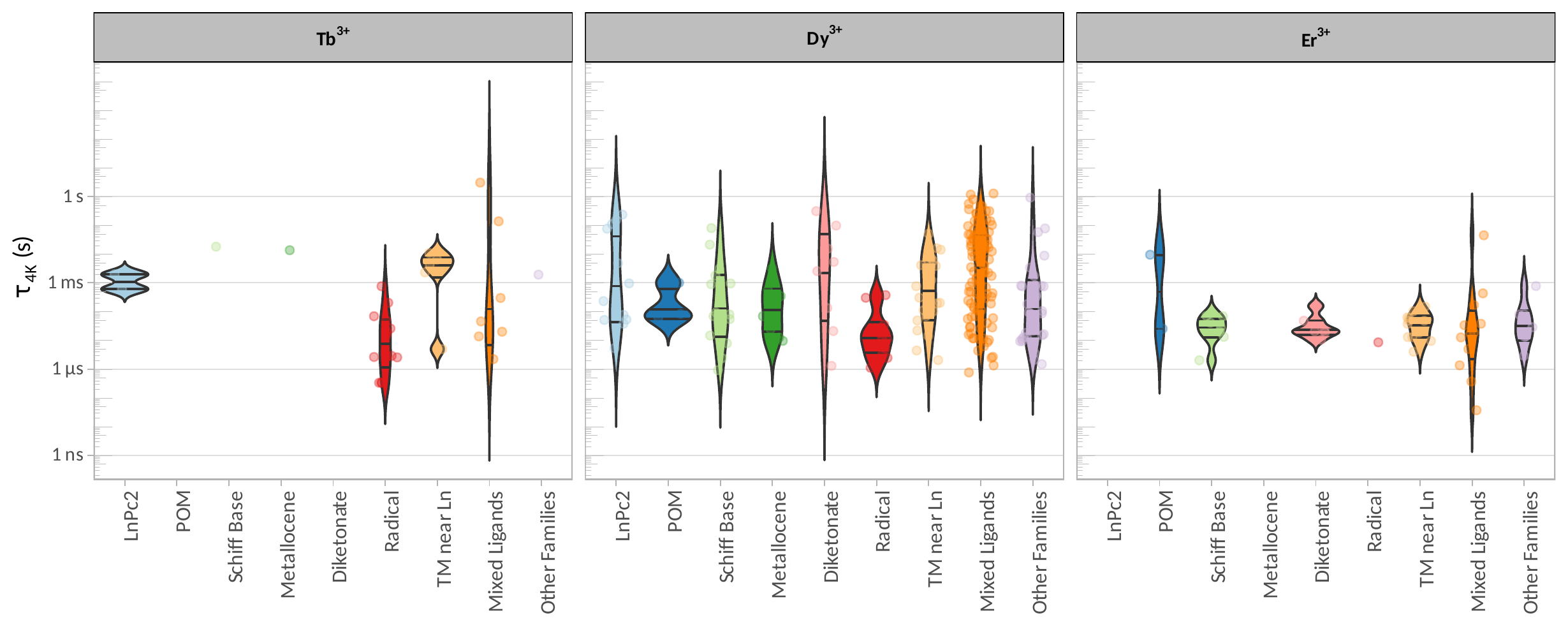}
  \caption{Estimated relaxation times at room temperature ($\tau_{300\mathrm{K}}$) \textcolor{black}{and at 4 K ($\tau_{4\mathrm{K}}$) for molecular nanomagnets based on} Tb$^{3+}$, Dy$^{3+}$, Er$^{3+}$ \textcolor{black}{represented as violin plots. Relaxation times are categorised} for the main chemical families considered in the SIMDAVIS dataset.\cite{Duan2022} \textcolor{black}{The Néel-Arrhenius equation was used to estimate this parameter. The violin plot outlines illustrate kernel probability density, i.e. the width of the coloured area represents the proportion of the data located there.} Further complementary representations employing other categorisation criteria are available in the SI.}
  \label{fig:screening}
\end{figure*}

\subsection{Experimental constraints and SIMDAVIS dataset screening for p-bit behaviour}



The most convenient setup, from the point of view of the technical requirements, involves working at room temperature (RT). This should also be a short-term goal for scalable technologies, especially if one takes energy consumption into consideration.
However, this seems yet out of the question with molecular nanomagnets employed either in traditional computation as bits or in quantum computation as spin qubits. For both applications, stochastic spin flips that happen at non-cryogenic temperatures are detrimental and need to be avoided. Then, working at RT would require long relaxation/coherence times, and this in turn would require a combination of an extremely high relaxation barrier and extremely weak spin-phonon coupling. These demands \textcolor{black}{have} not been achieved yet with molecular nanomagnets and it remains unclear whether they can ever be achieved. 

The situation is completely different for p-bits, where stochastic spin flips are expected and a mandatory part of the information processing. In this case \textcolor{black}{the electronic equipment needs to operate at least as fast as the relaxation time of the p-bit. If the equipment operates at significantly lower  frequency than the p-bit, only an average signal would be recorded and the stochasticity would be lost. Very slow p-bits are also not practical, since the overall operating speed of the device will be determined by the slowest among the p-bit relaxation time and the electronic equipment. The ideal is} an approximate match in terms of speed between the p-bits and the read-write capabilities of the electronic equipment one is using. If e.g. the electronics are able to read and write at 1~kHz, then one wants to work with a molecular nanomagnet presenting a relaxation time \textcolor{black}{$\tau\simeq 1$ ms at the working temperature}. 



The translation of these requirements to the parameters characterizing the spin dynamics of molecular nanomagnets is surprising. One does not actually need or desire particularly high values of the effective barrier $U_\mathrm{eff}$, which is, shockingly for the molecular nanomagnet community, not \textcolor{black}{necessarily} a key parameter. If one operates at 300~K, which is often close to the high temperature limit, spin relaxation time is given in a good approximation by \textcolor{black}{the attempt time,} $\tau_0$. \textcolor{black}{Assuming, that is, that Raman relaxation does not overtake the Orbach process (but why should it, given that Orbach's thermal dependence is stronger?).}  In that scenario,  \textcolor{black}{the goal is} to find highly processable molecular systems with a value of $\tau_0$ \textcolor{black}{that is compatible with} the available electronics\textcolor{black}{. Should one need to operate at lower temperatures, the goal will be to find the molecule presenting a response time in the desired time range.}

We \textcolor{black}{estimated} $\tau_\mathrm{300K}$ as the characteristic relaxation time according to \textcolor{black}{the Néel-Arrhenius} equation at 300~K for the 612 samples for which there is $U_\mathrm{eff}$ \textcolor{black}{and}~$\tau_0$ information in the SIMDAVIS dataset \textcolor{black}{(see Figure \ref{fig:screening} upper panel)}.\cite{Duan2022} \textcolor{black}{Here we} focus on the two most relevant chemical categories to classify molecular nanomagnets, namely the lanthanoid ion and the chemical family. We \textcolor{black}{show} with violin plots the fact that, while there is a large dispersion in $\tau_\mathrm{300K}$ values there is also an influence from the metal ion, from the chemical family, and from their combinations. \textcolor{black}{Still, it is evidenced that the three most popular ions for molecular nanomagnet design generally require very fast operating times to act as p-bits at room temperature.} 
\textcolor{black}{The analogous plot for $\tau_\mathrm{4K}$ can be seen in the Figure \ref{fig:screening} lower panel. In this case we discarded all samples with $U_\mathrm{eff}>50$~K to avoid plotting impractically (and in many cases, unrealistically) long relaxation times.}

A thorough study of the influence of different chemical variables on $\tau_\mathrm{300K}$ \textcolor{black}{and $\tau_\mathrm{4K}$} is available in the \textcolor{black}{Supplementary Information Sections S3 and S4. We found that} for the chemical design of molecular spin p-bits \textcolor{black}{that are slow enough to be operable at reasonable frequencies and }at room temperature, one could lean towards coordination spheres consisting of nine donor atoms in a mixture of Oxygen and Nitrogen, stemming from 3 ligands, \textcolor{black}{and probably choosing a} lanthanide ion \textcolor{black}{other than} Dy$^{3+}$, \textcolor{black}{Tb$^{3+}$ or Er$^{3+}$}.  Embodying each p-bit in larger ensembles of molecular spins would further facilitate working at manageable speeds and temperatures (see Supplementary Information Section S2).


As discussed in the previous section and depicted in Figure \ref{fig:pbitpair}(b), achieving a significant correlation between single spins at moderate magnetic fields such as $B=0.2$~T requires relatively low temperatures as in the order of 40 K. However, if one desires to work at RT, the signal will be too weak, but this is true in any case where one employs a single molecule. Much stronger signals, allowing one to pick up much weaker correlations, will arise from a large collection of molecular nanomagnets acting as a single p-bit. Indeed, note that Boltzmann populations are governed by the ratio between energy difference and working temperature. This means that, if one were to use similar sensitivity and sample sizes as in the case of commercial ac susceptometry, a $B=0.02$~T (3 orders of magnitude higher than regular ac field) would allow obtaining sufficient population difference at temperatures up to 3 orders of magnitude higher than in an ac experiment. In this case, would mean being able to quantify correlation at temperatures much higher than 300~K. Ultimately, the practical temperature limit and sample requirement will depend on the experimental implementation.

\section{Conclusions}

We \textcolor{black}{employed} STOSS, a microscopic simulator code for spin p-bits based on molecular nanomagnets, to explore this potential emergent technological application, distinct from conventional bits and from quantum qubits. \textcolor{black}{To test the software,} we reproduced the most characteristic macroscopic magnetic dynamics of molecular nanomagnets, namely magnetisation relaxation and in-phase, out-of-phase susceptometry, by simulating the individual states in a collective of spin p-bits. \textcolor{black}{With the help of STOSS,} we found that under realistic conditions it should be technologically possible to build small networks of p-bits based on molecular nanomagnets. \textcolor{black}{We found an inverse correlation between room temperature p-bit performance and molecular nanomagnets performance. This is unsurprising considering (a) $U_\mathrm{eff}$ is a good predictor of molecular nanomagnet behavior and (b) the approximate proportionality $\tau_0^{-1}\propto U_\mathrm{eff}^3$.\cite{Duan2022,abragam2012}} In this aspect, these Low Barrier Nanomagnets are proposed to build the basic neuronal units of Artificial Stochastic Neural Networks. Since molecular single ion magnets and spin qubits have been explored for decades,\cite{Chilton2022} this new potential application starts already with a wealth of molecular diversity, theoretical tools and systematically organised data facilitating statistical and machine learning studies.\cite{Duan2022,Nguyen2022,Lunghi2022} This gives magnetic molecules the potential to be screened, chosen, and adapted to the different technological constraints of different possible implementations. 
Here one needs to acknowledge that despite the continuous efficiency improvements in the information and communications technology sector, it has been shown that there is an increasing weight of computing-related emission in the climate crisis, a problem which cannot be addressed merely by technical advances based on futuristic modes of information processing.\cite{Knowles2021} However, we believe that research on molecular nanomagnets as probabilistic information carriers is of sufficient fundamental interest to merit further exploration.

\section*{Data availability}
All custom data generated and employed for this study are available at  \href{https://github.com/gerlizg/STOSS}{https://github.com/gerlizg/STOSS}.

\textcolor{black}{\section*{Code availability}
The code named STOSS (for STOchastic Spin Simulator) is available at \href{https://github.com/gerlizg/STOSS}{https://github.com/gerlizg/STOSS}. The instructions to reproduce all the graphic results are in the Supporting Information Section S5 and at \href{https://github.com/gerlizg/STOSS}{https://github.com/gerlizg/STOSS}.}

\section*{Author Contributions}
GMGF: (for all except 3.4) investigation, methodology, software and validation; (for 3.1 and 3.3) formal analysis; writing.
SGS: (for 3.2) software and validation; (for all except 3.4) visualisation.
ZH: (for 3.2) formal analysis; writing.
LER: conceptualisation; writing; (for 3.4) software, investigation, formal analysis and visualisation.
SCS: conceptualisation, supervision and writing.
AGA: conceptualisation and supervision, project administration, funding acquisition and writing.

\section*{Conflicts of interest}
There are no conflicts to declare

\section*{Acknowledgements}
L.E.R., S.G.S., S.C.S. and A.G.A. have been supported by the COST Action MolSpin on Molecular Spintronics (Project 15128). A.G.A. and Z.H. thank funding from European Union (EU) Programme Horizon 2020 (FATMOLS project), and A.G.A. also thanks Generalitat Valenciana (GVA) CIDEGENT/2021/018 grant. S.C.S acknowledges funding from the European Research Council (ERC) in EU Horizon 2020 grant agreement ERC-2017-AdG-788222 “MOL2D”. S.C.S., S.G.S. and L.E.R. gratefully acknowledge support from the Spanish Ministerio de Ciencia e Innovación (MICINN) grant PID2020-117264GB-I00, and A.G.A. thanks MICINN PID2020-117177GB-I00 grant (both MICINN grants co-financed by FEDER funds). S.G.S. acknowledges MICINN PRE2018-083350 grant related to MINECO CTQ2017-89528-P project. S.C.S. acknowledges Excellence Unit María de Maeztu CEX2019-000919-M funding. L.E.R. gratefully acknowledges support from GVA PROMETEO/2019/066. This study forms part of the Quantum Communication program and was supported by MICINN with funding from European Union NextGenerationEU (PRTR-C17.I1) and by GVA (QMol COMCUANTICA/010).

\clearpage










\balance


\bibliography{pbits} 
\bibliographystyle{pbits} 

\end{document}


\title{Supplementary Information of Lanthanide molecular nanomagnets as probabilistic bits}
\author{Gerliz M. Gutiérrez-Finol}
\affiliation{Instituto de Ciencia Molecular (ICMol), Universitat de Val\`encia, Paterna, Spain}
\author{Silvia Giménez-Santamarina}
\affiliation{Instituto de Ciencia Molecular (ICMol), Universitat de Val\`encia, Paterna, Spain}
\author{Lorena E. Rosaleny}
\affiliation{Instituto de Ciencia Molecular (ICMol), Universitat de Val\`encia, Paterna, Spain}
\author{Ziqi Hu}
\affiliation{Instituto de Ciencia Molecular (ICMol), Universitat de Val\`encia, Paterna, Spain}
\author{Salvador Cardona-Serra}
\affiliation{Instituto de Ciencia Molecular (ICMol), Universitat de Val\`encia, Paterna, Spain}
\author{Alejandro Gaita-Ariño}
\affiliation{Instituto de Ciencia Molecular (ICMol), Universitat de Val\`encia, Paterna, Spain}

\date{\today}

\maketitle

\tableofcontents

\newpage

\section{Microscopic spin p-bit modelling of macroscopic magnetization dynamics: isolated spins at constant field}

\textcolor{black}{We employ a (discrete time) Markov Chain Monte Carlo model for  each of the $N$ independent particles (in this case, effective spins $S=1/2$). The relative Markov chain probabilities for the spin flips between ground and excited spin states correspond to the relative Boltzmann populations of the two effective spin states $M_S=+1/2$, $M_S=-1/2$. Since each computational step has an associated natural time duration, the model allows one to follow $N$ independent time trajectories.}

In a first stage, our model intends to reproduce the macroscopic behavior of a \textcolor{black}{collection} of spins\textcolor{black}{, evolving at a constant external magnetic field. For this goal, the model employs relaxation} parameters obtained from fitted ac data of single molecule magnets.

\subsection{In the absence of magnetic field: recovering the overall relaxation exponent from individual stochastic spin flips}

The simplest case represented by our time-dependent model contains a chosen number of spin centers ($N$) that fluctuate during  a certain period of time ($t$) between two spin states (\textcolor{black}{labelled for simplicity} $\ket{0}$ and $\ket{1}$) that are degenerate for the whole duration of the calculation. These spin states correspond to the two opposite orientations of a single effective $S=1/2$ spin, which in the case of molecular nanomagnets is employed to describe the ground doublet in the absence of an external magnetic field. \textcolor{black}{In practice, these generally correspond to spin doublets such as $M_J=\pm\frac{15}{2}$, $M_J=\pm6$, $\ldots$}

We \textcolor{black}{are in the conceptual framework of an exponential magnetization decay of a collection of spins that is initially out of equilibrium and aiming to model the trajectory of each spin by calculating its probability for flipping, for a short time interval which corresponds to a  "time step" of length $t_s$. We will employ this spin flip probability to construct the Markov chain probabilities. Consequently, to evaluate this spin flip probability we implement a function of a power series, where the constants within the series take the value of 0, which is a special case of the Taylor expansion; this expression is known as the MacLaurin expansion which for a exponential function is described as the following equation:}

\begin{equation}\label{maclaurin}
e^x = 1 + \sum_{i=1}^{n}\frac{ x_{i}}{i!}
\end{equation}

In the limit of short $t_s$ \textcolor{black}{(here, small $x$), $e^x$ is well approximated by the linear term} $e^x=1+x$, meaning that for an exponential decay with $x<0$, the slope is just $x$\textcolor{black}{: the exponential function decays linearly at short times}. \textcolor{black} {This last idea is used for illustration}, but in the program we include the first 100 terms of the MacLaurin expansion. 

Let us examine what this means for a system with $N$ spins that are initially fully polarized "up" and which are decaying exponentially with a characteristic time $\tau$ down to a limit, at long times, of $N/2$ "up" and $N/2$ "down". The equation governing the number of spins pointing "up" $N_{\mathrm{up}}$ at a time $t$ is
\begin{equation}
    N_{\mathrm{up}}(t)=(N/2)+(N/2)\cdot e^{-t/\tau}
\end{equation}
or, in good approximation at short times \textcolor{black}{employing the MacLaurin expansion cut to the linear term with $x=-t/\tau$},
\begin{equation}
    N_{\mathrm{up}}(t)=(N/2)+(N/2)\cdot (1-{t/\tau})
\end{equation}
{\color{red}
\begin{equation}
    N_{\mathrm{up}}(t)=(N\cdot (1-{t/(2\tau)})
\end{equation}}
this means that, from the total of $N$ spins the fraction that decays after a time step $t_s$ is $N\cdot\frac{t_s}{2\cdot\tau}$. \textcolor{black}{More generally, this macroscopic decay fraction $\frac{t_s}{2\cdot\tau}$ per time step $t_s$ can be equated to a microscopic spin flip probability $\frac{t_s}{2\cdot\tau}$ or $100\cdot\frac{t_s}{2\cdot\tau}\%$}.

In other words, 
\textcolor{black}{
one can determine the probability corresponding to a desired time step:}
{\color{red}
\begin{equation}\label{eqtime}
    p = \frac{t_s}{2\cdot\tau}
\end{equation}
}

\textcolor{black}{Conversely, }
if we want to employ time steps $t_s$ with a predefined, constant and small probability $p$ for a spin flip, the time step duration of the simulation is unequivocally defined by using the relaxation time of the magnetic moment ($\tau$) for a particular molecule and the desired probability $p$ by using the relation eq. \ref{eqtime}
{\color{red}
\begin{equation}\label{eqtime}
    t_s = 2\cdot\tau \cdot p
\end{equation}
}

According to this, the model applies, for time steps of $\tau/50$, equal probabilities of 1\% for each spin of experiencing a spin flip either from $\ket{0}$ to $\ket{1}$ or from $\ket{1}$ to $\ket{0}$. The model gives essentially the same result if time steps are much shorter, e.g. $\tau/5000$, and the probabilities for spin flips are proportionally smaller e.g. 0.001\%. \textcolor{black}{Again, here we work with simple numbers for illustration but the code works with the first 100 terms of the MacLaurin expansion, not just with the first one.}


As seen in the main text, the minimum least-square fitting of \textcolor{black}{this kind of simulation} offers a value of $\tau$ which is compatible with the initial value used for the prediction. However, as discussed below the correspondence with the experimental behavior becomes worse at very low temperatures. 

In principle, \textcolor{black}{the MacLaurin expansion} would allow us to resort to longer timesteps, saving computational time. However, note that our aim is not merely the effective simulation of the macroscopic behavior but also the correct simulation of the individual probabilities. For this, we want to observe each individual spin flip, since each such event is significant in a p-bit network. This requires employing short-time steps to reduce the number of missed spin flips.

As a limitation of our model and perspective for future improvements, note that simulating the collective behavior by a sum of independent individual evolutions, in terms of spin relaxation over time, is possible only assuming an exact exponential decay. In the exponential function, the slope $dy/dx$ at any point $x_0$ is proportional to the value $y$ at $x_0$, and this is essential for us to be allowed to apply the collective equation to the individual units. In intuitive terms, the individual spin does not need knowledge of the state of the rest of the spins to evolve according to exponential decay. The situation is fully analogous to first-order reactions in chemistry, or the decay of radioactive isotopes. 

In contrast, reproduction of stretched exponentials
\begin{equation}
    M(t) = M_\mathrm{eq} + (M_0 - M_\mathrm{eq})\mathrm{exp}[-(t/\tau)^\beta]
\label{Decay}
\end{equation}
with $\beta\neq1$, where the slope depends on how far one is from the y-axis, would require the probability for every single spin to be affected by this in some way. This is  feasible as an extension of our current methodology and indeed \textcolor{black}{it is being done to simulate experimental results} at very low temperatures \textcolor{black}{where the dipolar coupling cannot be neglected}.

Simulated curves in the main text are scaled for comparison with experimental values as every first value is normalized to the first point in the measurements extracted from Guo et al.

\subsubsection{Relation with p-bit concepts: dwell times and lock signals}

To characterize p-bits it is usual to measure their characteristic dwell times $t_{\mathrm{spin}\uparrow}$,$t_{\mathrm{spin}\downarrow}$, i.e., the set of individual continuous time periods where the p-bit preserves the same value. 
In our case, the evolution of the microscopic dwell times $t_{\mathrm{spin}\uparrow}$ and $t_{\mathrm{spin}\downarrow}$ has the same mathematical form as the overall magnetisation decay, so it serves no practical purpose to plot them separately. Similarly, for p-bits it is important to quantify the maximum ``lock'' signal $R_{lock}$, i.e. the maximum achievable contrast between the two extreme states 
that can be obtained by applying an input bias to ``lock'' one of the states. For spin p-bits, the bias will be either a magnetic field or a spin-polarized current, and the signal will be the average magnetisation of the p-bit. This can be estimated as a "dwell time ratio", or ratio between the total time spent in spin states "up" and "down": 
\begin{equation}
    R_{lock}(T)=\frac{\sum{t_{\mathrm{spin}\uparrow}}}{\sum{t_{\mathrm{spin}\downarrow}}}
\end{equation} 
This is equivalent to the Boltzmann distribution as simulated in the next section.

\subsection{In the presence of magnetic field: recovering the Boltzmann distribution by biasing the transition probabilities}

In the second stage, we start from the setup studied in the previous stage and lift the condition of spin degeneracy. In this more complex case, the effect of an external magnetic field is added, by modifying the probabilities of observing a transition between magnetic states depending on the starting state. Thus,
\begin{equation}\label{eqprob}
    P(\ket{0} \mathrm{to} \ket{1}) \neq P(\ket{1} \mathrm{to} \ket{0})
\end{equation}
which theoretically corresponds to the Zeeman effect by stabilizing and destabilizing either spin state respectively \textcolor{black}{and reaching different equilibrium populations}. In this case, we have imposed a few external conditions to allow the system to be directly comparable to the previous case. The most important of these are: \\

a)	$P$($\ket{0}$ to $\ket{1}$) and $P$($\ket{1}$ to $\ket{0}$) are modified to match the ratio of the expected population ratio within a Boltzmann distribution. \\

b)	The sum of both probabilities is kept constant (2$\%$ in the case of timesteps of length $\tau/50$). \\

c) The applied magnetic field could be constant or changeable (following a cosine function) in the course of the experiment. \\

Once the magnetic field is fixed, the energy must be calculated through the expression:

\begin{equation} \label{Hamiltonian}
E_{Zeeman} = g \mu_B B M_J
\end{equation}

For calculating the probabilities of the spin being in a specific Zeeman state, we use Boltzmann distribution equation which input parameters are temperature and energy. A series of random numbers are then obtained and, at each time step, each random number is compared to the state's probability of spin flip (which means changing from the lower state energy to the higher one, and vice versa). In addition, it is important to notice that due to the split of these spin states under a constant external field, the flip probability is unique for each level and proportional to the energy difference. This is illustrated in Figure S1. Here the populations of the excited state is set to 100$\%$ at t=0  the system is left to thermalize under the effect of an opposite direction external magnetic field ($B$) in the range between 0T and 5T. For each curve, the obtained population of spin levels at thermal equilibrium recovers Boltzmann distribution.

\begin{figure*}[h]\ContinuedFloat
\centering
\includegraphics[width=0.95\columnwidth, valign=t]{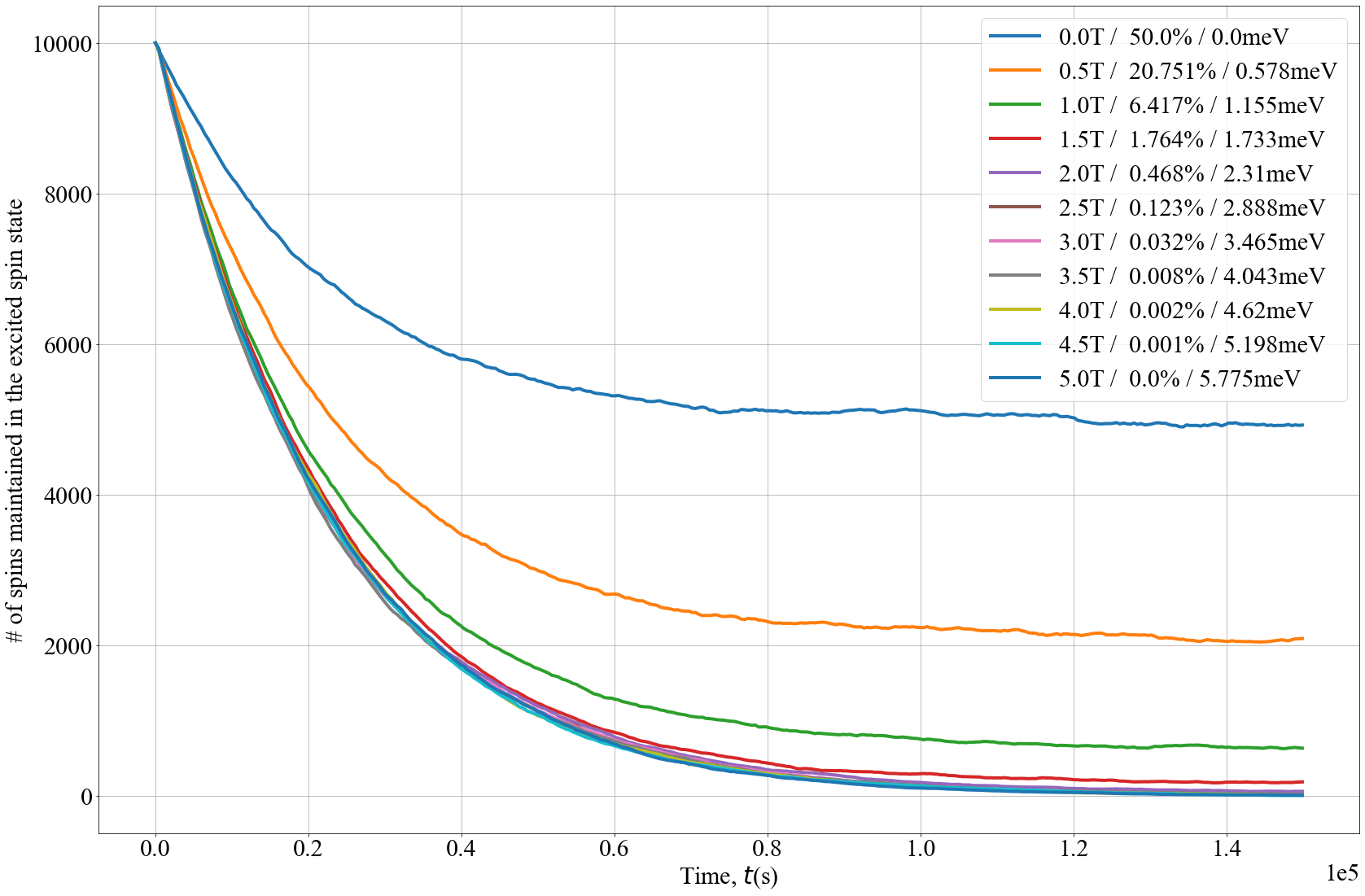}
\caption{Relaxation curves of a two-level spin system assuming that a constant magnetic field is applied contrary to the starting spin alignment. Each curve represents the spin evolution at 5K. The legend corresponds to: Magnetic Field (Tesla)/ Population of the Excited State (total percent) / Energy Difference (meV).}
\label{figSI:fit4}
\end{figure*}

\clearpage

\begin{figure*}[h]
\includegraphics[width=0.3\columnwidth]{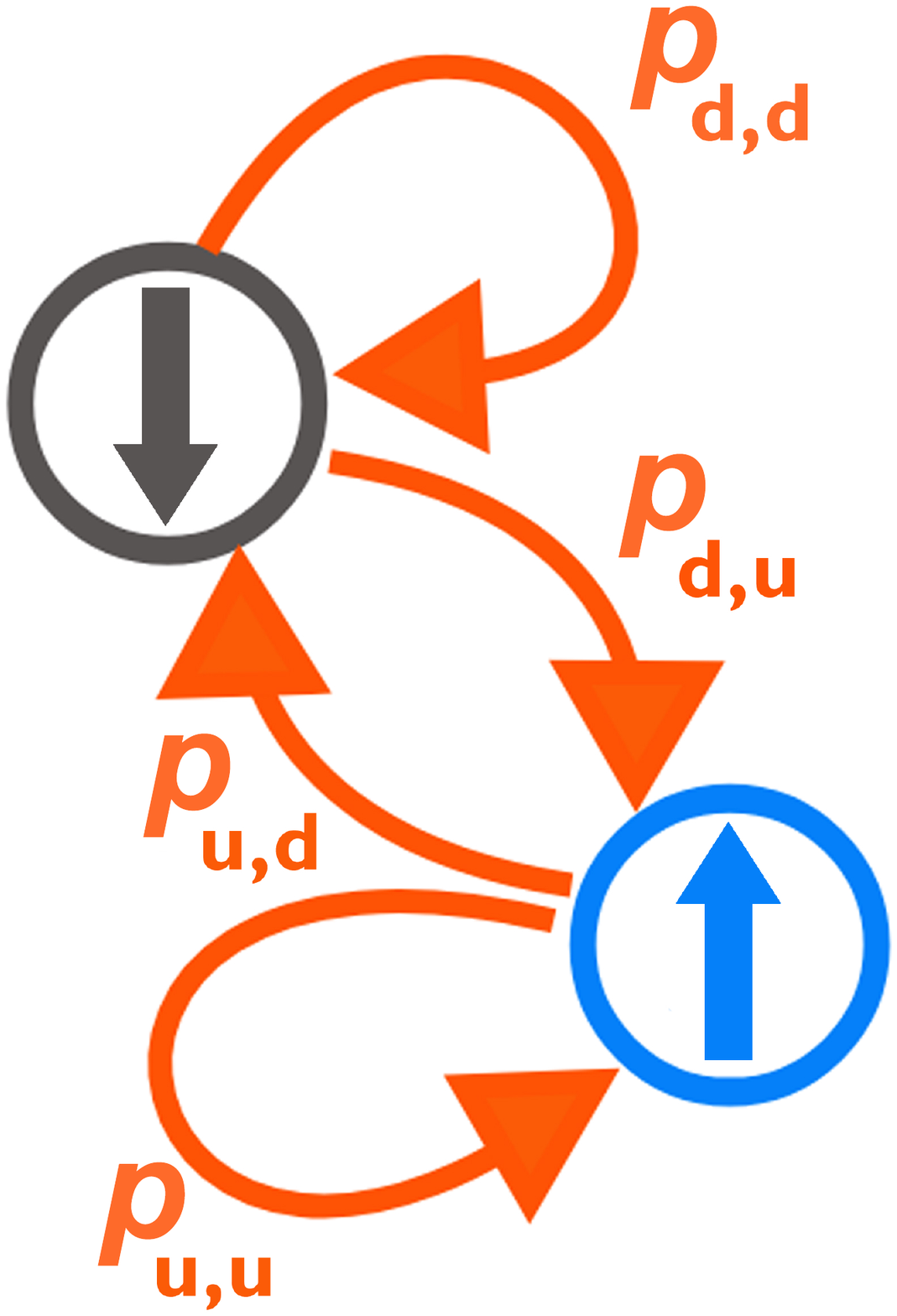}
\caption{\textcolor{black}{Markov Chain diagram with two states: spin down (grey) and spin up (blue). At each time step, depending on the state, one finds up to four distinct probabilities, to keep the state ($p_\mathrm{d,d}$,$p_\mathrm{u,u}$), or to flip the state ($p_\mathrm{d,u}$,$p_\mathrm{u,d}$). Since at each step each spin needs to either flip or not, $(p_{d,d}+p_{d,u})=1$ and $(p_{u,d}+p_{u,u})=1$.} }
\label{figSI:Markovkate}
\end{figure*}

\textcolor{black}{
\subsection{STOSS: Stochastic Spin Simulator}
}
\textcolor{black}{
All the models in the present work have been developed and implemented in the program named STOSS (STOchastic Spin Simulator), freely available in the repository: \href{https://github.com/gerlizg/STOSS}{https://github.com/gerlizg/STOSS}. 
STOSS is based on a Markov Chain Monte Carlo algorithm, where Markov probabilities of the stochastic behavior of a spin at each time step ($p_{d,d}$, $p_{d,u}$, $p_{u,d}$ and $p_{u,u}$, see Figure~\ref{figSI:Markovkate}) are calculated as detailed below by employing the Boltzmann distribution of $N$ particles at a given time, and, optionally in the presence of a magnetic field.}
 
\textcolor{black}{Two crucial features in this particular case are (a) we simulate independent particles, meaning we work with $N$ identical and independent Markov chains and (b) each calculation step has an associated natural time in real time units, which is taken into consideration for the calculation of the transition probabilities. }

\textcolor{black}{The spin flip probabilities $p_{d,d}$, $p_{d,u}$, $p_{u,d}$ and $p_{u,u}$ are calculated by considering that the sum $(p_{d,u}+p_{u,u})$ corresponds to the total spin flip probability calculated by using the Taylor expansion as in eq. \ref{maclaurin} and that the ratio $\frac{p_{d,u}}{p_{u,d}}$ is defined by the ratio of populations calculated as the Boltzmann distribution at a given temperature and magnetic field. $p_{d,d}$ and $p_{u,u}$ are obtained simply by difference since $p_{d,d}+p_{d,u}=1$ and $p_{u,d}+p_{u,u}=1$. }

\begin{figure*}[h]
\includegraphics[width=1\textwidth]{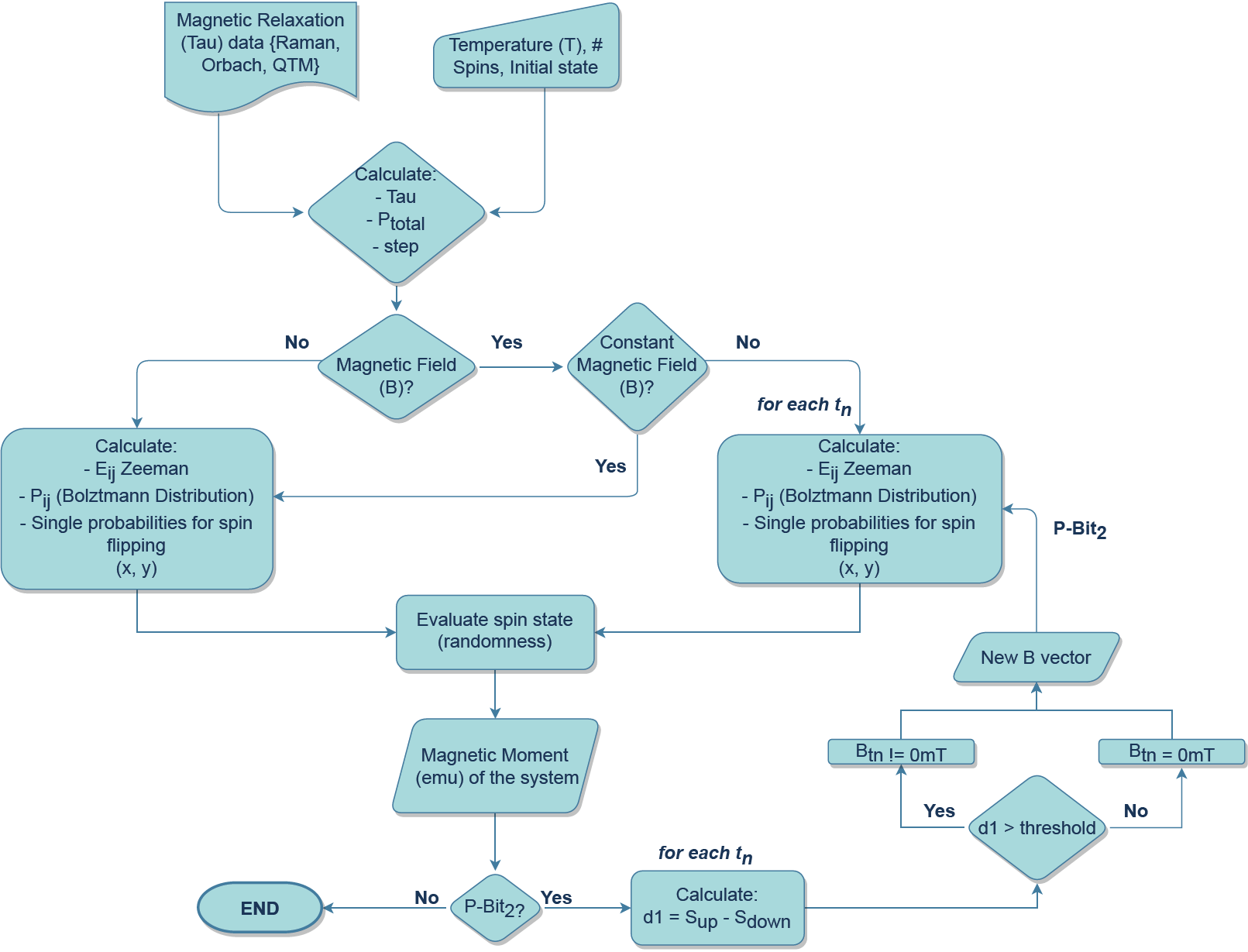}
\caption{\textcolor{black}{Flow chart of STOSS code.}}
\label{figSI:flow}
\end{figure*}

\textcolor{black}{STOSS is parametric and thus empirical. The dynamics are estimated relying on relaxation parameters that have been determined experimentally, in this case from the Single Ion Magnet [(Cp$^\mathrm{iPr5}$)Dy(Cp$^\mathrm{*}$)]$^+$ in reference \cite{Guo2018} 
More specifically, crucial parameters employed by STOSS are $U_{eff}$, $\tau_0$, $\tau_{QTM}$ and Raman parameters $C$ and $n$.}
\textcolor{black}{A detailed scheme for the spin simulator STOSS is shown in 
 \ref{figSI:flow}}

\textcolor{black}{STOSS has been employed for the rationalization of the magnetization dynamics -in particular magnetization relaxation and magnetic hysteresis- of an endohedral metallofullerene.}[Hu et al, submitted]

\clearpage


\section{Modelling dynamically driven spin p-bits}

Once the modelling is confirmed to offer the expected results for well-known cases, we will apply this methodology to a more complex several p-bits coupling experiment. In this further step, closer to the experimental simulation, we allow the p-bit system to couple with an oscillating system, being this (A) an external ac field or (B) another p-bit.

\subsection{Synchronization with a periodic drive: in-phase vs out-of-phase ac susceptometry}

In a first step we run the program with only a single p-bit under the influence of a sinusoidal ac magnetic field. In this case, the number of variables increases notably:
\begin{itemize}
    \item[-] Molecular relaxation parameters (which, following equation (1) in the main text, govern $\tau$).
    \item[-] Temperature, which combined with the relaxation parameters also controls $\tau$ but additionally governs the equilibrium Boltzmann distribution. Which in the model means the relative probabilities of spin flip depending on whether the spin is parallel or antiparallel to the external field.
    \item[-] AC magnetic field amplitude, which, combined with the temperature, controls the Boltzmann distribution and relative spin flip probabilities; note that this determines how many statistics one needs to detect an ac signal, whether in-phase or out-of phase.
    \item[-] AC frequency, which, combined with the $\tau$ resulting from the relaxation parameters and the operating temperature, results in the magnetic response.
    \item[-] Total size of the system (total number of spins $N$), which, when large enough, allows one to detect subtler magnetic responses due to a stronger statistical power.
\end{itemize}





\begin{figure*}[h]
\includegraphics[width=1\textwidth]{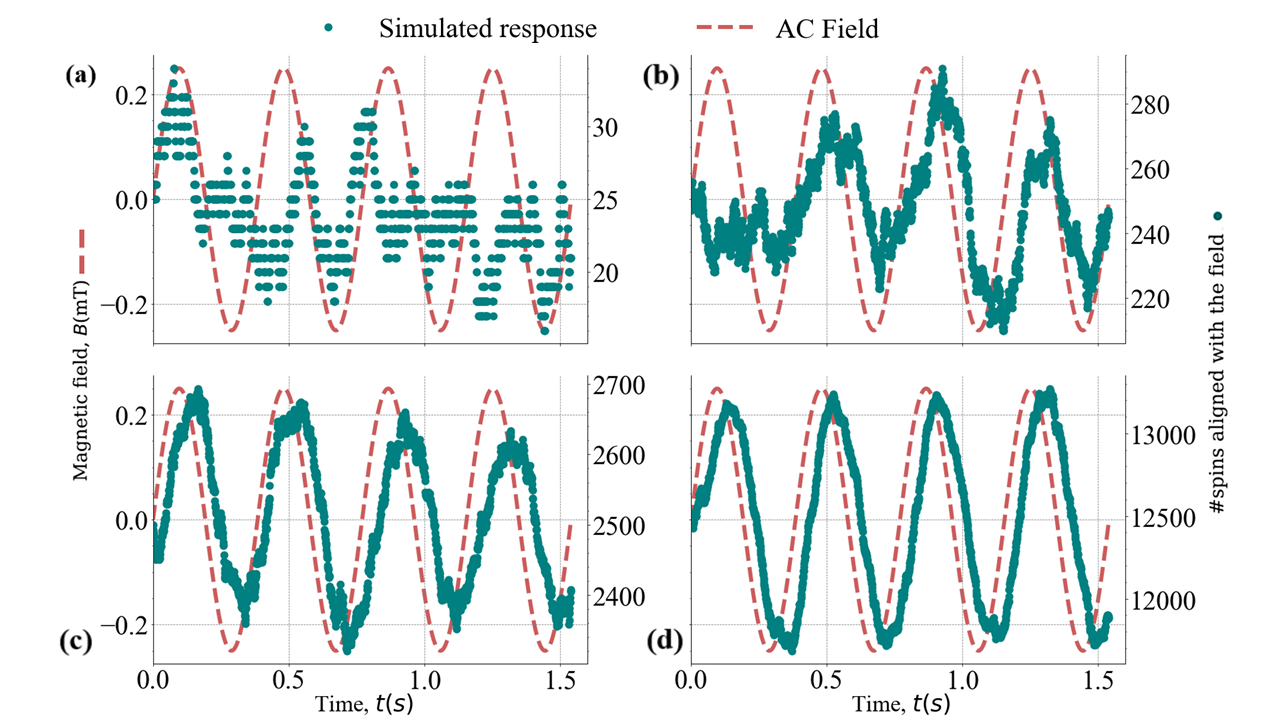}
\caption{\textcolor{black}{Periodically driven p-bits behave exactly as spins in an ac susceptometry experiment calculated time evolution of stochastic spins described by the parameters corresponding to [Dy(bath)(tcpb)\textsubscript{3}], at 20~mK and under an external ac field of amplitude $B_\mathrm{max}=0.25$~mT and frequency {$2~\mathrm{Hz}$.}}}
\label{figSI:field}
\end{figure*}

\textcolor{black} {As a first example, we simulated 4 AC cycles with a period being an order of magnitude longer than $\tau_\mathrm{QTM}$, this is shown in the main text (see Figure 3). Nevertheless, to present a more intuitive insight of the response signal from the collective behaviour of an increasing number of spins considered, we present four different simulations (see Figure \ref{figSI:field}) with N = 50 (a), 500 (b), 5000 (c), and 25000 spins (d), respectively}. 

\textcolor{black}{We ran simulations with $N=50000$ spins and an array of 57 frequencies between 60~mHz and 0.8~kHz to numerically evaluate the dynamical response of [Dy(bath)(tcpb)$_3$] at 20~mK. As indicated in the main text, tcpb = 1-(4-chlorophenyl)-4,4,4-trifluoro-1,3-butanedione and bath = 4,7-diphenyl-1,10-phenanthroline, where $\tau_\mathrm{QTM}=$ 0.067 s, $n$ = 4.90, $C = 7.80 \times 10^{-2}$ s$^{-1}$ K$^{-4.90}$, $\tau_0=$ $2.83 \times 10^{-9}$ s, and an effective energy barrier of $U_\mathrm{eff}$ = 116.07 cm$^{-1}$ (167.87 K). The results of the simulation are plotted in the extended figure~\ref{fig:fit4}.} 

We fitted every simulation to the function :
\begin{equation}\label{fitting}
    y = a \cdot \mathrm{sin}(2\pi x/b ) + c \cdot \mathrm{cos}(2\pi x/b )
\end{equation}
where \textcolor{black}{$x$ is the time, $y$ is the difference between spins pointing up and down and} $a$, $b$ and $c$ are the coefficients to be determined. For all the simulations, the median value \textcolor{black}{of the magnetic response} is normalized to zero, and accordingly, the rest of the curve is normalized. \textcolor{black}{The results of the fits are summarized on table~\ref{tab:my-table}.}

The in-phase ($\chi'$) and out-of-phase ($\chi''$) magnetic susceptibilities are fitted by the generalized Debye model:

\begin{equation}\label{inphase}
    \chi' = \chi_\mathrm{S} + (\chi_\mathrm{T} - \chi_\mathrm{S})\frac{1+(\omega \tau)^{1-\alpha}\mathrm{sin}(\pi \alpha /2)}{1+2(\omega \tau)^{1-\alpha}\mathrm{sin}(\pi \alpha /2)+(\omega \tau)^{2-2\alpha}}
\end{equation}

\begin{equation}\label{outphase}
    \chi'' = (\chi_\mathrm{T} - \chi_\mathrm{S})\frac{(\omega \tau)^{1-\alpha}\mathrm{cos}(\pi \alpha /2)}{1+2(\omega \tau)^{1-\alpha}\mathrm{sin}(\pi \alpha /2)+(\omega \tau)^{2-2\alpha}}
\end{equation}

where $\chi_\mathrm{S}$ and $\chi_\mathrm{T}$ are adiabatic and isothermal susceptibilities, respectively; $\omega$ is the frequency of the applied ac field; $\alpha$ is the parameter that accounts for the distribution of relaxation time ($\tau$). When $\alpha=0$ only one relaxation process is present, which is precisely the case in our fitting. \\

\begin{table}[h]
\centering
\caption{\textcolor{black}{Table with the resulting fitted coefficients in equation \ref{fitting} at the series of ac frequencies as seen in Figure~\ref{fig:fit4}.}}
\label{tab:my-table}
{\color{red}
\begin{tabular}{@{}cccc@{}}
\toprule
\# & $a$         & $b$         & $c$      \\ \midrule
0     &    2097.616977    &      17.101356    &      -50.301424    \\
1     &    2098.586218    &      13.680927    &      -67.373180    \\
2     &    2094.969960    &      10.935857    &      -99.650923    \\
3     &    2092.859711    &      8.762656     &     -87.817678    \\
4     &    2099.057654    &      7.007205     &     -133.152632    \\
5     &    2082.053650    &      5.597633     &     -192.559778    \\
6     &    2079.119279    &      4.482594     &     -191.764436    \\
7     &    2053.201495    &      3.587807     &     -178.925707    \\
8     &    2055.124874    &      2.874749     &     -214.460719    \\
9     &    2016.227910    &      2.297003     &     -340.746882    \\
10    &     2033.644434   &       1.838993    &      -383.266044    \\
11    &     1945.933595   &       1.471531    &      -484.556903    \\
12    &     1799.726030   &       1.175807    &      -600.148924    \\
13    &     1775.924831   &       0.943197    &      -644.375862    \\
14    &     1585.204025   &       0.753025    &      -894.470213    \\
15    &     1496.122169   &       0.604707    &      -863.210919    \\
16    &     1326.192299   &       0.485813    &      -916.467128    \\
17    &     1158.771351   &       0.391122    &      -818.364456    \\
18    &     906.208451    &      0.310944     &     -877.845110    \\
19    &     713.442211    &      0.248871     &     -808.485852    \\
20    &     551.341192    &      0.199543     &     -658.922602    \\
21    &     418.095123    &      0.160768     &     -575.683509    \\
22    &     281.278325    &      0.127345     &     -525.991499    \\
23    &     242.051024    &      0.103008     &     -443.788717    \\
24    &     182.107440    &      0.082325     &     -368.570423    \\
25    &     71.872871     &     0.064903      &    -310.586080    \\
26    &     46.330230     &     0.051786      &    -239.134850    \\
27    &     52.664461     &     0.041196      &    -190.407783    \\
28    &     48.181093     &     0.033898      &    -167.553994    \\
29    &     6.526565      &    0.026320       &   -118.327416    \\
30    &     34.463395     &     0.021685      &    -97.408877    \\
31    &     -2.710737     &     0.016799      &    -81.013154    \\
32    &     25.118267     &     0.014050      &    -69.968705    \\
33    &     33.697677     &     0.011390      &    -50.564741    \\
34    &     -23.838708    &      0.008327     &     -36.269602    \\
35    &     -7.147336     &     0.006800      &    -41.035925    \\
36    &     20.708293     &     0.005748      &    -14.225108    \\
37    &     16.051382     &     0.004388      &    -29.285808    \\
38    &     -15.309341    &      0.003222     &     -8.927968    \\
39    &     -12.380109    &      0.002600     &     -6.506677    \\
40    &     5.845430      &    0.002224       &   -14.870431    \\
41    &     12.102697     &     0.002394      &    12.488710    \\
42    &     -5.696184     &     0.001400      &    -12.577812    \\
43    &     -2.317193     &     0.001101      &    -8.627777    \\
44    &     1.700090      &    0.000920       &   -4.479264    \\
45    &     0.893586      &    0.000752       &   -6.418913    \\
46    &     -1.163013     &     0.000575      &    -7.109187    \\
47    &     0.376856      &    0.000443       &   -1.829769    \\
48    &     4.410450      &    0.000417       &   -2.022805    \\
49    &     -0.095197     &     0.000259      &    1.293996    \\
50    &     -2.759055     &     0.000304      &    -0.758321    \\
51    &     -1.435935     &     0.000176      &    -0.424882    \\
52    &     -2.228339     &     0.000169      &    -1.235490    \\
53    &     -2.066526     &     0.000122      &    -0.955109    \\  \bottomrule
\end{tabular}
}
\end{table}

\begin{figure}[h!]
\centering
  \includegraphics[width=\columnwidth]{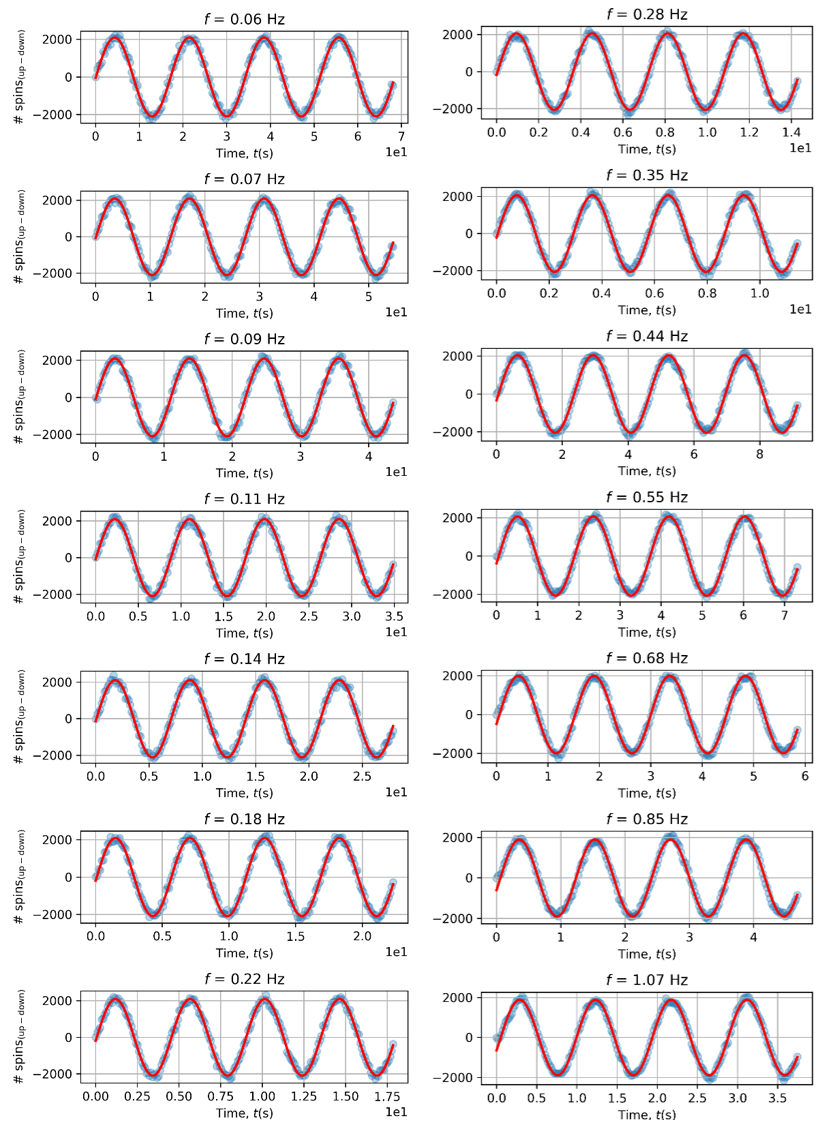}
  \label{fig:fit1}
\end{figure}
\clearpage

\begin{figure}[h!]
\centering
  \includegraphics[width=\columnwidth]{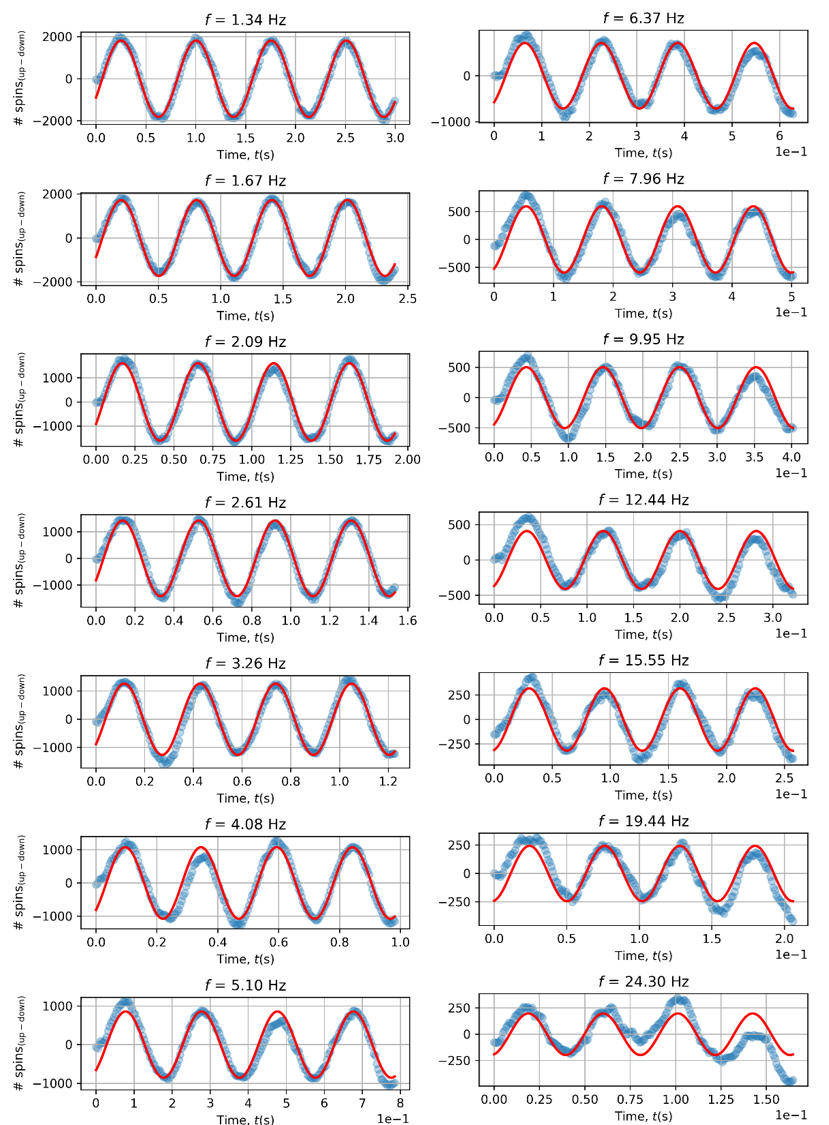}
  \label{fig:fit2}
\end{figure}
\clearpage

\begin{figure}[h!]
\centering
  \includegraphics[width=\columnwidth]{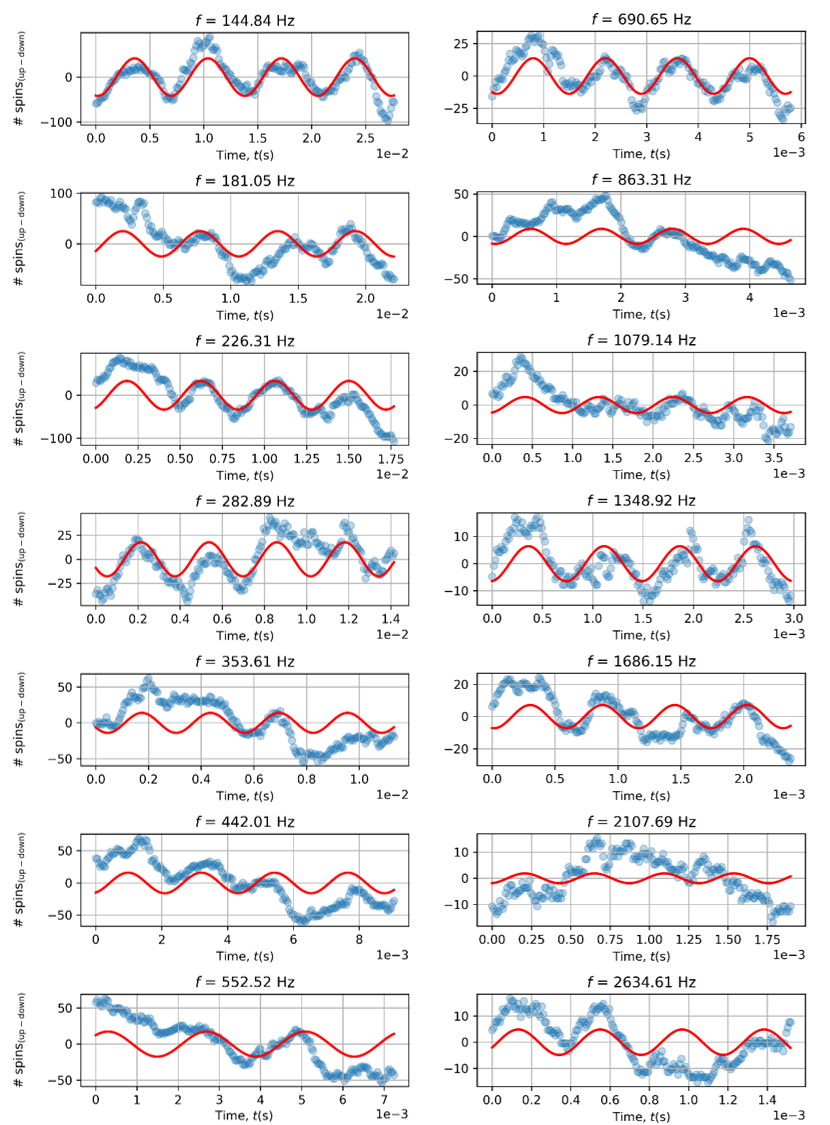}
  \label{fig:fit3}
\end{figure}
\clearpage

\begin{figure}[h!]
\centering
  \includegraphics[width=\columnwidth,left]{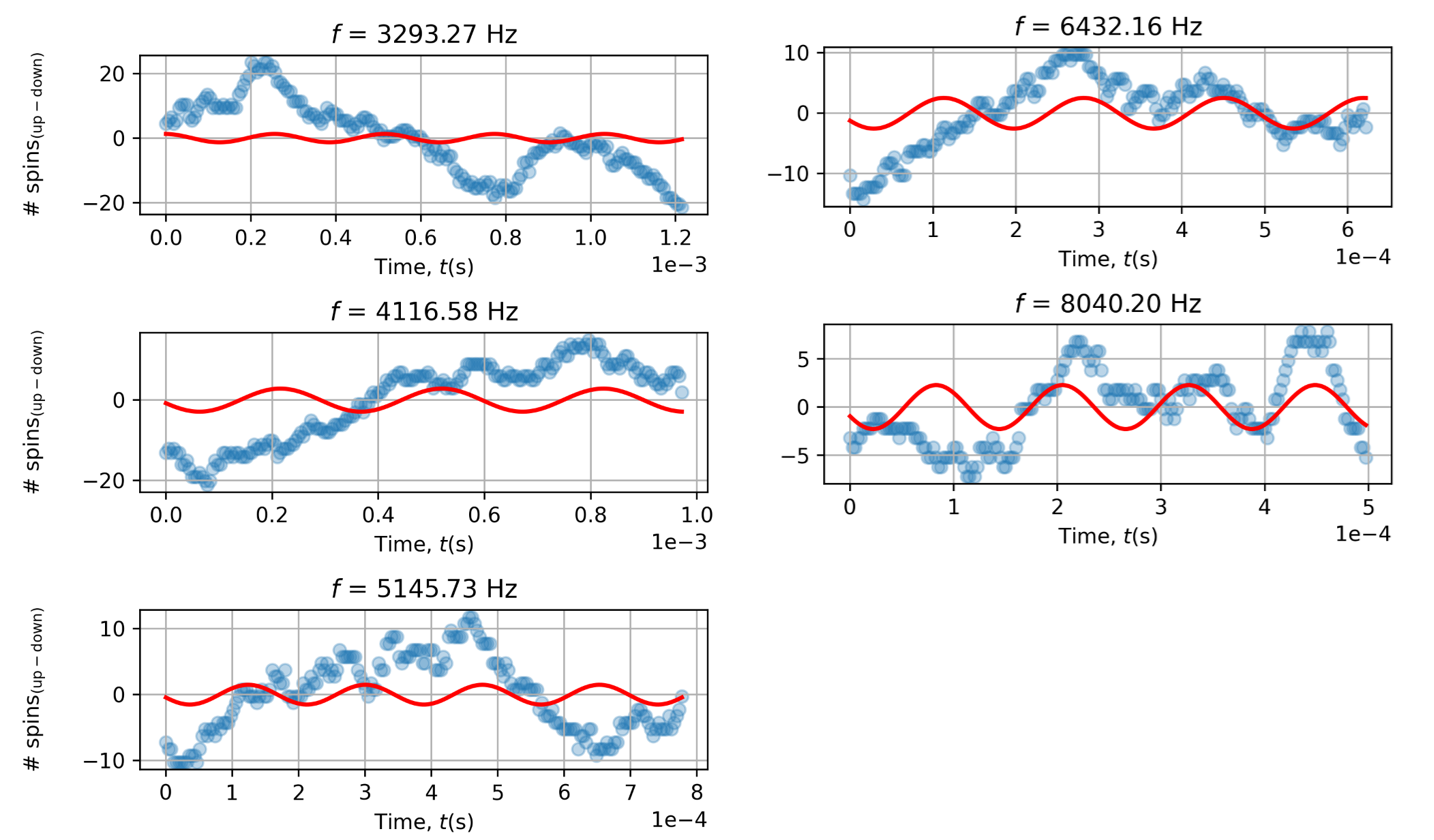}
  \caption{Single p-bit evolution under the influence of a sinusoidal ac magnetic field at \textcolor{black}{47} different frequencies \textcolor{black}{between $f=0.06$~Hz (60 mHz) and $f=8042$~Hz (8 kHz}. In each graph, the points correspond to the simulated response of N=\textcolor{black}{5}0000 spins and the red line to the fitted \textcolor{black}{linear combination of sine and cosine employing the parameters in Table~\ref{tab:my-table}}.}
  \label{fig:fit4}
\end{figure}
\clearpage




\subsection{\textcolor{black}{Effects of employing a single vs multiple magnetic molecules per p-bit}}
\label{multispin}

\begin{figure*}[h!]
\centering
  \includegraphics[width=\textwidth]{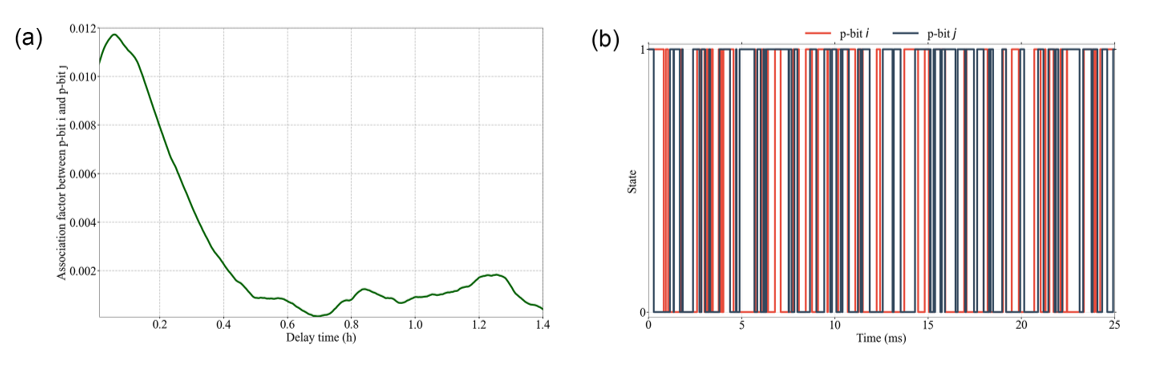}
  \caption{\textcolor{black}{Scheme for a toy network of associated p-bits based on the collective behavior of [(Cp$^\mathrm{iPr5}$)Dy(Cp$^\mathrm{*}$)]$^+$ molecules at 
$T=40~\mathrm{K}$, $1$ spin per p-bit; where the state of p-bit $i$ controls a magnetic field $B=0.2~\mathrm{T}$ acting on p-bit $j$.(a) Association factor $\phi$ comparing the state of p-bit $i$ at a certain time with state of p-bit $j$ after a certain delay time \textcolor{black}{where the x-axis is the delay time}  (b) State of p-bit $i$ and p-bit $j$ vs time; the states are seemingly stochastic }}
  \label{fig:pbitnetwork}
\end{figure*}

\textcolor{black}{The smallest p-bit would be of nanometric scale, a single spin, for example on a single molecule. An exploration of this case is explored in Figure \ref{fig:pbitnetwork}. As one can see, the association, in this case, is minimal, meaning even if one achieves the feat of measuring the state of a single spin, the logical operation would be possible but severely limited. This is compared with a collective of $10^6$ spins where the observed association is much more clear. In this case, we chose the record-holding Single Ion Magnet (SIM) to illustrate the extremely slow associated operating times that result from working at temperatures below the thermal blocking.}

\textcolor{black}{
Of course, detecting a single molecular nanomagnet is extremely challenging, and not an attractive prospect for scalable devices. Fortunately, in contrast with most molecular spin qubit proposals, one does not need to equate individual nanomagnets with individual p-bits. Each p-bit \textcolor{black}{can} consist of a large molecular ensemble rather than a single molecule, and indeed there are rapid advances both in the interaction of thin ensembles of spins with electronic circuits\cite{Ebel2021,Serrano2022} and in chemical paths for protecting monolayers of spins from disruption by a substrate.\cite{Tesi2023} As detailed below, working with a monolayer of $M$ molecules has the following major implications:
\begin{enumerate}
    \item Higher signal detection, approximately proportional \textcolor{black}{to $M$ in the limit of highly polarized spins or} to $\sqrt{M}$ \textcolor{black}{in the limit of zero or low fields.} 
    \item A much stronger association.
    \item Slower \textcolor{black}{randomization} dynamics, with the characteristic time being modified by a factor approximately proportional to $\sqrt{1/M}$ (see below for details).
    \item The possibility of employing a continuous output rather than a binary one, a feature that would allow the application of \textcolor{black}{this} hardware for neural networks.
    \item As a limitation, if one wants to read binary signals from spin ensembles, there will be either a bias among the p-bit states, as in the example detailed in the main text or a technical detection threshold 
    below which the signal will be in an undefined state. But if this is a technical limitation of multi-spin p-bits, it means it would be impossible to work with single spins.
\end{enumerate}
}

\textcolor{black}{Let us say we can't measure the state of a single spin and instead employ, as a single p-bit state, a ``spin excess'', defined as the sign of the difference (number of spins up minus number of spins down) of a large number $M$ of molecular spins. For example, if the ``spin excess'' of spins pointing up is above a certain threshold which we can detect, the state of the p-bit is 1, else it is 0. Compared with the case of a single spin per p-bit, the two main magnitudes that are affected by choosing a higher or lower number for $M$ are the signal amplitude and the magnetization dynamics. Both the signal amplitude and the dynamics may depend on the magnetic field. So to explore the effect of employing multiple magnetic molecules per p-bit let us distinguish the situations of (a) very weak (infinitesimal) magnetic fields, relevant for the generation of random binary numbers and (b) a finite magnetic field, for information processing in general.}

\textcolor{black}{In the case of a very weak magnetic field, the equilibrium magnetic signal of the sample is close to zero, and it is very useful to think of the problem as a random walk. As a starting point, let us imagine a system with $M/2$ molecular spins pointing up and $M/2$ pointing down (zero “spin excess”) and let us estimate the evolution at short times. The expected situation after a short time step, in particular for a time step corresponding to a spin flip probability $p$ for each spin, can be approximated by an unbiased random walk of $p\cdot M$ steps. This is so because one expects $p\cdot M$ spin flips and initially there is no bias between up$\rightarrow$down or down$\rightarrow$up flips. This means that the expectation distance from zero magnetization at short times will be proportional to $\sqrt{M}$. Indeed, for samples with very different values of $M$, the expected amplitude of the magnetic noise is expected to scale as the square root of the number of spins. In practical terms, this means that the detection of random binary signals will improve with $\sqrt{M}$.}

\textcolor{black}{Still in the case of a very weak magnetic field, the dynamics will also be similar to a (biased) random walk. Note, crucially, that as soon as the numbers of spins up and down are different, the expected numbers of up$\rightarrow$down vs down$\rightarrow$up spin flip events will also be unequal, meaning the direction of the walk will be biased. There will be a certain tendency to "rebound" to the initial state. This bias will rise as the relative spin excess (spin excess divided by total number of spins), since a given absolute spin excess will be more noticeable for a smaller total number of spins, and have no effect if the total number of spins is close to infinity. In the extreme case of $M=\infty$, the mapping to the random walk is even more apt, and, as for a random walk, it is likely that the system never returns to its initial unbiased state. In the opposite limit, for $M=1$ each spin flip necessarily means a change in the spin sign.  Indeed, if one expects the absolute spin excess to be proportional to $\sqrt{M}$, the expected bias should be proportional to $\sqrt{M}/M=1/\sqrt{M}$. So, p-bits presenting lower values of $M$ will rapidly reach a higher bias, thus a statistically faster "rebound" and a more frequent change in sign of spin excess, and vice versa, higher values of $M$ will maintain a low bias for a long time, thus statistically slower "rebound" and a less frequent change in sign of spin excess. This means that, for purely stochastical reasons, larger collectives of spins are expected to flip the sign of their collective spin sum slower, even if each of the spins are behaving as independent paramagnets. In practical terms, for our purposes this means that the speed of random binary signals that are detectable will scale as $1/\sqrt{M}$: stronger random signals will necessarily also present slower randomization times.}

\begin{figure}[hbtp]
\includegraphics[width=0.4\textwidth]{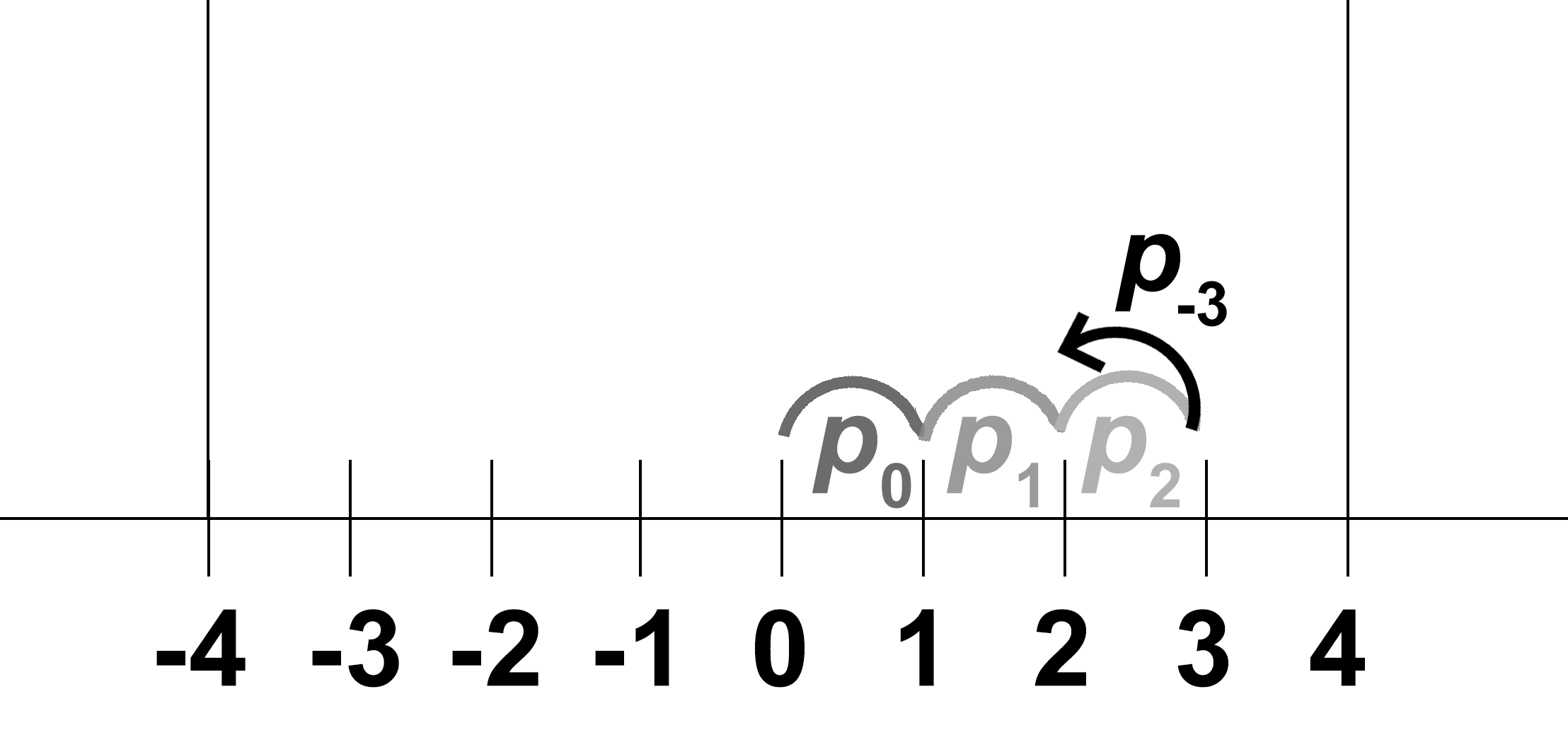}
\caption{\textcolor{black}{The spin excess is limited by the total number of magnetic molecules $M$. If we map this onto a random walk, there will be "walls" restricting the maximum amplitude of the walk. Moreover, for equal probabilities $p_{\mathrm{u,d}}$,$p_{\mathrm{d,u}}$ (see Supplementary Figure~\ref{figSI:Markovkate}), the global probability to advance will become progressively smaller, with the probability to "rebound" becoming progressively larger as the walk gets near one of the walls.}}
\end{figure}

\textcolor{black}{In the case of a finite magnetic field, the equilibrium magnetic signal grows proportionally with $M$ since we will be on the linear magnetic response regime. In practical terms, this will facilitate detection of the p-bit state. On the other hand, the characteristic response time is a molecular property, as verified in ac experiments where the maximum in out-of-phase signal is independent of the amount of sample. For our purposes, this means that the response times are not significantly slowed down by $M$.}

\newpage
\section{Screening the SIMDAVIS dataset for p-bit operating speed at room temperature}

\textcolor{black}{This section explores potential p-bits via the analysis of molecular nanomagnets in the SIMDAVIS dataset \cite{Duan2022}, in particular at room temperature (300 K). If we are to work with conventional electronic equipment, it is reasonable to focus on the molecules that have an estimated relaxation time ($\tau_{300\mathrm{K}}$) of close to 1 ms, since this will be a realistic requirement if the electronic equipment has to operate at least as fast as the p-bit relaxation time. Importantly, the electronic device materials which will be used to control and monitor the p-bit state have to be time-compatible with the SIM molecules that will embody the p-bits. The observation of the following plots where we represent the relaxation time as a function of several chemical and physical parameters can point us to the most favorable ones in order to achieve p-bits working at a reasonable operating speed.}

\textcolor{black}{Here note that estimating spin dynamics of molecular nanomagnets suffers from two fundamental limitations at room temperature, both stemming from the fact that their parameterisation is based on the ground doublet of spin states, and from experimental data at low temperatures. (A) Assuming that the magnetic signal from each molecule is given by its ground doublet, even at room temperature, gives an exaggerated Ising-like character to its magnetism. In practice, the magnetic moment will not be as axial as it is at low temperatures. (B) Additionally, further spin dynamics will be active at high temperatures that are not considered in equation 6 in the main text.}

\textcolor{black}{Let us address these concerns. About (A), one has to recall that, while $U_\mathrm{eff}$ is often below 100 K, for most molecular nanomagnets the total Crystal Field splitting is at least of the order of room temperature. This means that at 300 K the system will not be strictly axial, but it also will not be strictly isotropic, and there will still be a preferential magnetisation axis. About (B), note that even if higher spin levels participate in Orbach relaxation at high temperatures, relaxation will in general take place by the route with the lowest $U_\mathrm{eff}$.} 

We start by a comparison of the behavior of $\tau_{300\mathrm{K}}$ (Figure 5, upper panel, in the main text depicting $\tau_{300\mathrm{K}}$) and $\tau_0$ (see Supplementary Figure \ref{figSI:screeningtau0}), in both cases as a function of metal ions and chemical families. \textcolor{black}{The N\'eel-Arrhenius equation (\ref{OrbachArrhenius}) was employed to estimate $\tau_{300\mathrm{K}}$ using the $U_\mathrm{eff}$ and $\tau_0$ parameters contained in the SIMDAVIS dataset for 612 samples.} 

\begin{equation}
    \tau = \tau_0\cdot\mathrm{e}^{U_\mathrm{eff}/T}
\label{OrbachArrhenius}
\end{equation}

One can see that \textcolor{black}{$\tau_{300\mathrm{K}}$ and $\tau_0$,} while not strictly identical, \textcolor{black}{can correlate quite well, especially when the temperature is much higher than $U_\mathrm{eff}$, so that $\tau_{300\mathrm{K}}$ is approximately $\tau_0$.} This is to be expected since $\tau_0$ is the limit of $\tau$ at infinite temperature.

\begin{figure*}[h]
\centering
  \includegraphics[width=\textwidth]{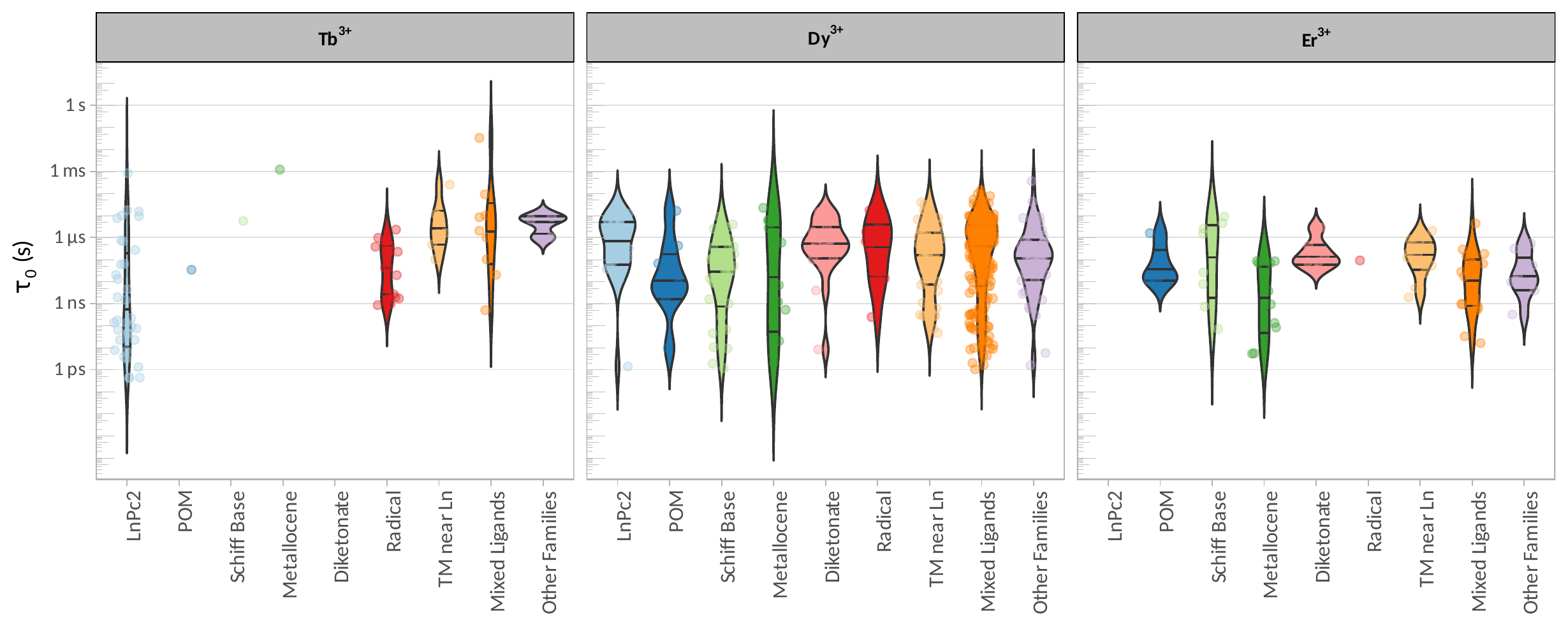}
  \caption{$\tau_{0}$ for \textcolor{black}{molecular nanomagnets based on} Tb$^{3+}$, Dy$^{3+}$, Er$^{3+}$, in each case \textcolor{black}{categorised} for the main chemical family considered in the \textcolor{black}{SIMDAVIS} dataset.}
  \label{figSI:screeningtau0}
\end{figure*}

To allow for an easier visualization of the relation between different ways of classifying Ln-based SIMs and their typical operating speeds as p-bits at room temperature, we classified each sample according to its $\tau_{300\mathrm{K}}$ range, as:
\begin{itemize}
\item fast: $ \tau_{300\mathrm{K}} < 0.5 \mu \mathrm{s}$
\item slow: $ \textcolor{black}{50 \mu \mathrm{s}} < \tau_{300\mathrm{K}} < 0.5 \mu \mathrm{s} $
\item slowest: $\tau_{300\mathrm{K}} >\textcolor{black}{50 \mu \mathrm{s}} $
\end{itemize}
Within this scheme, \textcolor{black}{about 60 \%} of the samples in the SIMDAVIS dataset are classified as ``fast'', i.e. one would require electronics that are capable of operating faster than \textcolor{black}{2} MHz (in many cases, faster than 1 GHz) to read individual molecules. A \textcolor{black}{significant} group are classified as ``slow'', where electronics being able to operate around the 1 MHz frequency scale would be adequate. \textcolor{black}{Below 5 \% of the samples} are classified as ``slowest'', where, given enough sensitivity, even single molecules could be read by electronics operating at speeds much smaller than 1 MHz (in some cases, down to 1 kHz). \textcolor{black}{As discussed above (see Supplementary Section \ref{multispin}), larger ensembles of $M$ molecules would have a collective magnetic moment that responds to a magnetic field as fast as the single molecule, but which in absence of stimulus drifts with $1/\sqrt{M}$ speed, which means they could be detected without requiring fast electronics. Additionally, p-bits made of $M$ magnetic molecules would be expected to produce signals that are at least $\sqrt{M}$ stronger compared with unimolecular p-bits.}

Note that this is not a full statistical study, but rather a data \textcolor{black}{exploration}. However, some insights seem immediate. We will start by the coordination sphere (which species of atoms it contains, the coordination number, the total number of ligands \textcolor{black}{and coordination geometry}), then revise the lanthanide ion (and its Kramers vs non-Kramers character as well as its oblate/prolate character), and finally compare with the known behavior, in terms of hysteresis and of ac susceptometry.

\subsection{Coordination sphere}
In terms of the coordination sphere, let us start with the coordination elements, i.e. the donor atoms. Coordination spheres that consist of a mixture of Oxygen and Nitrogen are overrepresented in the ``slow'' category, in comparison with all the other possibilities (only Oxygen, only Nitrogen, only Carbon and "others"). The data distribution in the boxplots allows an alternate visualization of the same tendency, with the median $\tau_{300\mathrm{K}}$ of Oxygen+Nitrogen coordination being substantially higher than the rest (see Supplementary Figure \ref{figSI:tauvscoordele}).

\begin{figure*}[h]
\centering
  \includegraphics[width=0.67\textwidth]{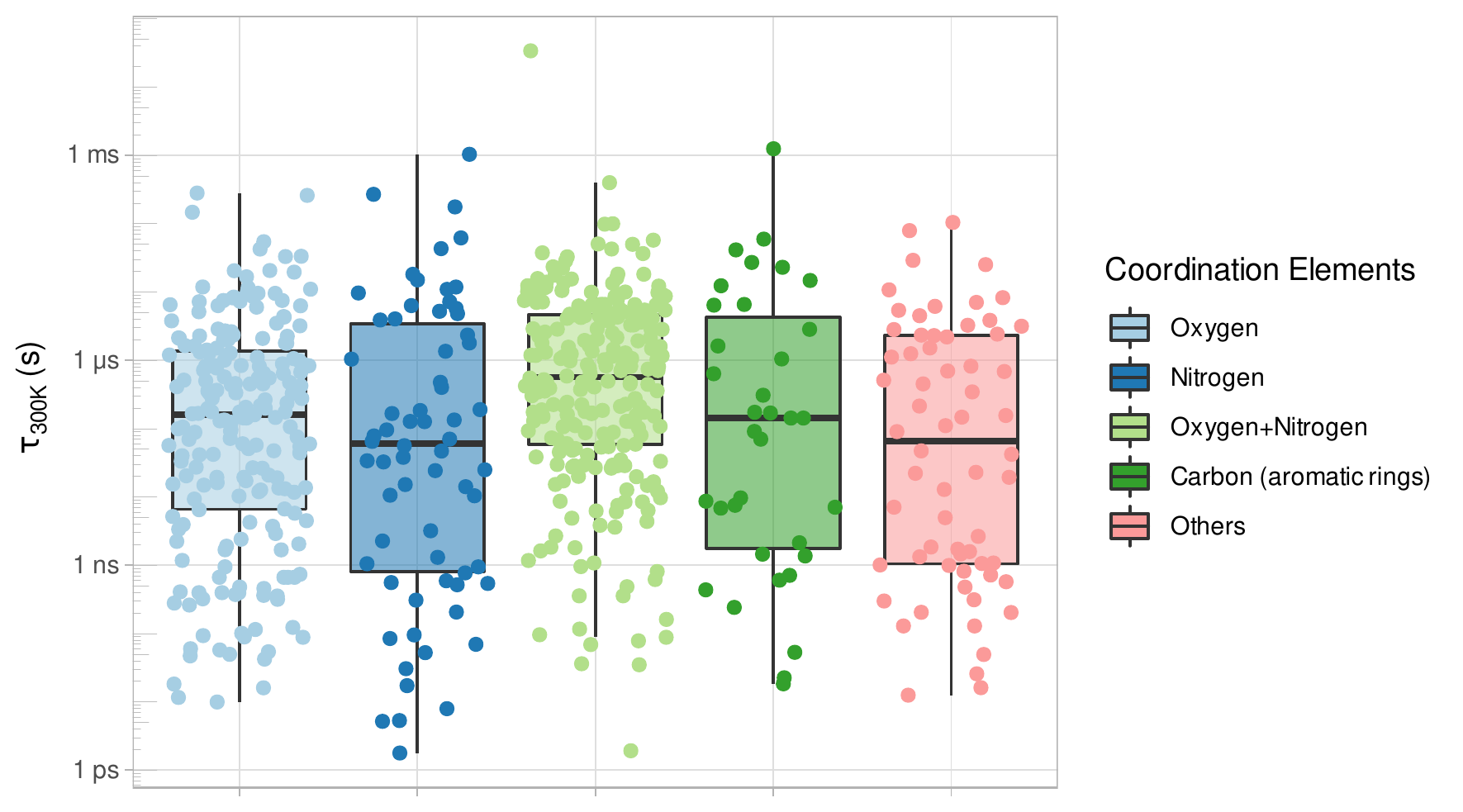}
  \includegraphics[width=0.48\textwidth]{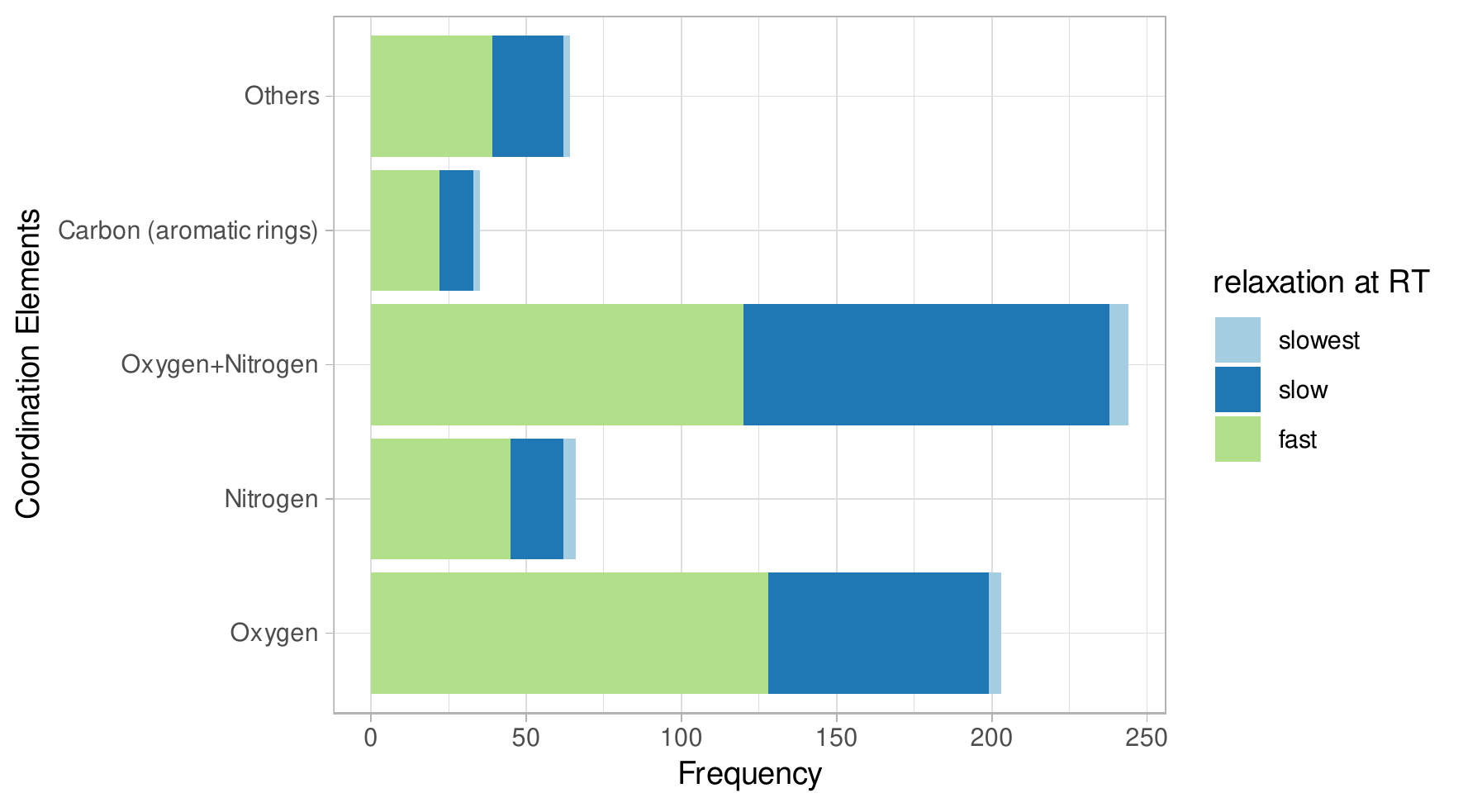}
  \includegraphics[width=0.48\textwidth]{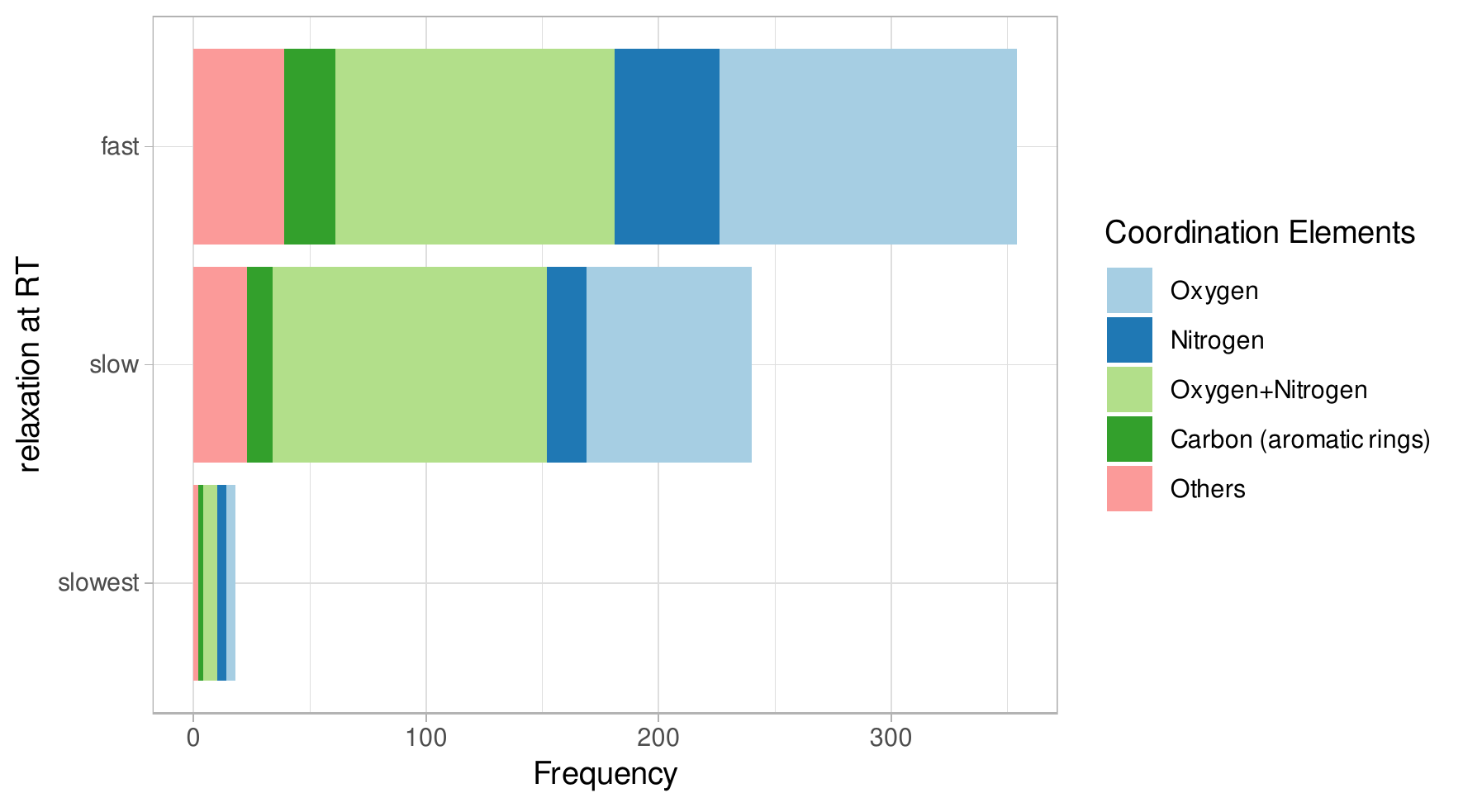}
  \caption{Relationship between the coordination elements and $\tau_{300\mathrm{K}}$ \textcolor{black}{ within the SIMDAVIS dataset}.}
  \label{figSI:tauvscoordele}
\end{figure*}

Moving on to how many atoms are directly coordinated to the metal, coordination numbers CN=8 and especially CN=9 are overrepresented both among the ``slow'' and ``slowest'' categories, and the boxplot representation allows seeing that a similar tendency is presented by CN=6 (see Supplementary Figure \ref{figSI:tauvscoordnum}). Something similar can be said of complexes with a small number of ligands: complexes consisting of \textcolor{black}{3, 4, 5 and 6} ligands are overrepresented both among the ``slow'' and ``slowest'' categories
; in the same representation, one can appreciate the extraordinarily fast relaxation times at room temperature of most complexes with seven ligands 
(see Supplementary Figure \ref{figSI:tauvsnumligands}). As found in the original SIMDAVIS study, this behavior is also driven by the unique contribution of complexes with a pentagonal bipyramid shape, which present extremely fast $\tau_{300\mathrm{K}}$ behavior (see Supplementary Figure \ref{figSI:tauvspoly}).

\begin{figure*}[h]
\centering
  \includegraphics[width=0.67\textwidth]{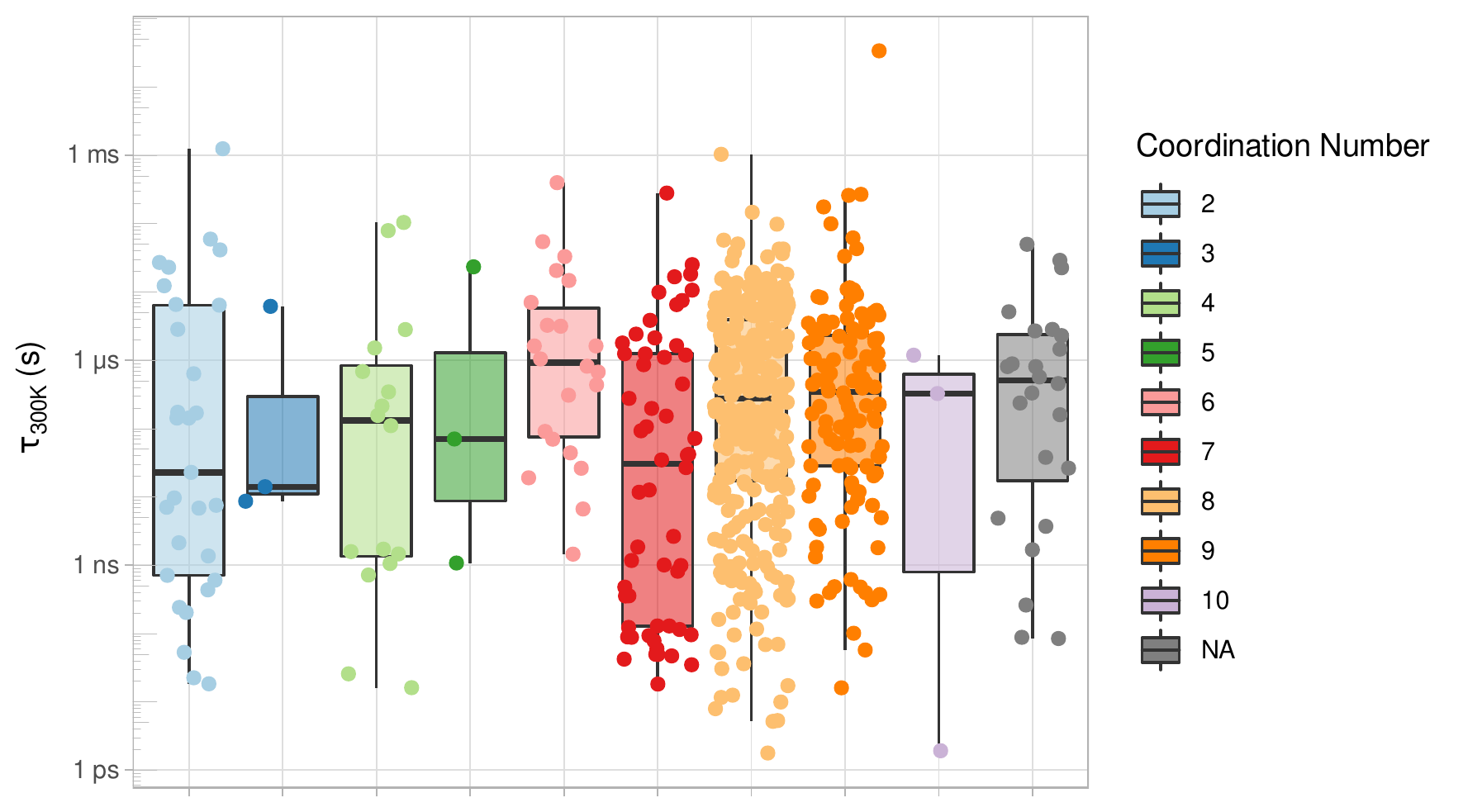}
  \includegraphics[width=0.48\textwidth]{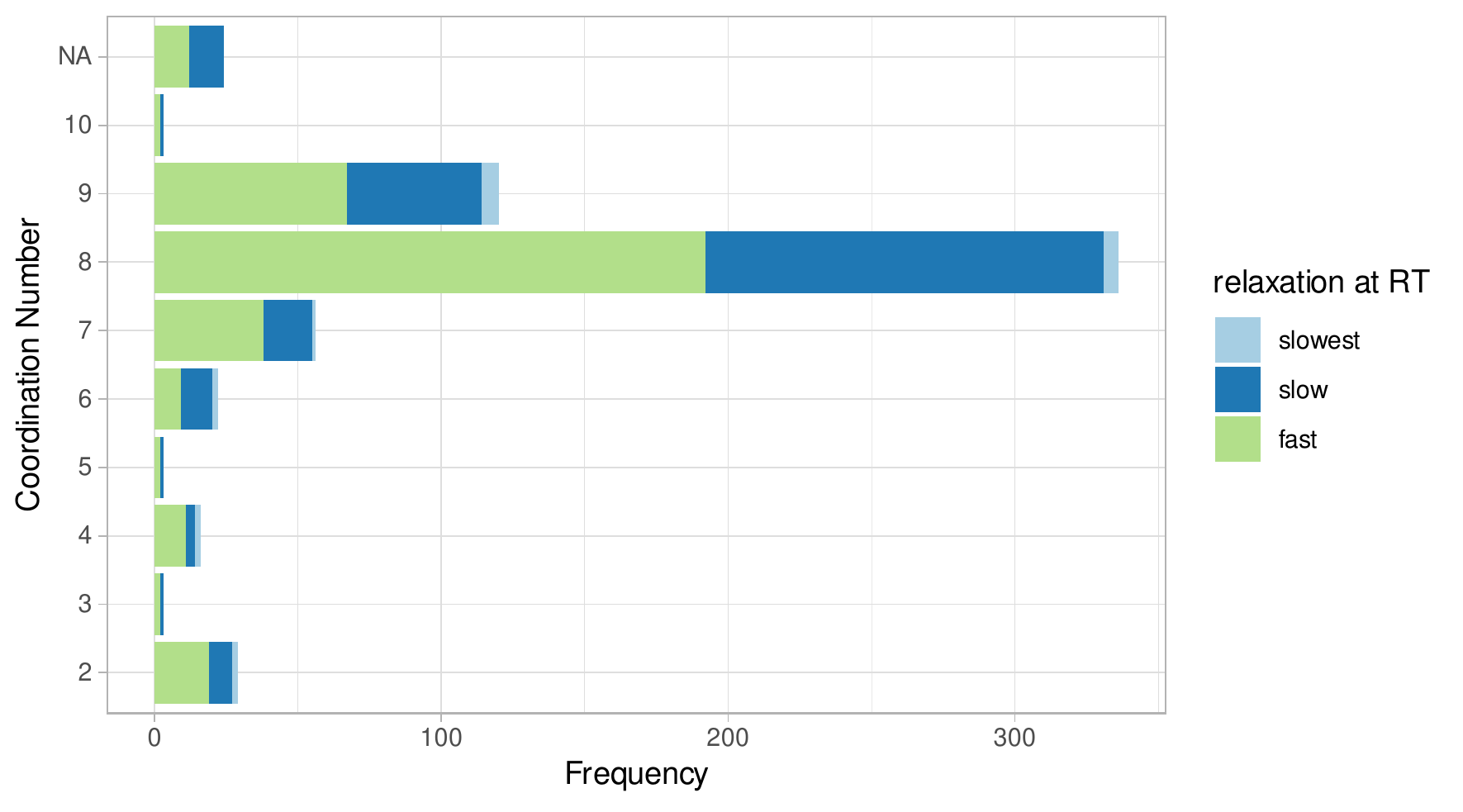}
  \includegraphics[width=0.48\textwidth]{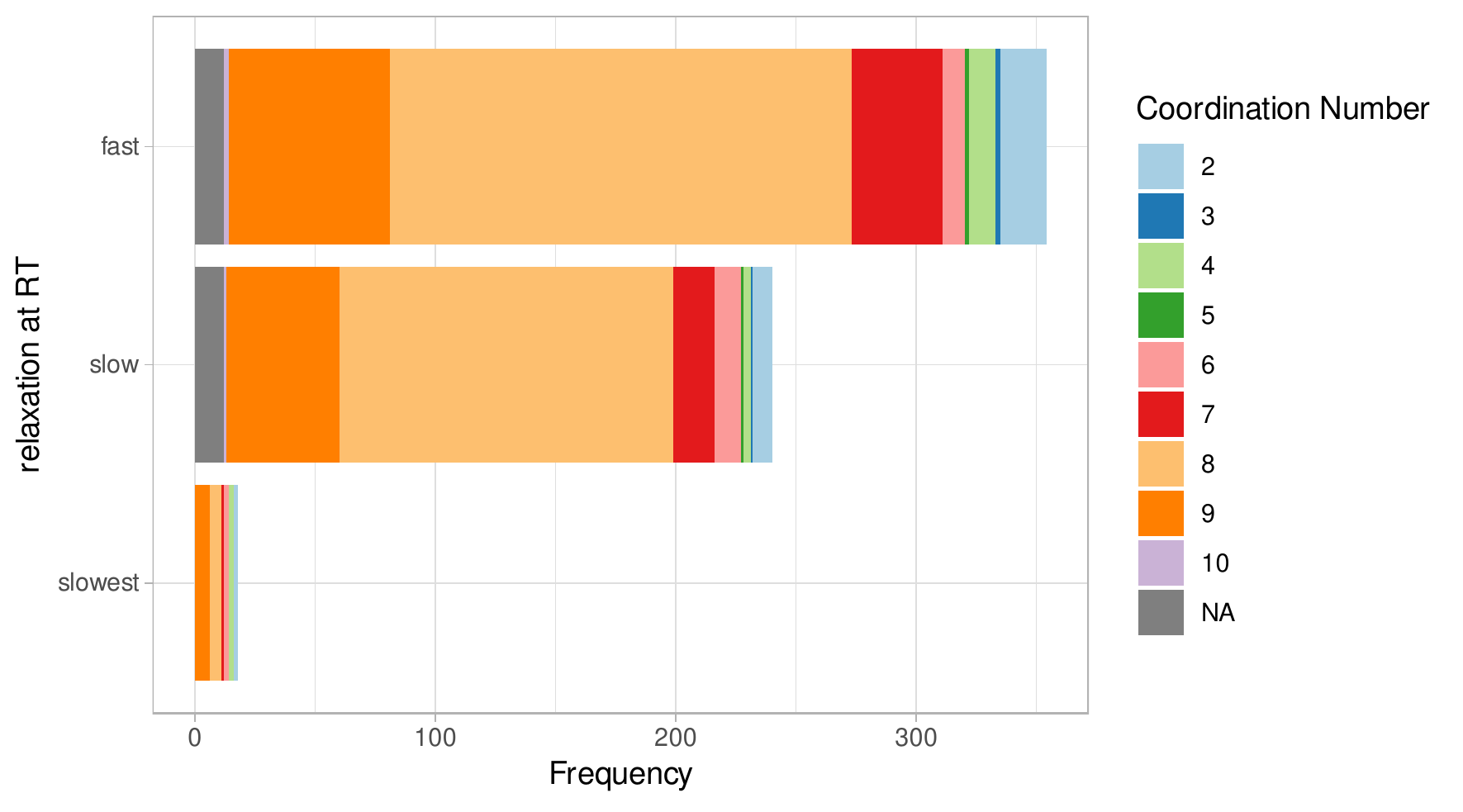}
  \caption{Relationship between the coordination number and $\tau_{300\mathrm{K}}$ \textcolor{black}{ within the SIMDAVIS dataset}.}
  \label{figSI:tauvscoordnum}
\end{figure*}

\begin{figure*}[h]
\centering
  \includegraphics[width=0.67\textwidth]{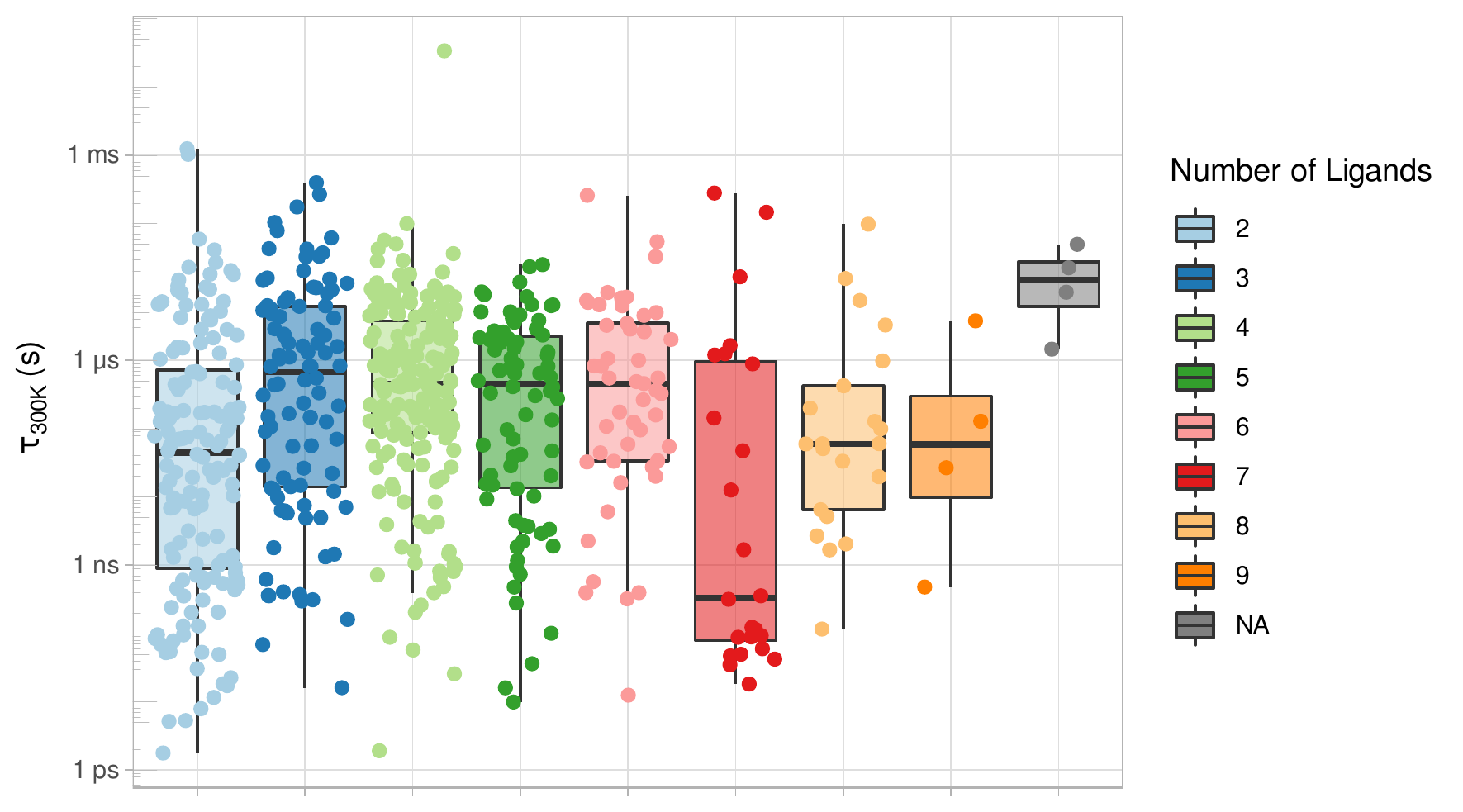}
  \includegraphics[width=0.48\textwidth]{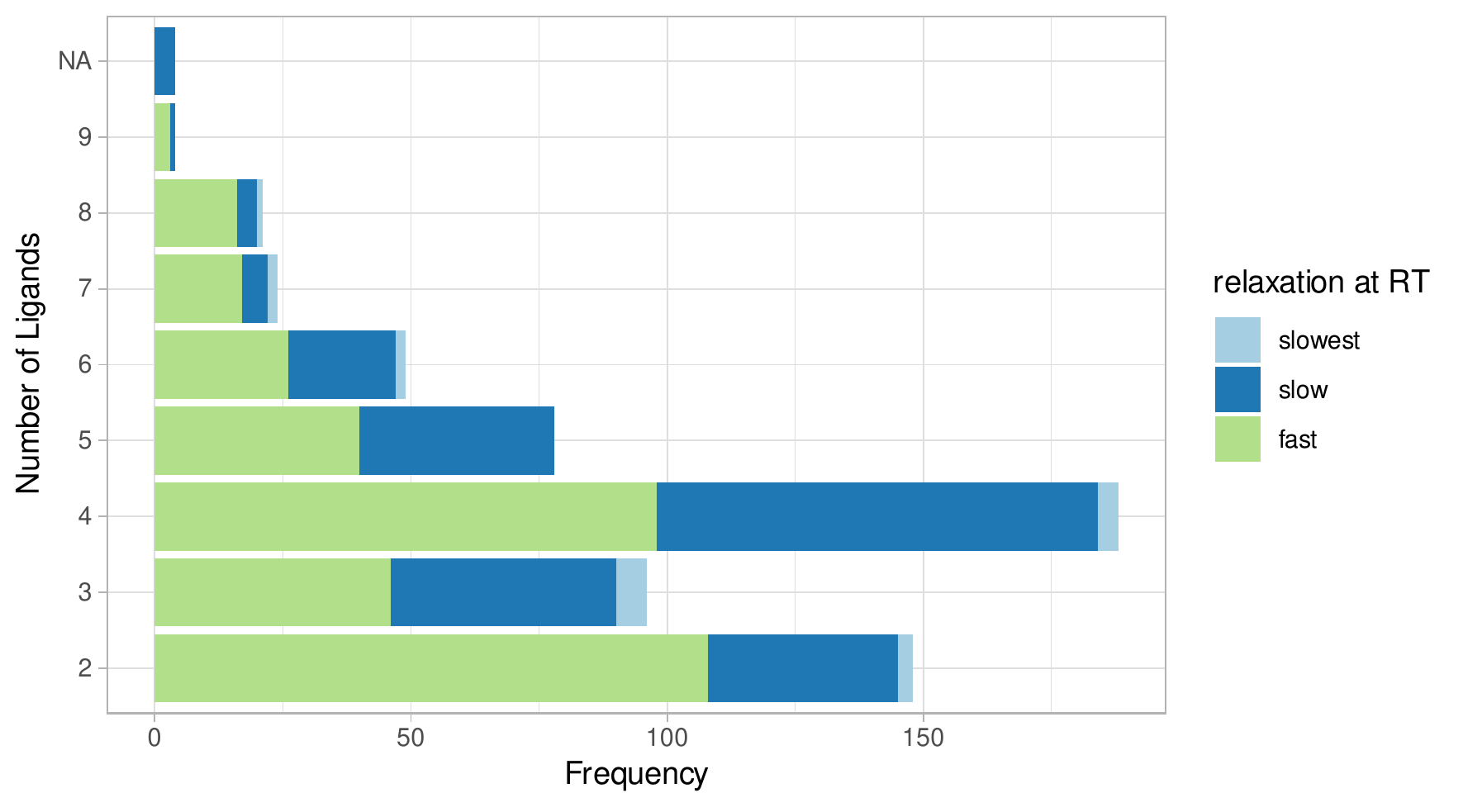}
  \includegraphics[width=0.48\textwidth]{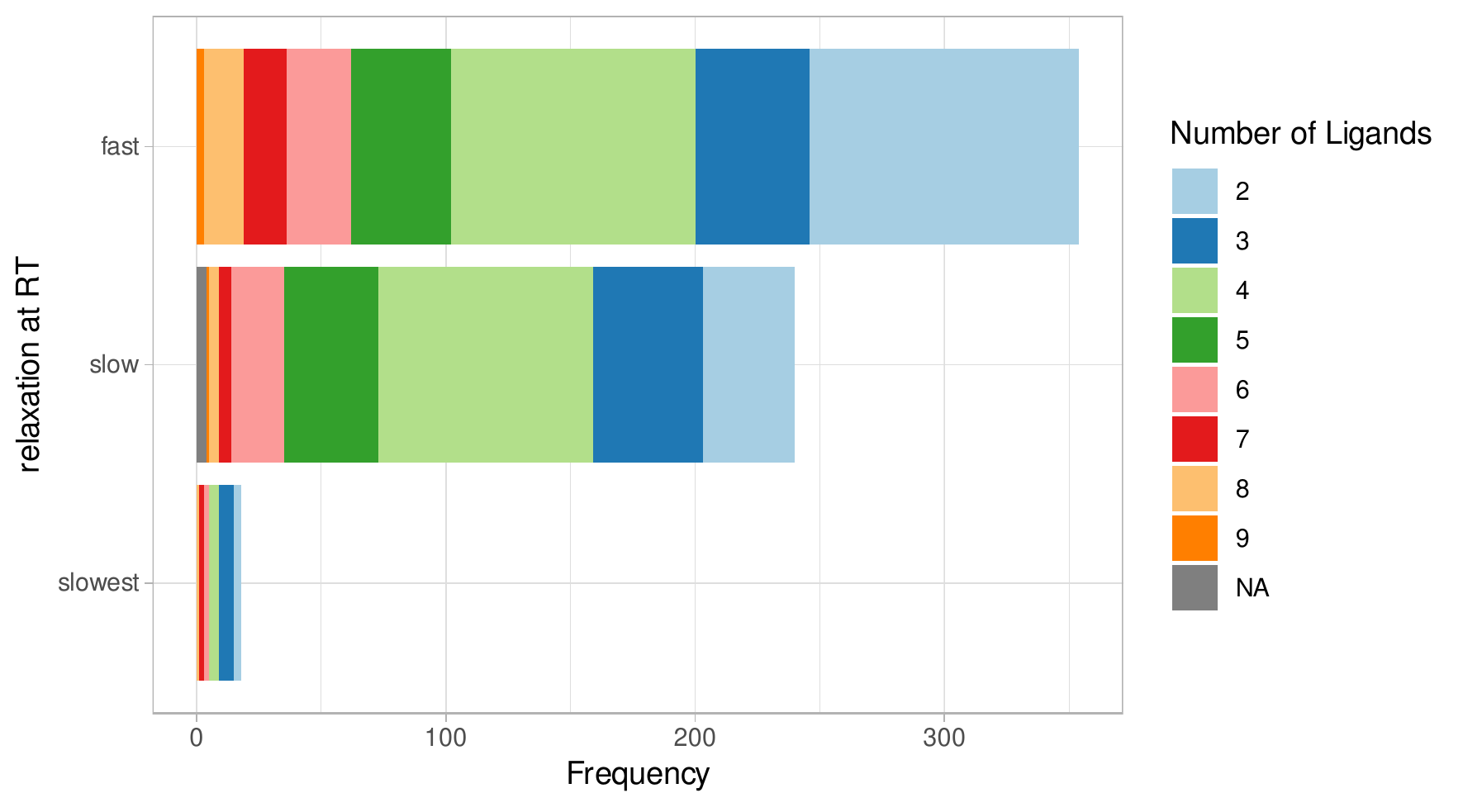}
  \caption{Relationship between the number of ligands and $\tau_{300\mathrm{K}}$ \textcolor{black}{ within the SIMDAVIS dataset}.}
  \label{figSI:tauvsnumligands}
\end{figure*}

\begin{figure*}[h]
\centering
  \includegraphics[width=0.67\textwidth]{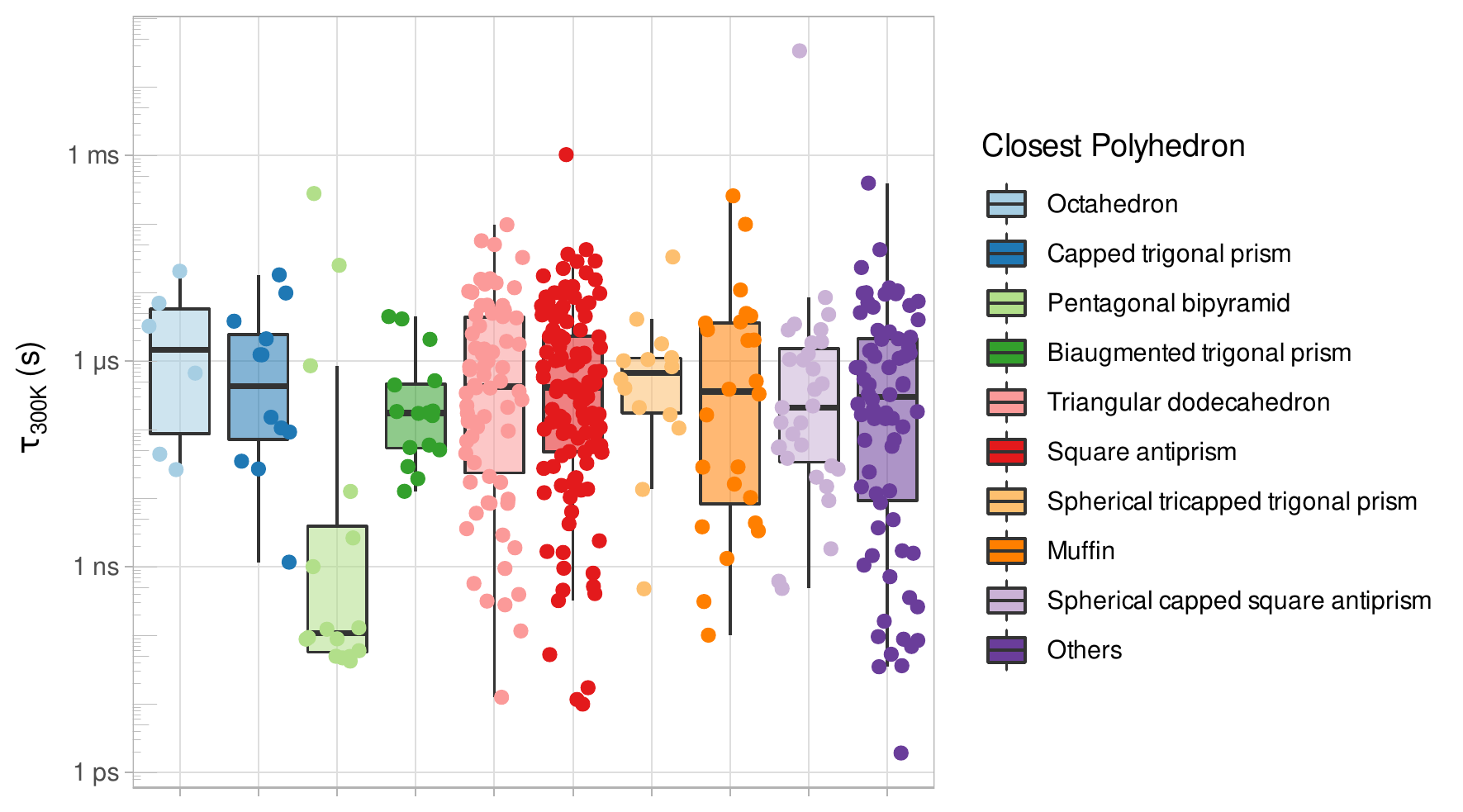}
  \includegraphics[width=0.48\textwidth]{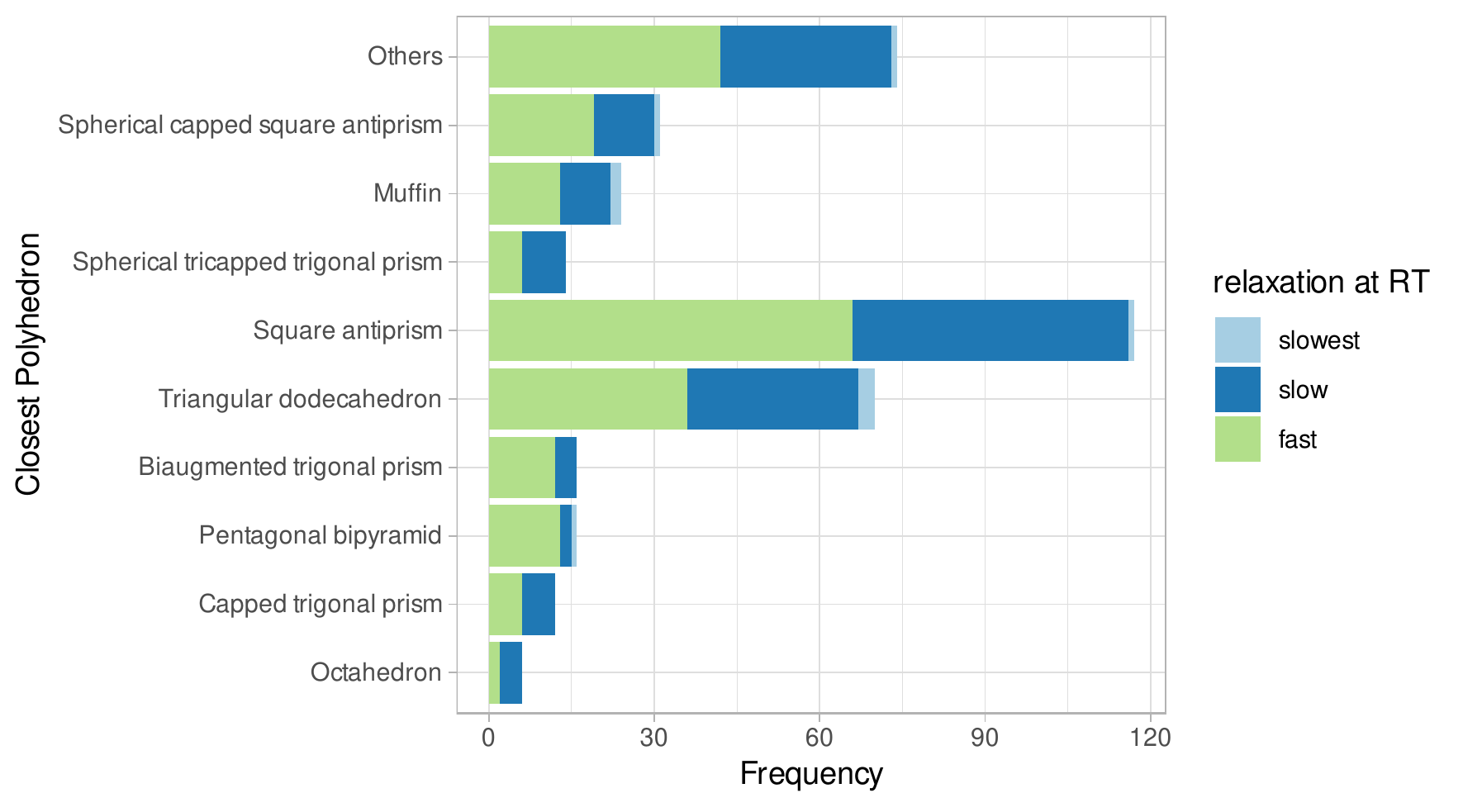}
    \includegraphics[width=0.48\textwidth]{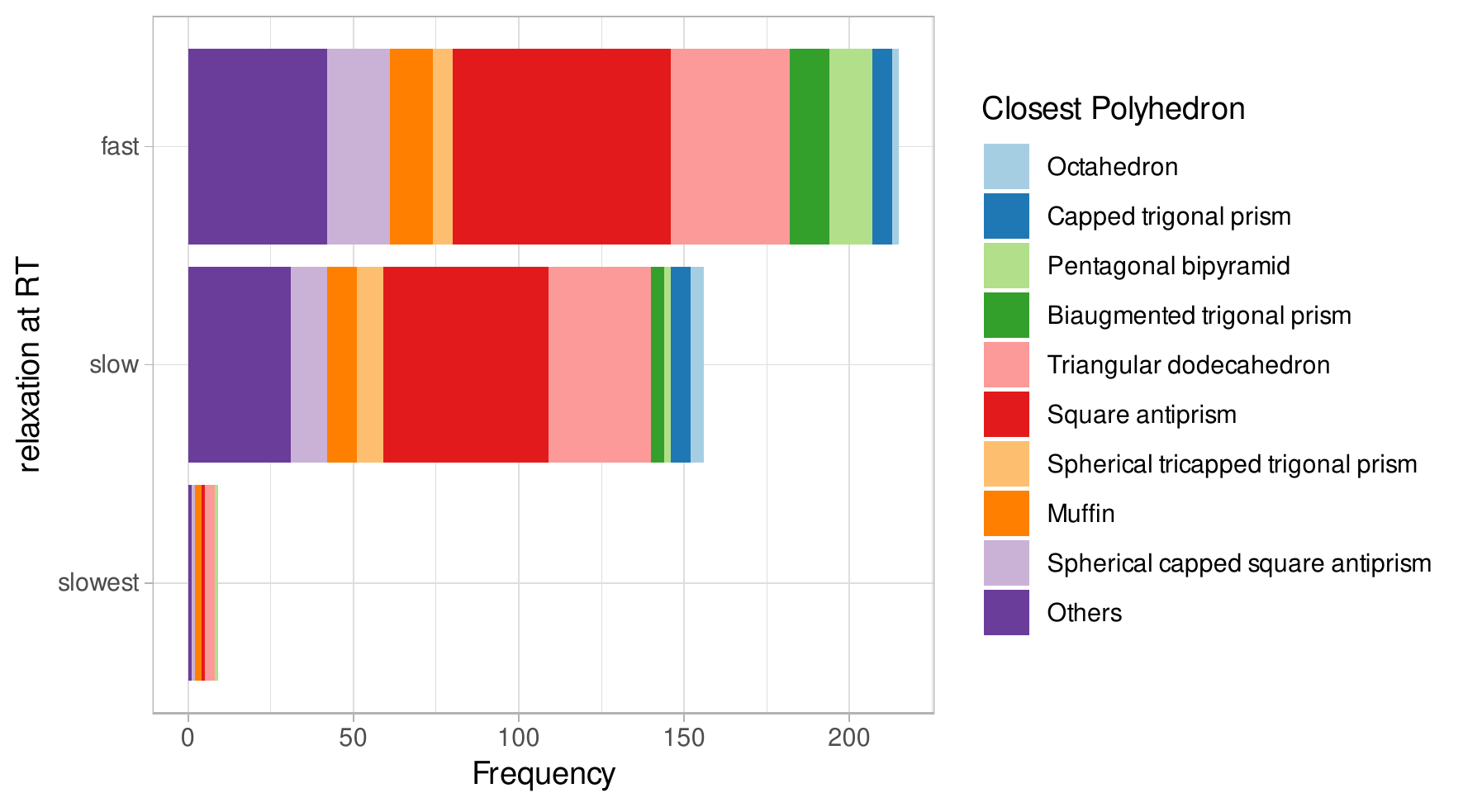}
  \caption{Relationship between the coordination polyhedron and $\tau_{300\mathrm{K}}$ \textcolor{black}{ within the SIMDAVIS dataset}.}
  \label{figSI:tauvspoly}
\end{figure*}

So, chemically, coordination spheres with a high coordination number like 9, a small number of ligands like 3 and a mixture of Oxygen+Nitrogen in the coordination sphere seem like a good recipe if one is aiming for obtaining a slow characteristic time $\tau_{300\mathrm{K}}$, compatible with relatively slow electronics.  \textcolor{black}{Such slow operating speeds are crucial to enable
switching on and off the relatively high magnetic fields required to create detectable signals at room temperature}. 

\clearpage

\subsection{Lanthanide ion}

In terms of the lanthanide ion and their Kramers vs non-Kramers character or their \textcolor{black}{anisotropy} (oblate/prolate character), there are also some general tendencies to be observed. \textcolor{black}{Oblate ions have a slower} median $\tau_{300\mathrm{K}}$. Additionally, \textcolor{black}{all} samples in the ``slowest'' category \textcolor{black}{are oblate ions} (see Supplementary Figure \ref{figSI:tauvsLnani}). The difference is much less remarkable in the comparison between Kramers and non-Kramers ions (see Supplementary Figure \ref{figSI:tauvsLnKra}). The contrast between the two categorization schemes means that this is not some artifact driven by just a single lanthanide, which would influence the two categories with the same intensity.
In terms of individual lanthanides, \textcolor{black}{the quest for p-bits operating at room temperature should favour any Ln ion other than the three most popular ones for SIMs (Dy$^{3+}$, Tb$^{3+}$, Er$^{3+}$), since these three present generally faster $\tau_\mathrm{300K}$ (see Supplementary Figure \ref{figSI:tauvsLnion}). As was the case with the coordination sphere, this is because of the inverse correlation between $\tau_0$ and $U_{\mathrm{eff}}$: good SIMs tend to present high $U_{\mathrm{eff}}$ whereas practical p-bits require long $\tau_0$.} 

\begin{figure*}[h]
\centering
  \includegraphics[width=0.67\textwidth]{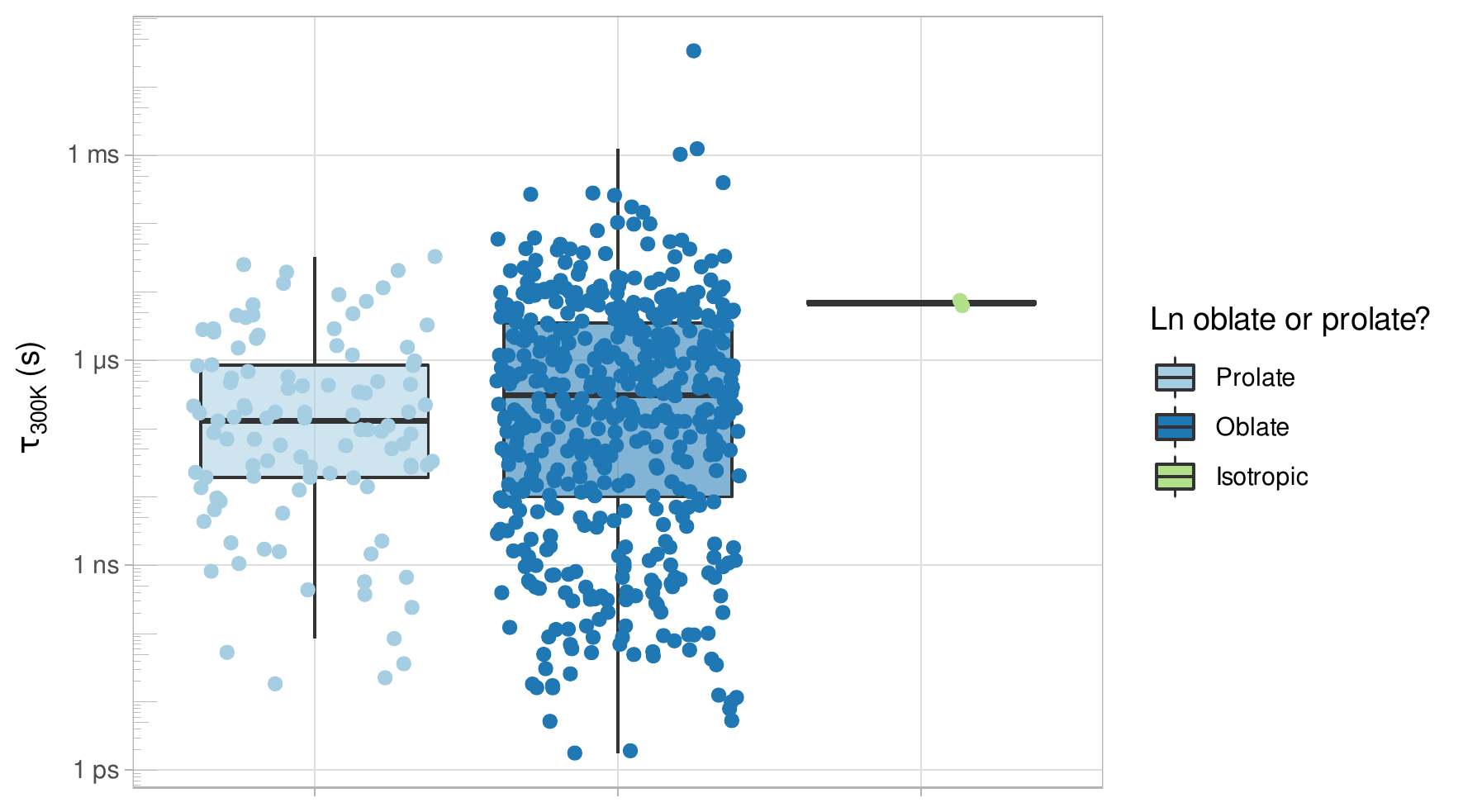}
  \includegraphics[width=0.48\textwidth]{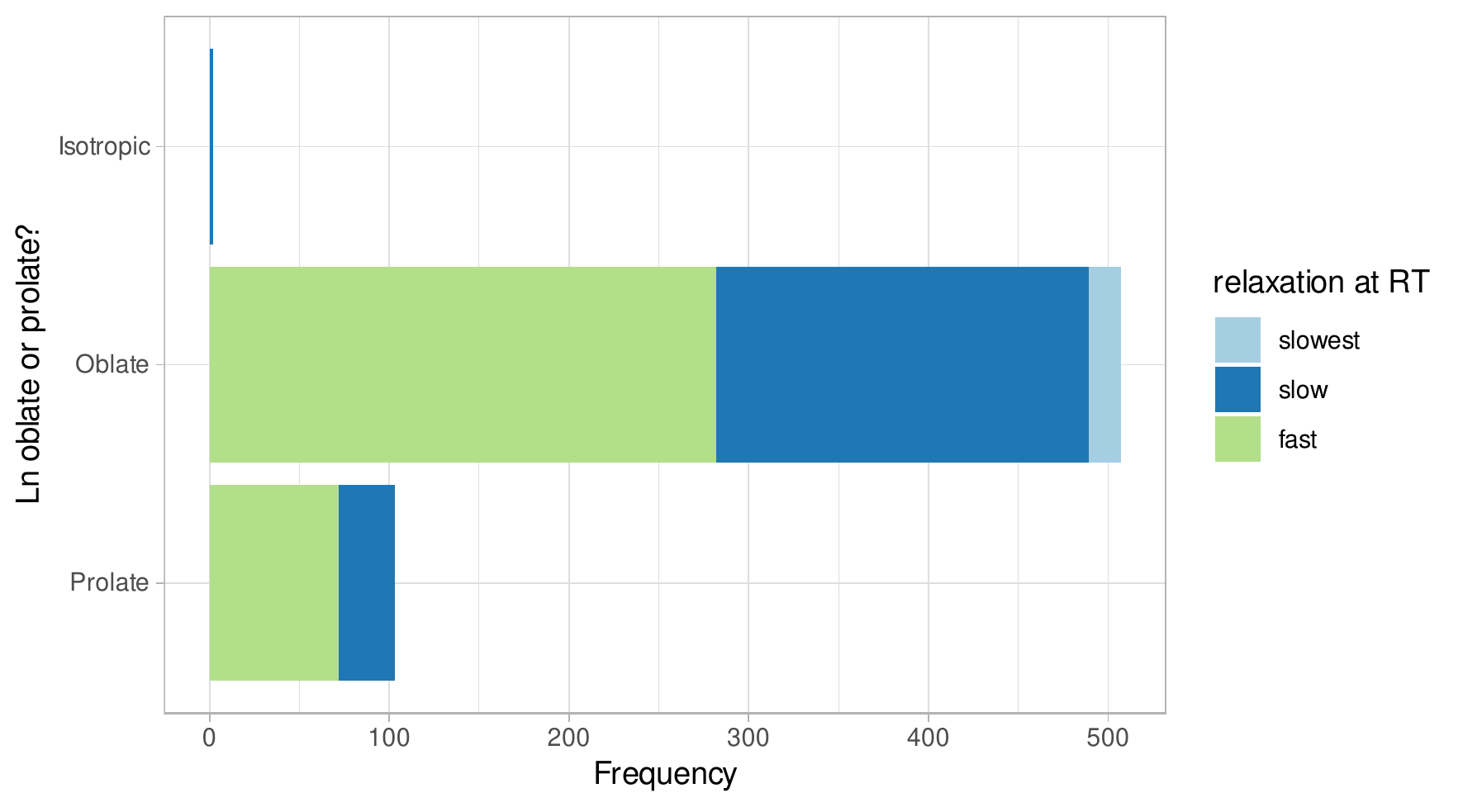}
  \includegraphics[width=0.48\textwidth]{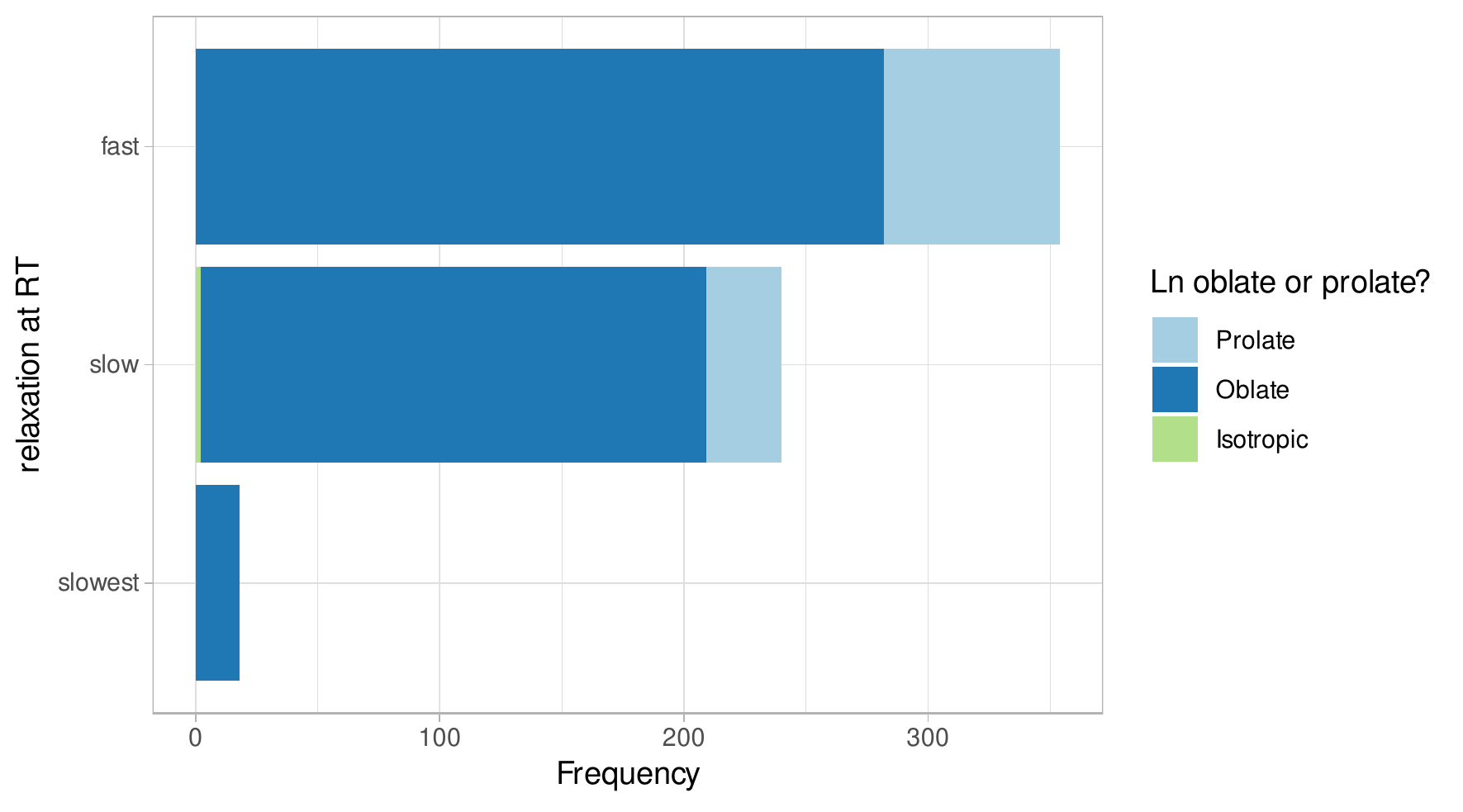}
  \caption{Relationship between the lanthanide ion's anisotropy and $\tau_{300\mathrm{K}}$ \textcolor{black}{ within the SIMDAVIS dataset}.}
  \label{figSI:tauvsLnani}
\end{figure*}

\begin{figure*}[h]
\centering
  \includegraphics[width=0.67\textwidth]{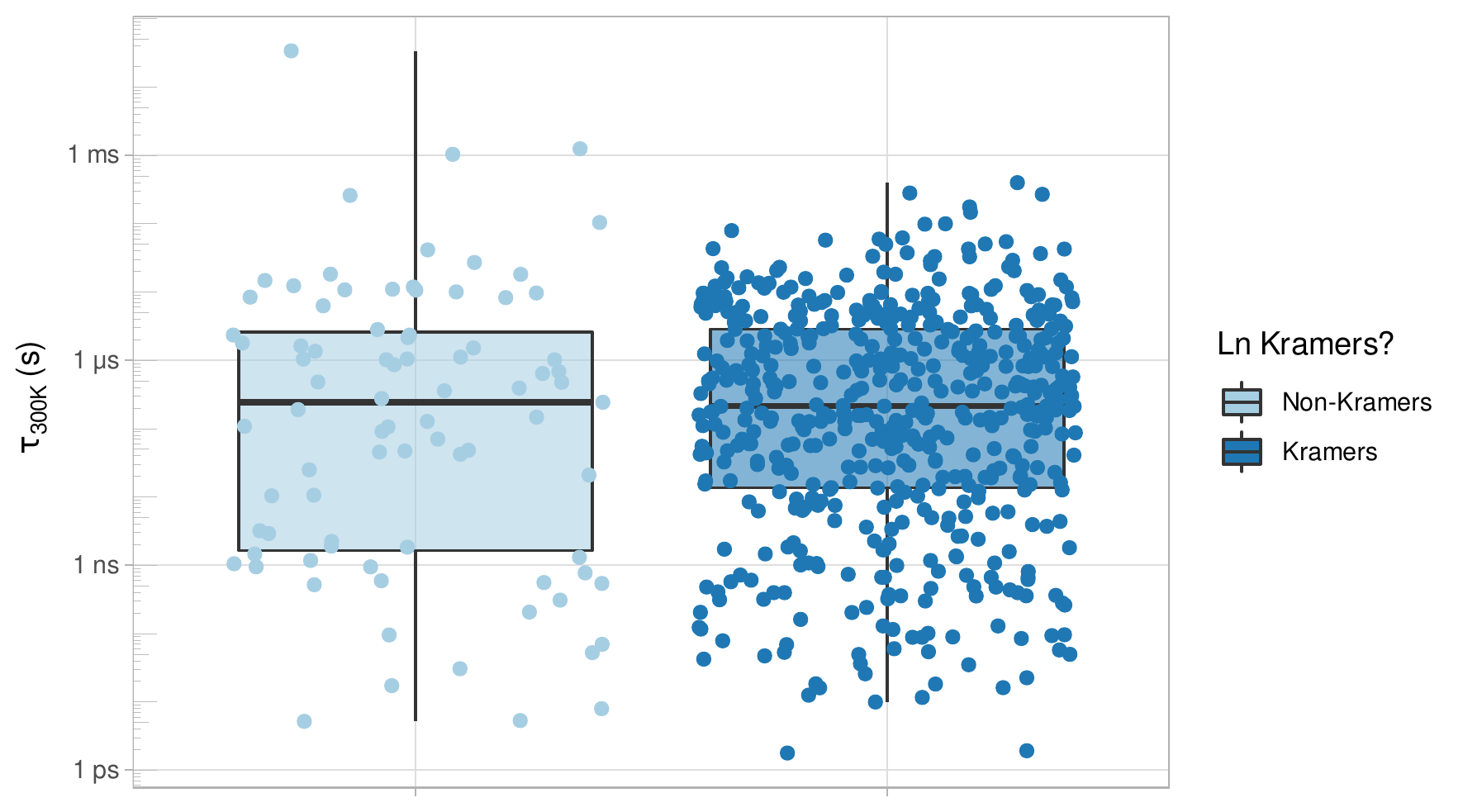}
  \includegraphics[width=0.48\textwidth]{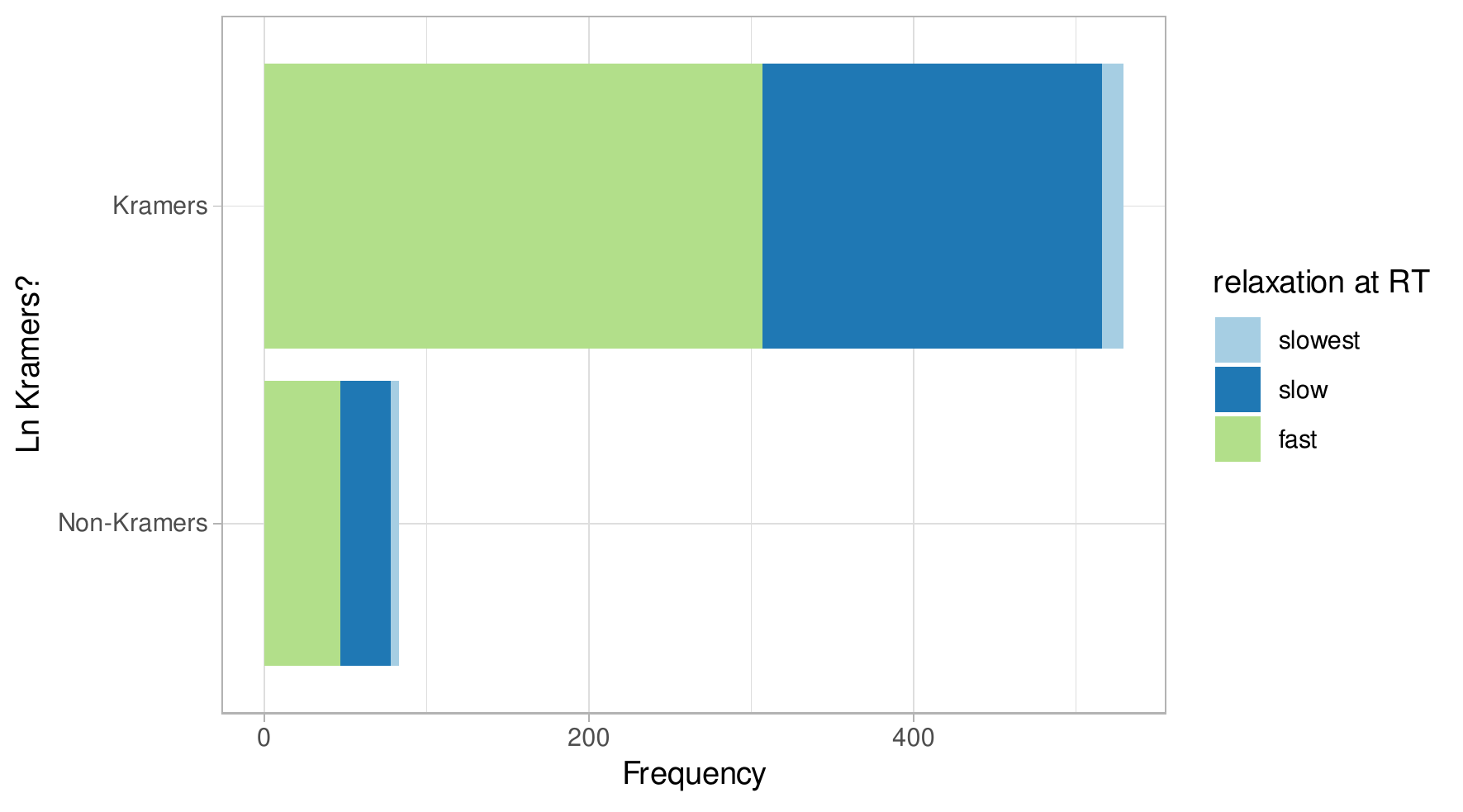}
  \includegraphics[width=0.48\textwidth]{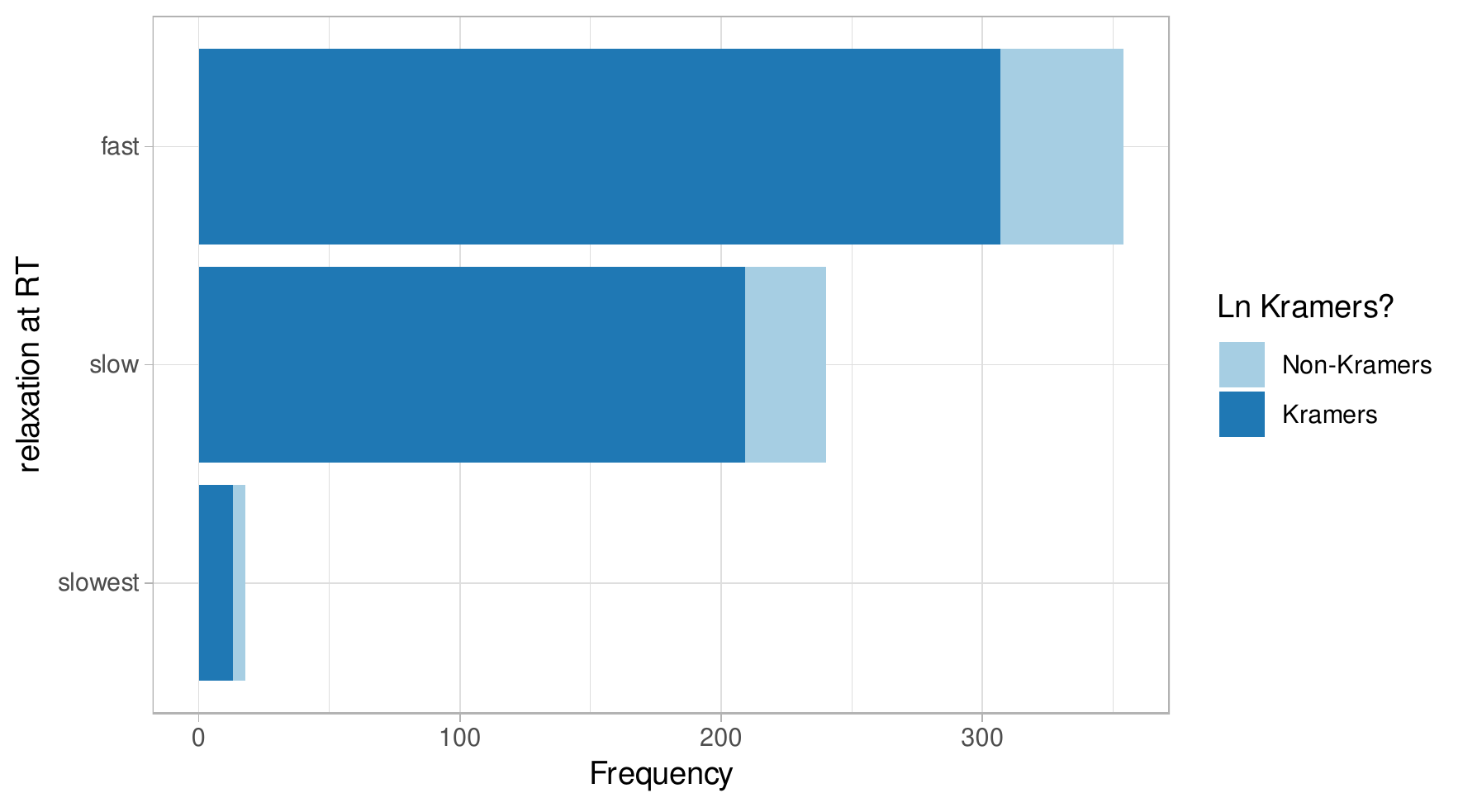}
  \caption{Relationship between the lanthanide ion Kramer's or non-Kramer's character and $\tau_{300\mathrm{K}}$ \textcolor{black}{ within the SIMDAVIS dataset}.}
  \label{figSI:tauvsLnKra}
\end{figure*}

\begin{figure*}[h]
\centering
  \includegraphics[width=0.67\textwidth]{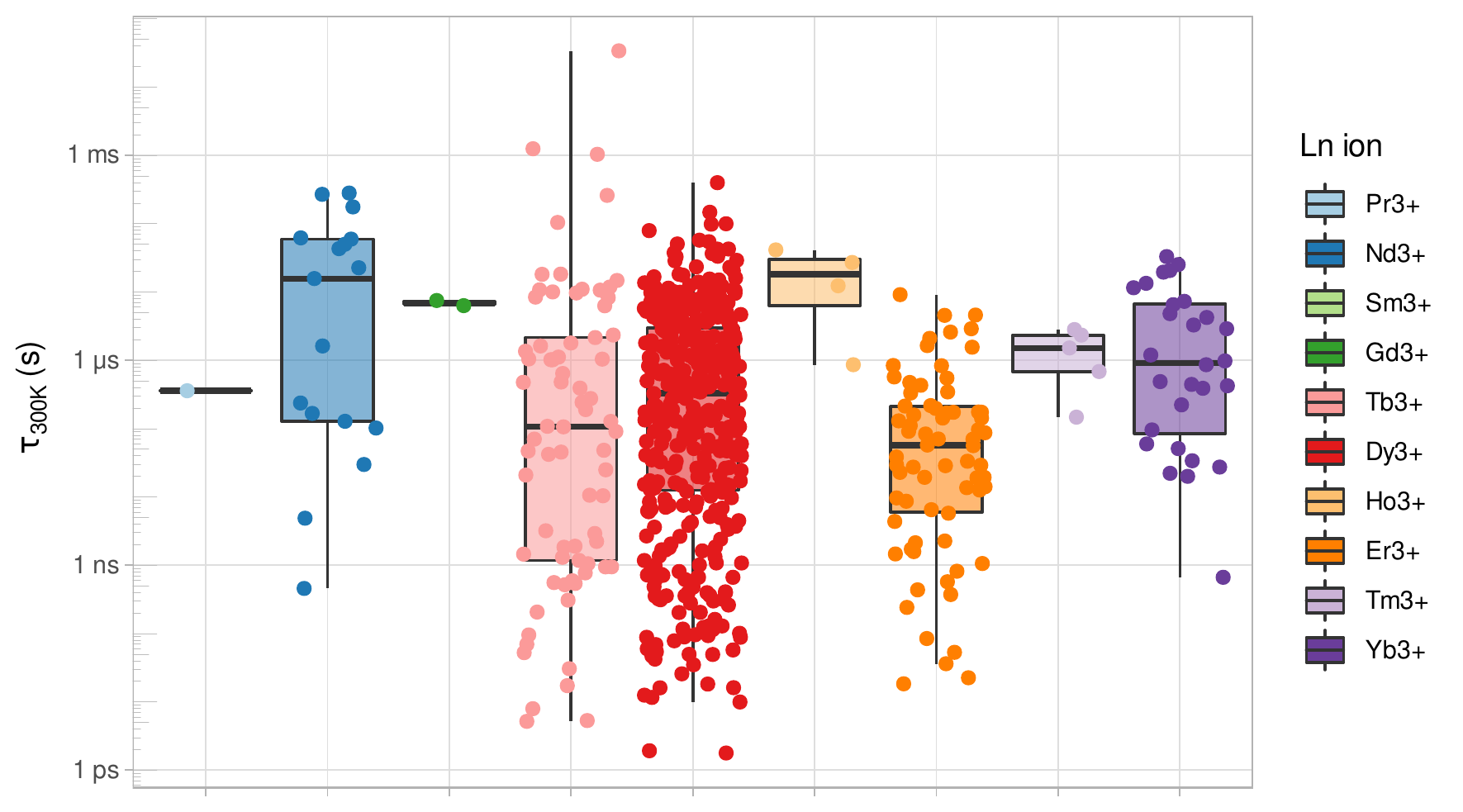}
  \includegraphics[width=0.48\textwidth]{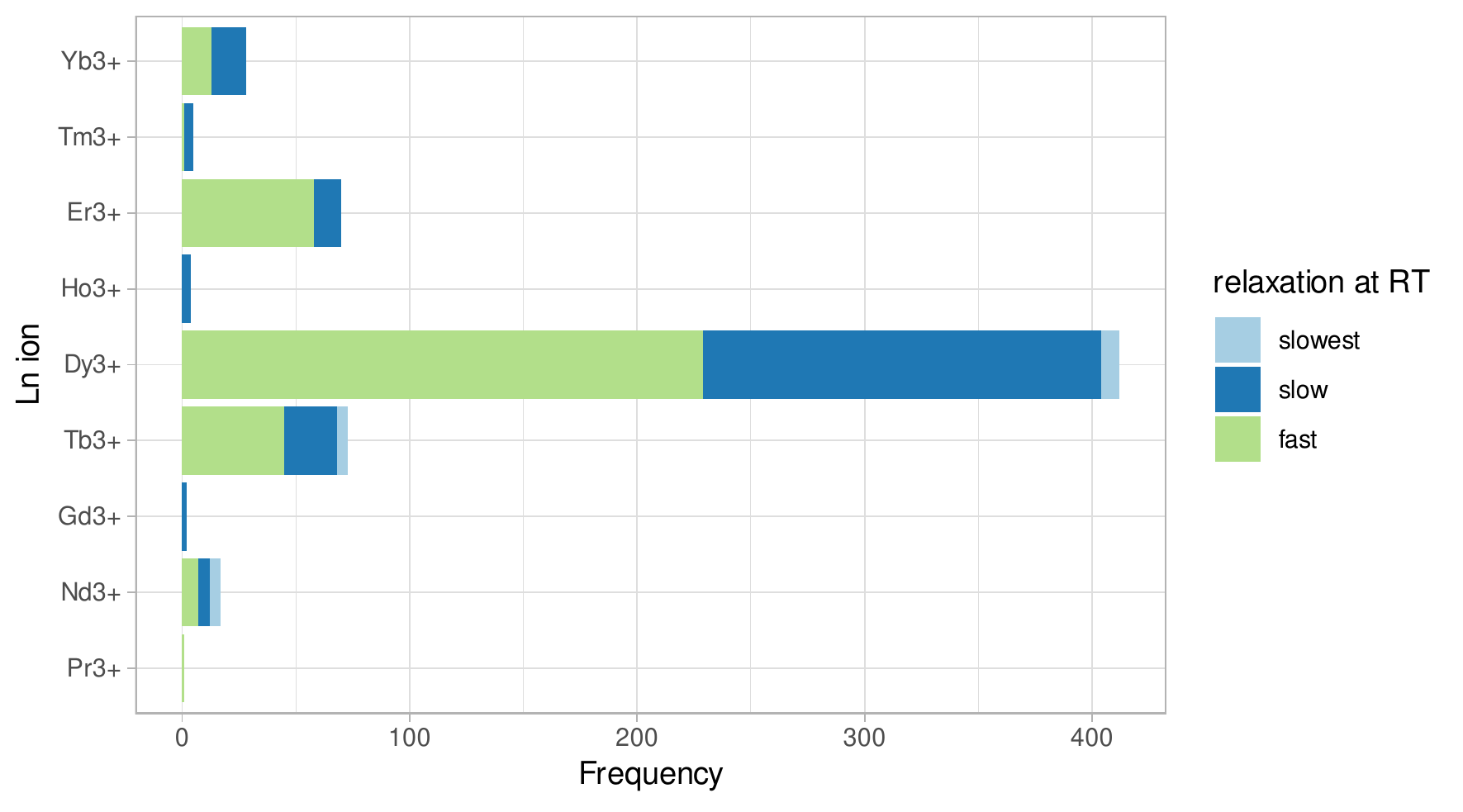}
  \includegraphics[width=0.48\textwidth]{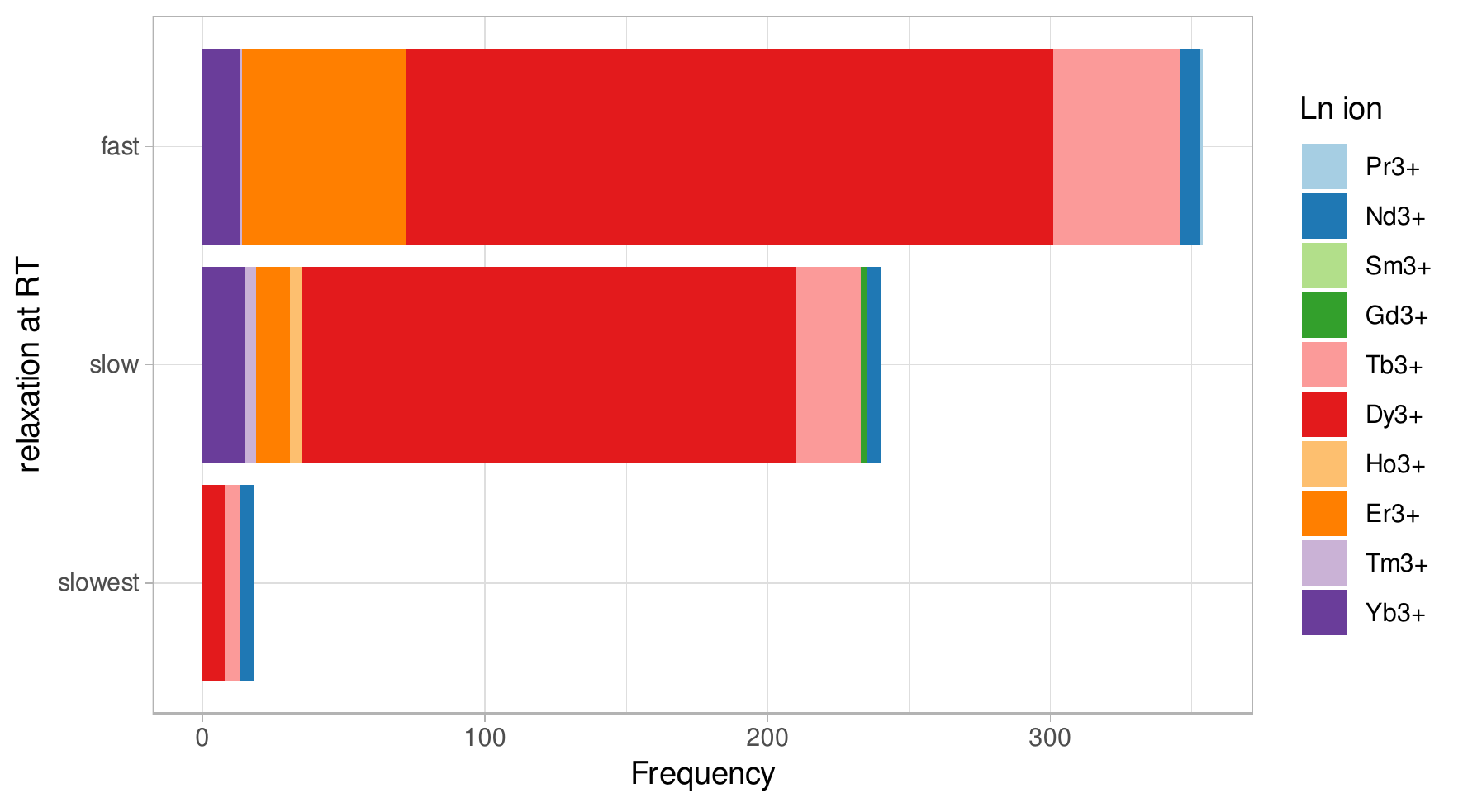}
  \caption{Relationship between the lanthanide ion and $\tau_{300\mathrm{K}}$ \textcolor{black}{ within the SIMDAVIS dataset}.}
  \label{figSI:tauvsLnion}
\end{figure*}

\clearpage

\subsection{Relaxation behavior}

Comparing $\tau_{300\mathrm{K}}$ with the observed behavior in terms of hysteresis and ac susceptometry can also help us gain some intuition. The samples presenting \textcolor{black}{no} hysteresis tend to present \textcolor{black}{slower} $\tau_\mathrm{300K}$ times, this is clear both from the bar charts and from the boxplots (see Supplementary Figure \ref{figSI:tauvsHyst}). \textcolor{black}{This inverse correlation between SIM performance and p-bit performance} is less intense for ac susceptometry (see Supplementary Figure \ref{figSI:tauvsXim}).

As stated above, the contradiction between what the field of SIMs considers ``slow'' systems (typically with high $U_\mathrm{eff}$ barriers) and what we are characterizing as ``slow'' systems (low $\tau_{300\mathrm{K}}$, in practice meaning low $\tau_0$) is absolutely expected from the generally observed behavior between $U_\mathrm{eff}$ and $\tau_0$. 

\begin{figure*}[h]
\centering
  \includegraphics[width=0.67\textwidth]{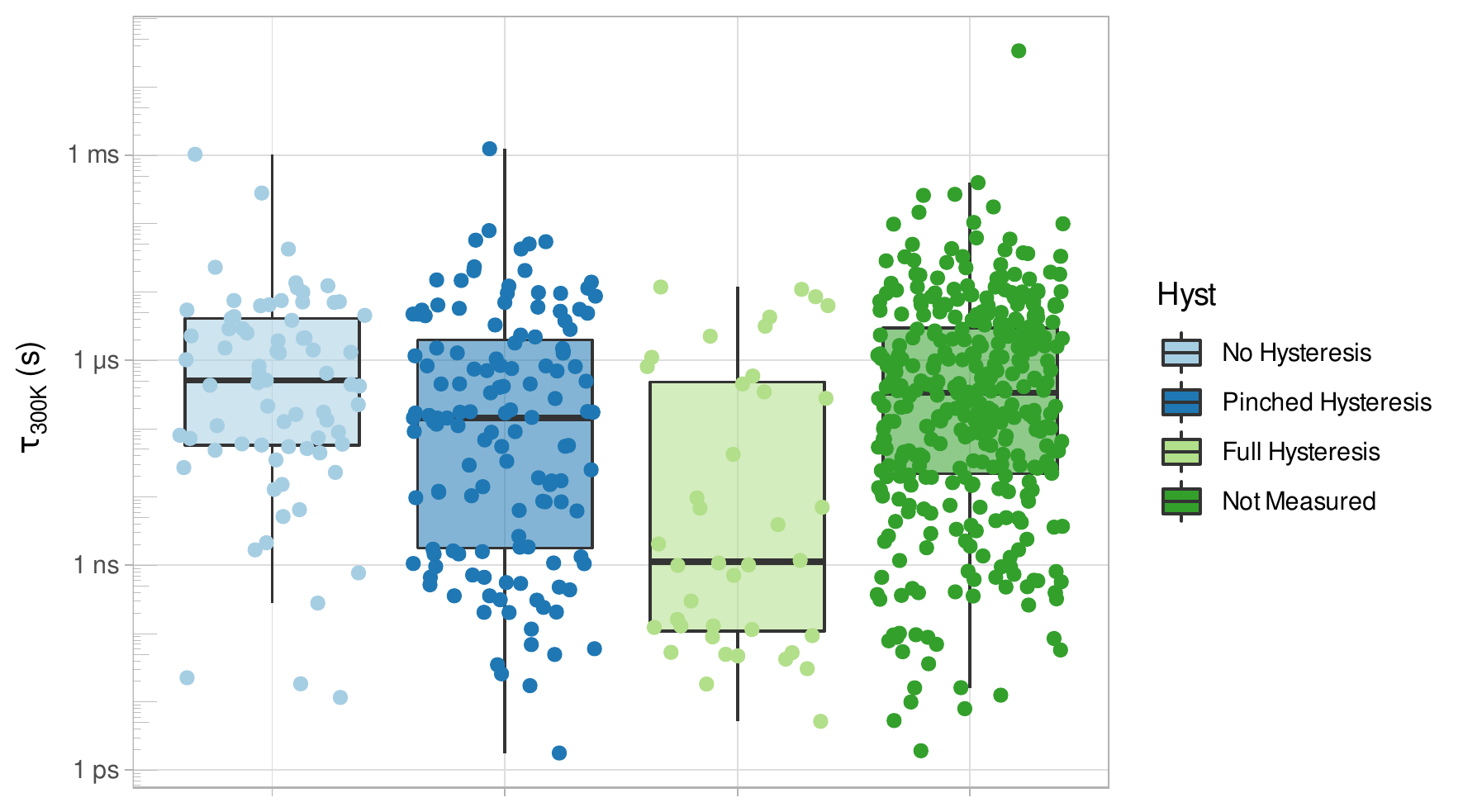}
  \includegraphics[width=0.48\textwidth]{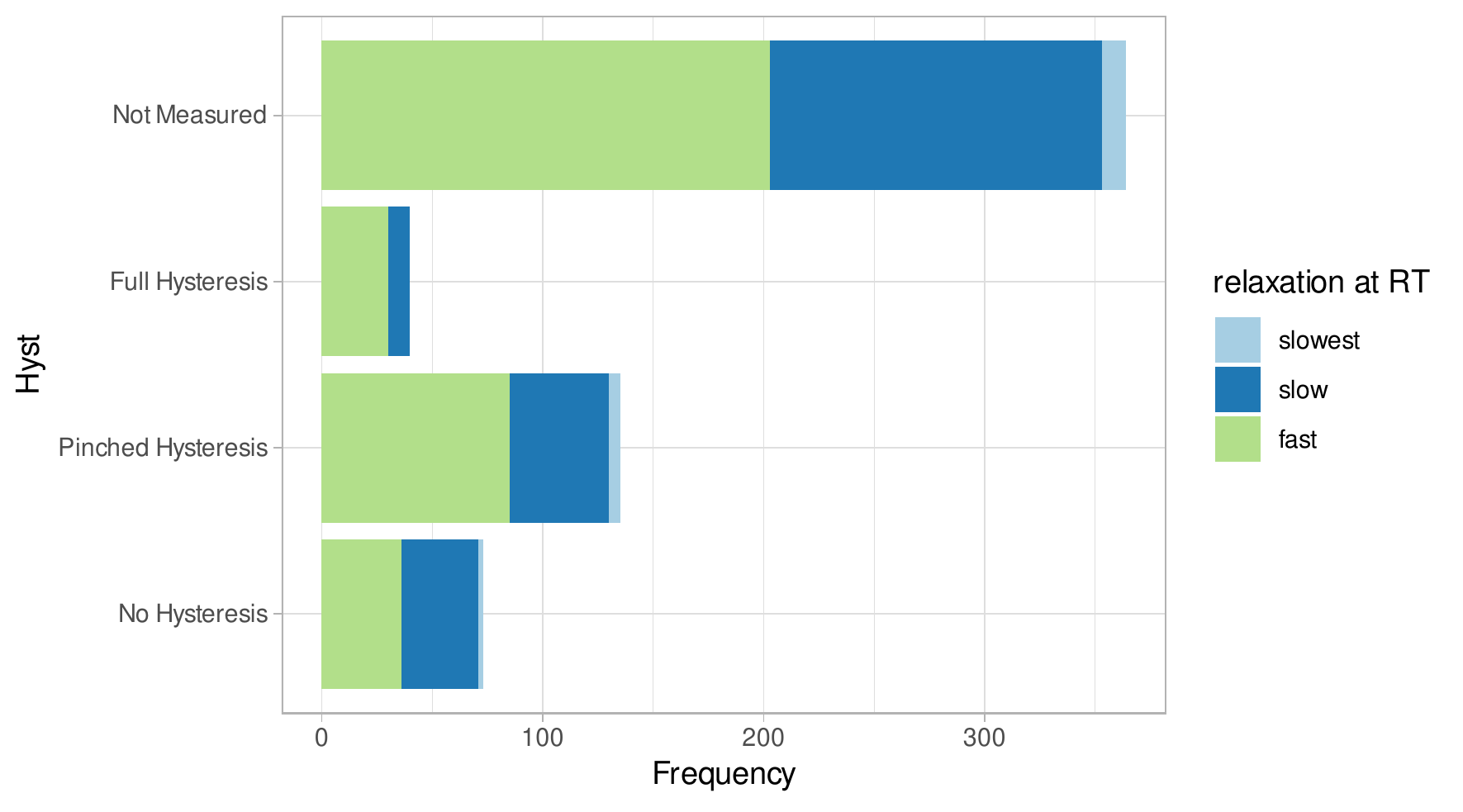}
  \includegraphics[width=0.48\textwidth]{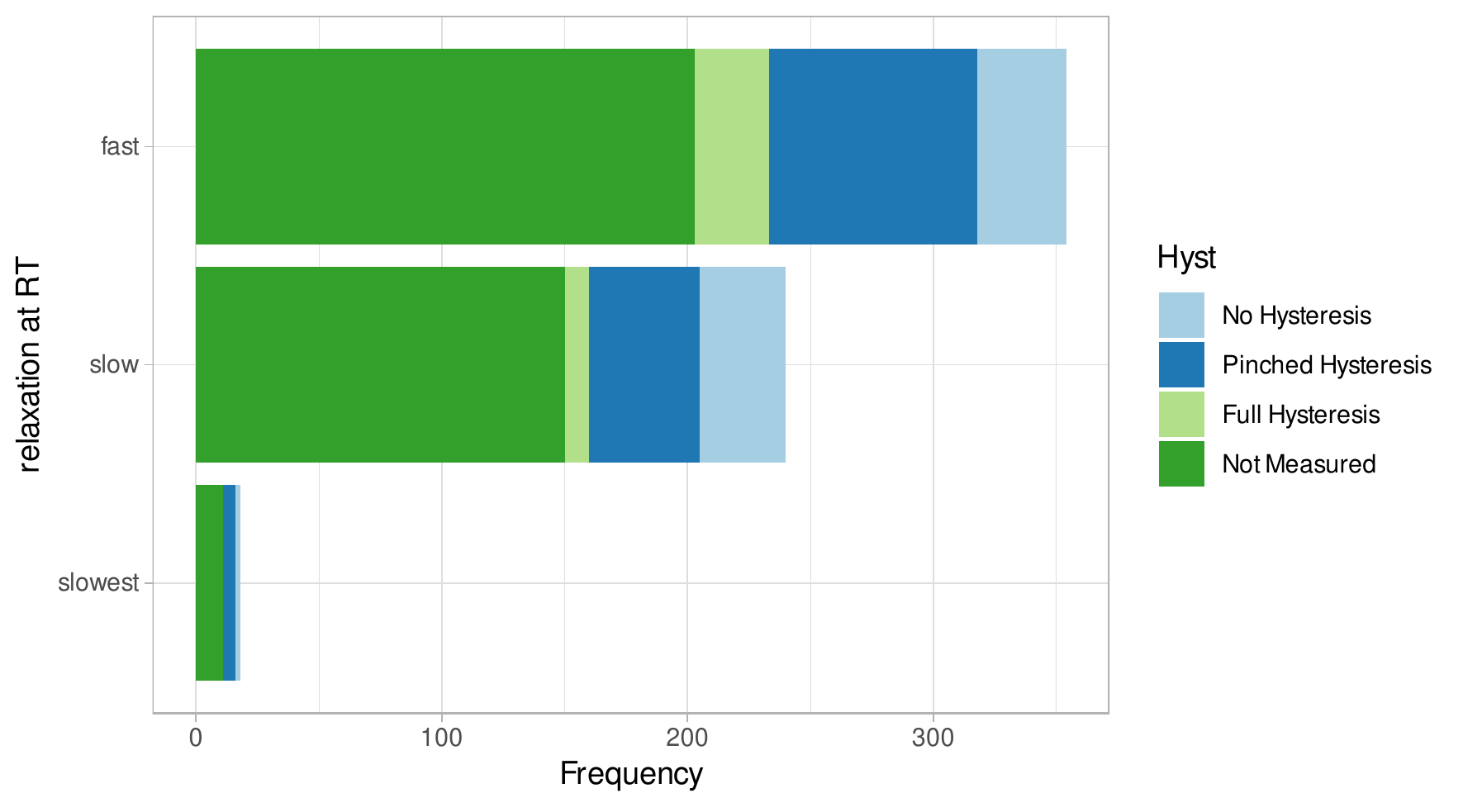}
  \caption{Relationship between the hysteresis behavior and $\tau_{300\mathrm{K}}$ \textcolor{black}{ within the SIMDAVIS dataset}.}
  \label{figSI:tauvsHyst}
\end{figure*}

\begin{figure*}[h]
\centering
  \includegraphics[width=0.67\textwidth]{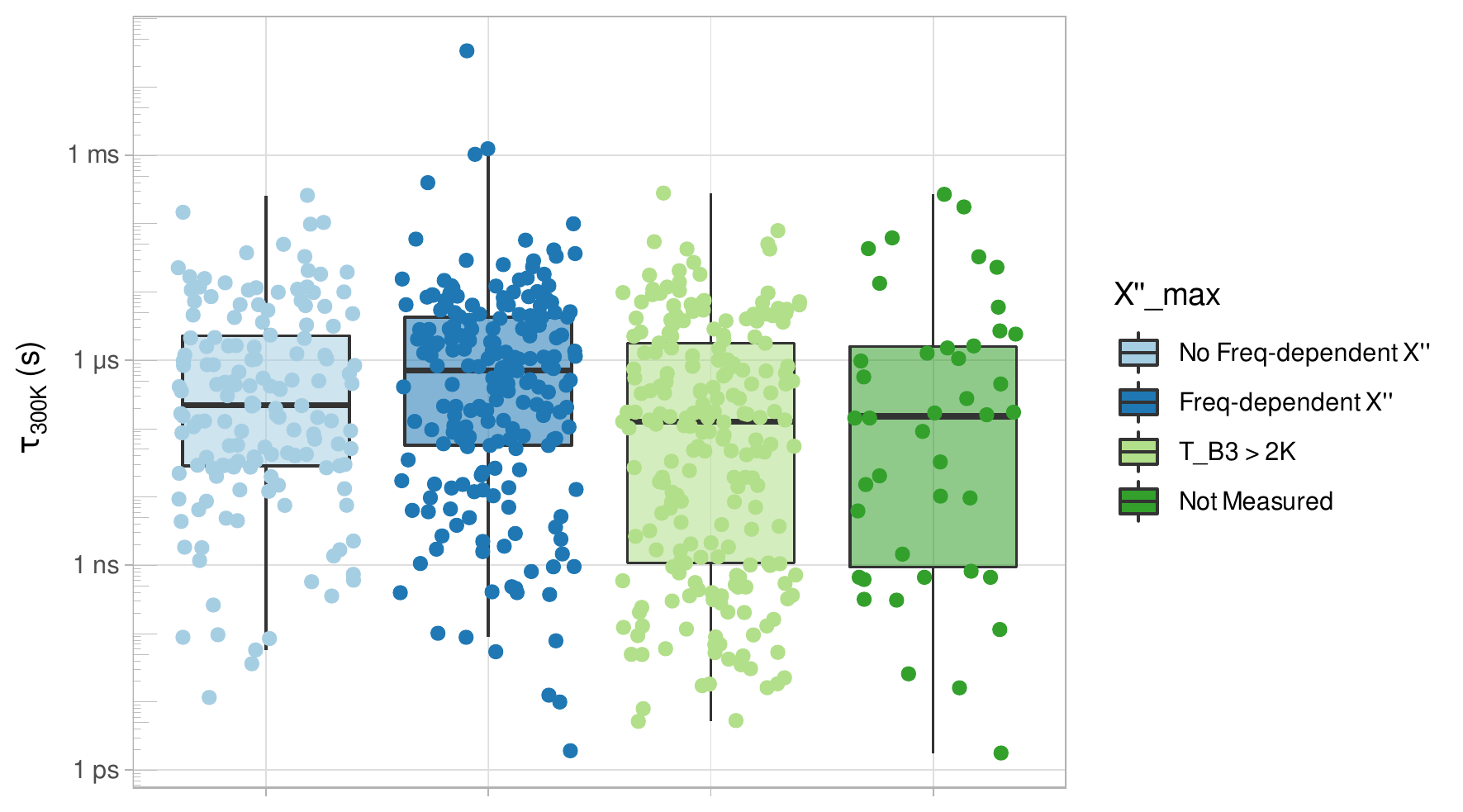}
  \includegraphics[width=0.48\textwidth]{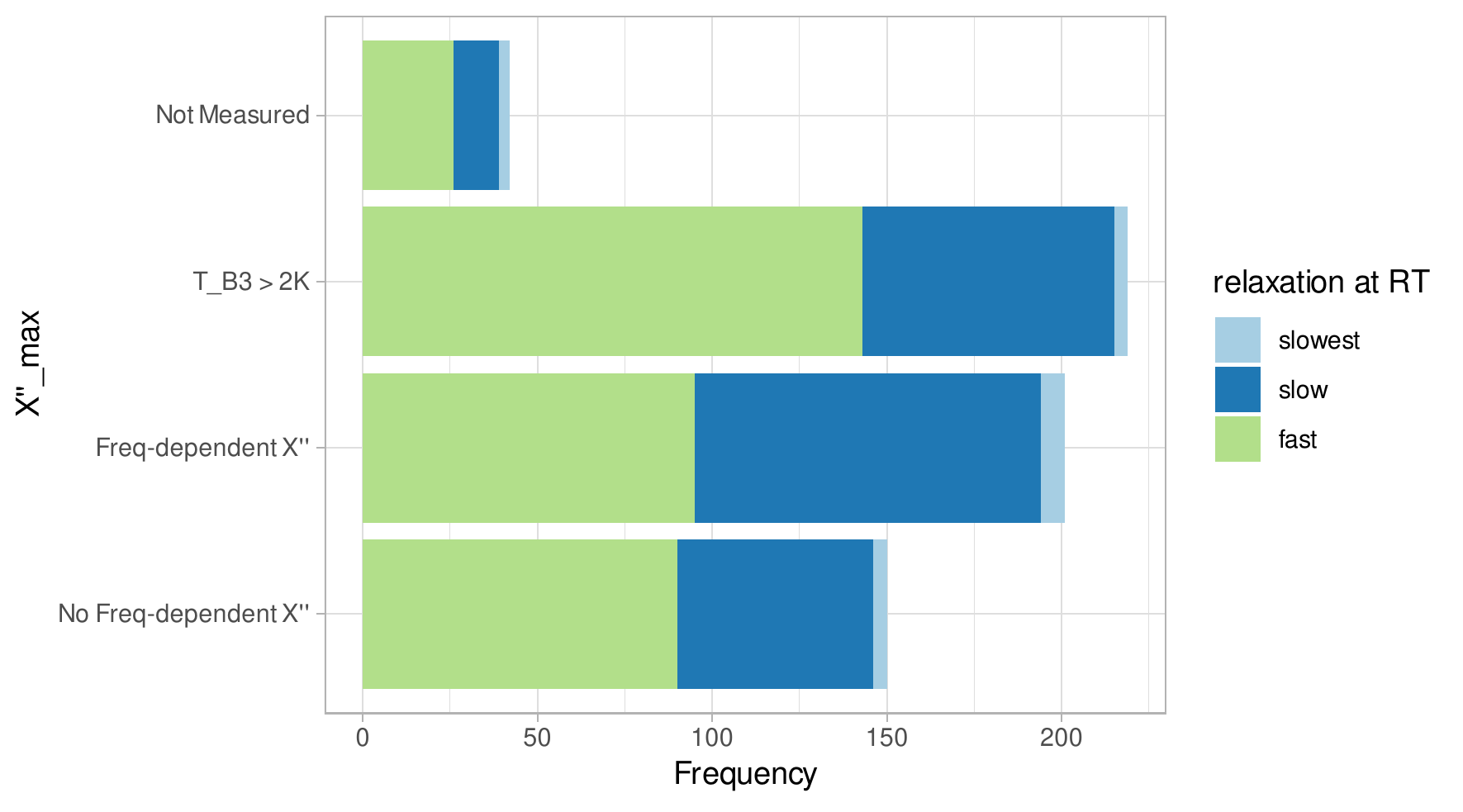}
  \includegraphics[width=0.48\textwidth]{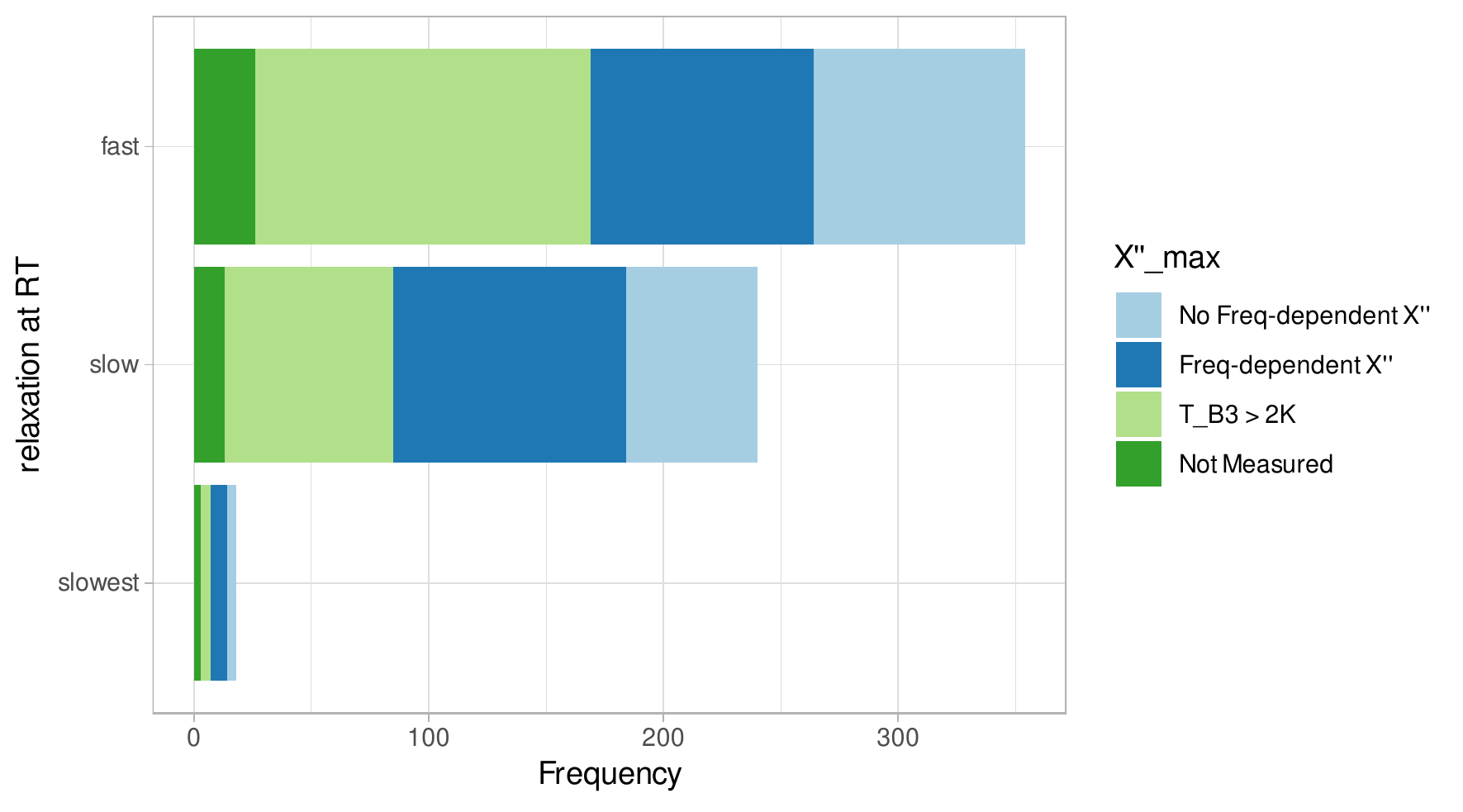}
  \caption{Relationship between the ac out of phase behavior and $\tau_{300\mathrm{K}}$ \textcolor{black}{ within the SIMDAVIS dataset}.}
  \label{figSI:tauvsXim}
\end{figure*}

\clearpage
\textcolor{black}{\section{Screening the SIMDAVIS dataset for p-bit operating speed at $T=4$ K}}

\textcolor{black}{We applied the same criteria as above, except that we filter out all samples where $U_\mathrm{eff} >$ 50 K since either (a) at 4 K this would result in operationally unpractical long times or, often, (b) the fitting data for these cases has generally been obtained at high temperature, neglecting Quantum Tunneling of the Magnetization and Raman processes, one of which would actually be the responsible for the behavior in the limit of low temperature. This means we worked here with a reduced dataset. Even with this caution, lowering the temperature dramatically affects the relaxation times, as it is obviously expected for a thermally activated process: the overall timescale has been slowed down by 3 orders of magnitude.} 

\textcolor{black}{As a main point to consider in this analysis is the fact that at high $T$ the behavior is mostly governed by $\tau_0$ whereas at low $T$ it should be mostly be governed by $U_\mathrm{eff}$. One needs to recall that there is a (weak) negative correlation between $\tau_0$ and $U_\mathrm{eff}$, meaning that in the cases where opposite trends are observed for high and low $T$ these can be rationalized as being due to this negative correlation between $\tau_0$ and $U_\mathrm{eff}$, whereas in the cases where the same trends are observed for high and low $T$, these happen despite this negative correlation between $\tau_0$ and $U_\mathrm{eff}$ and are therefore more noteworthy. As we will see below, the expected opposite behaviors for high and low $T$ are observed in the case of the coordination number, number of ligands and closest polyhedron. In contrast, similar trends for high and low $T$ are observed for the coordination elements, lanthanide ion (independently including oblate/prolate and Kramers/non-Kramers character).}


\subsection{\textcolor{black}{Coordination sphere}}

\textcolor{black}{A first striking difference here is the absence of Carbon in the coordination sphere. This is a result of the exclusion of samples with $U_\mathrm{eff} >$ 50 K. Other than that, in the boxplot representation one can see that, again, lanthanides coordinated by Nitrogen relax slightly faster in average, and samples coordinated by a combination of Oxygen and Nitrogen atoms present a slighly slower relaxation. This is confirmed in the barcharts, where the Nitrogen+Oxygen coordination is overrepresented in the "slowest" category at $T=4$ K, just like it was overrepresented in the "slow" category at $T=300$ K.}

\begin{figure*}[h]
\centering
  \includegraphics[width=0.67\textwidth]{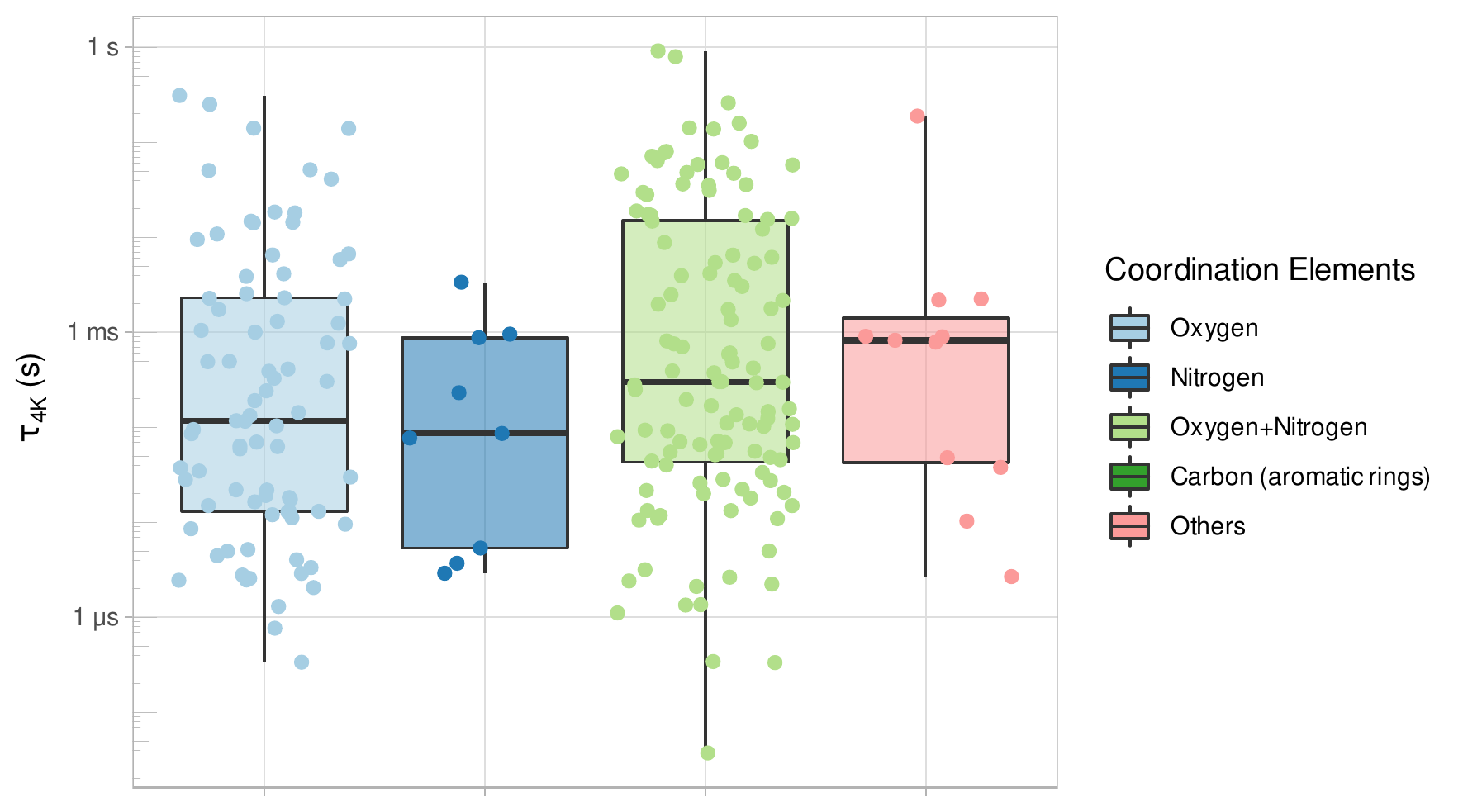}
  \includegraphics[width=0.48\textwidth]{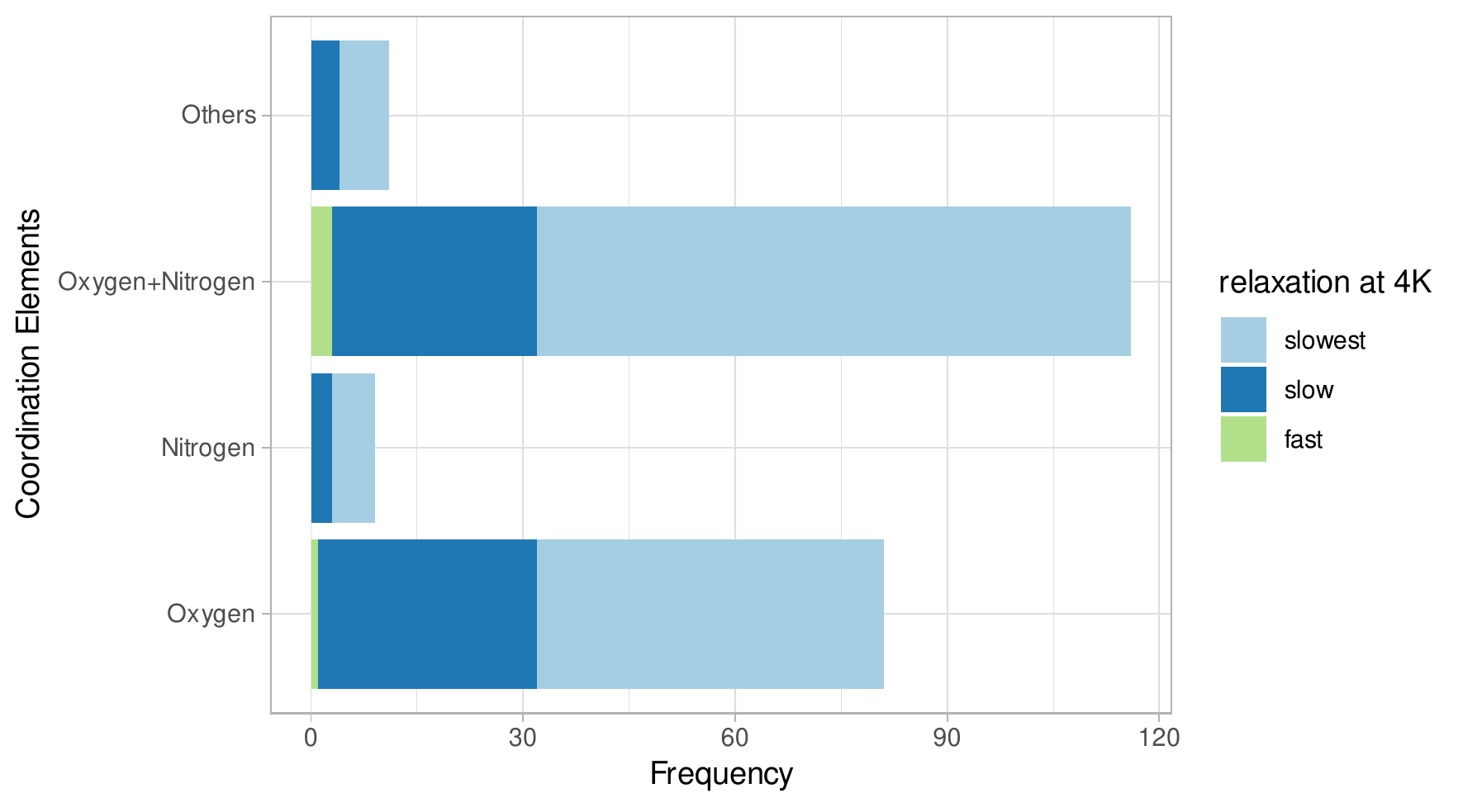}
  \includegraphics[width=0.48\textwidth]{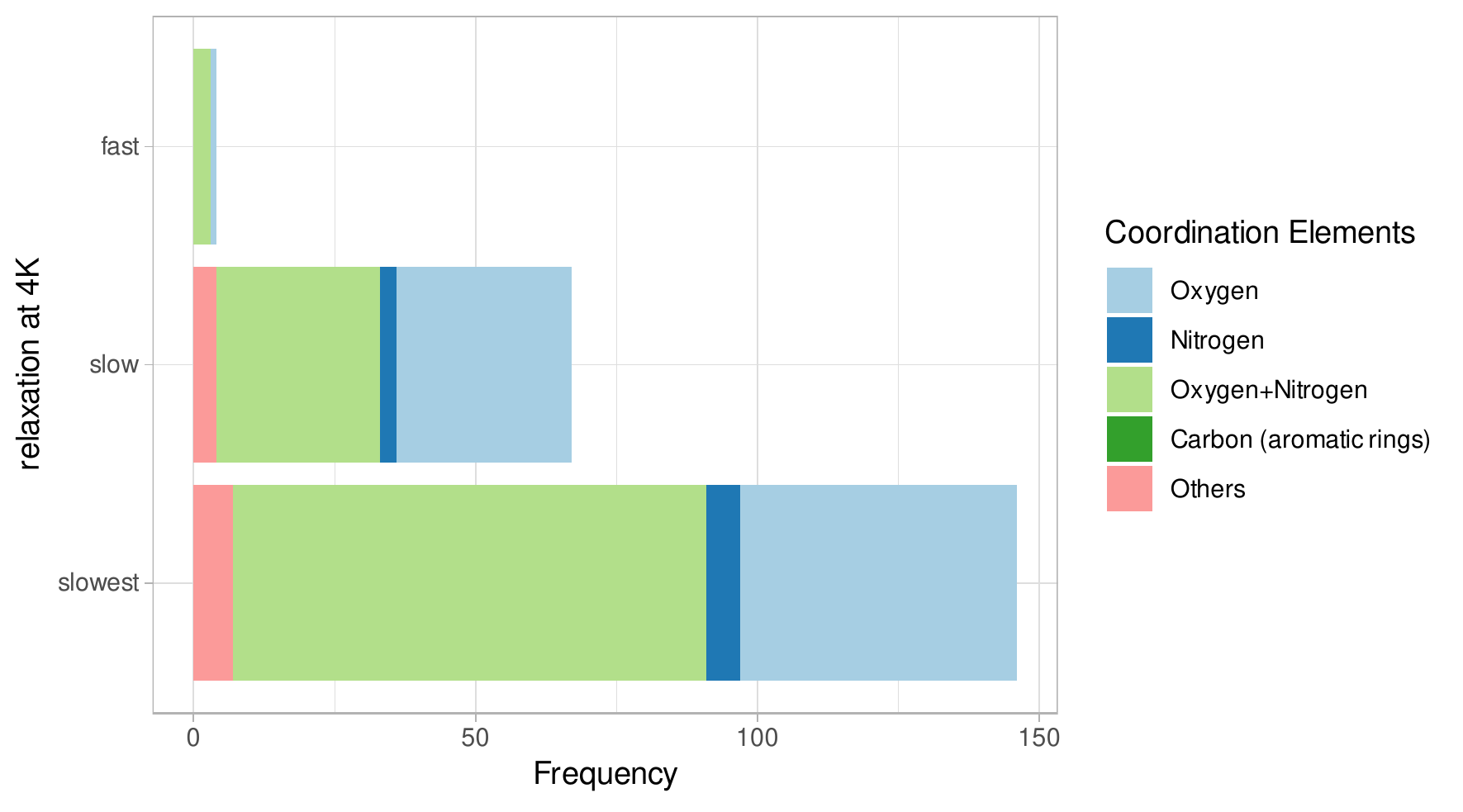}
  \caption{\textcolor{black}{Relationship between the coordination elements and $\tau_{4\mathrm{K}}$ within the SIMDAVIS dataset}.}
  \label{figSI:tauvscoordele4k}
\end{figure*}

\textcolor{black}{In contrast, we see a difference between the high-T and the low-T behavior in the case of the coordination number. Here it seems that the median $\tau_{4\mathrm{K}}$ decreases with increasing coordination number, while the opposite behavior was observed for $\tau_{300\mathrm{K}}$. For the number of ligands, the tendency is less clear but again we can see that the tendency is not necessarily mantained, with ions coordinated by seven ligands, and in particular for pentagonal bipyramid coordination, being at $T=4$ K by no means especially faster relaxing compared with others with lower or higher ligand number, as was the case at $T=300$ K.}

\begin{figure*}[h]
\centering
  \includegraphics[width=0.67\textwidth]{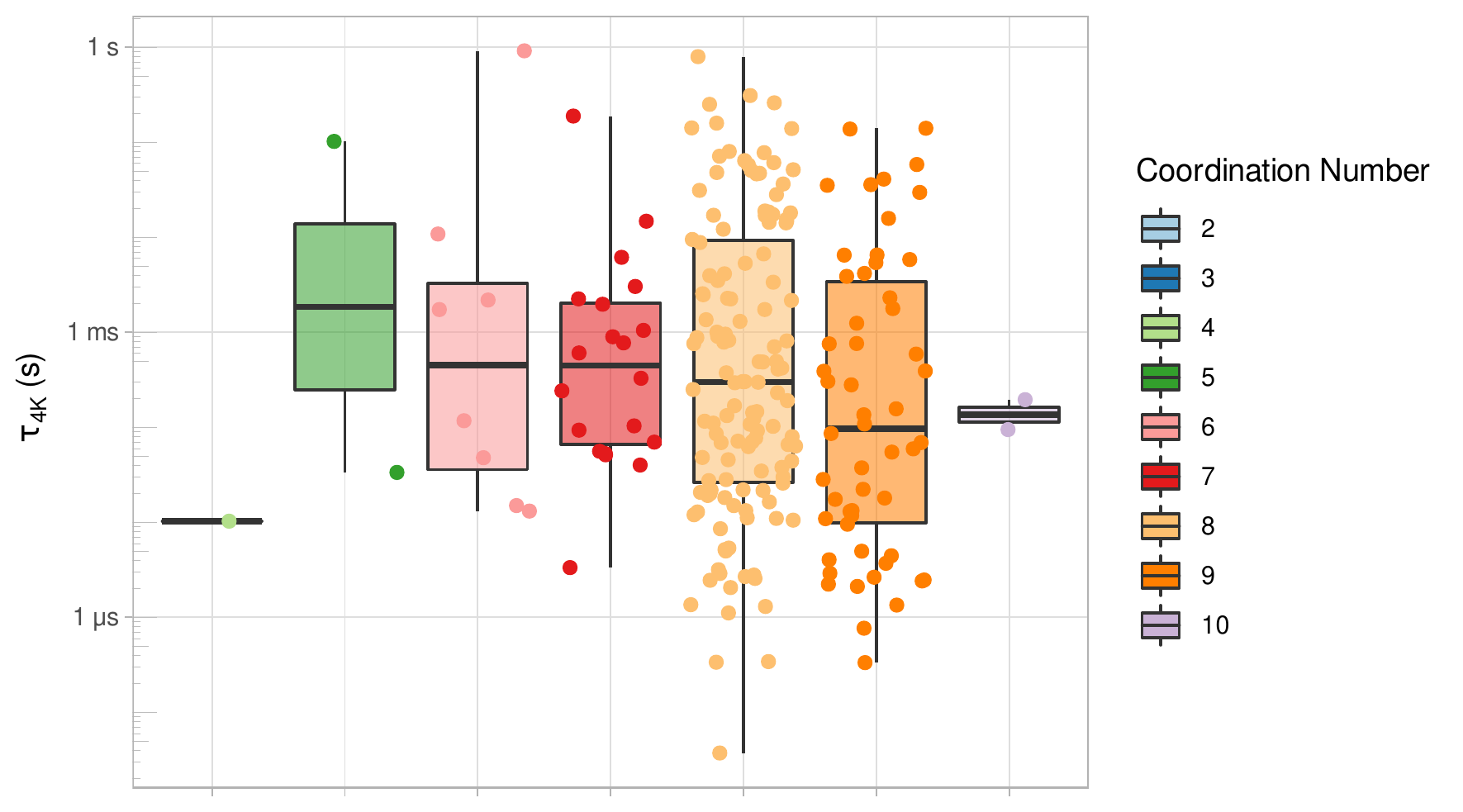}
  \includegraphics[width=0.48\textwidth]{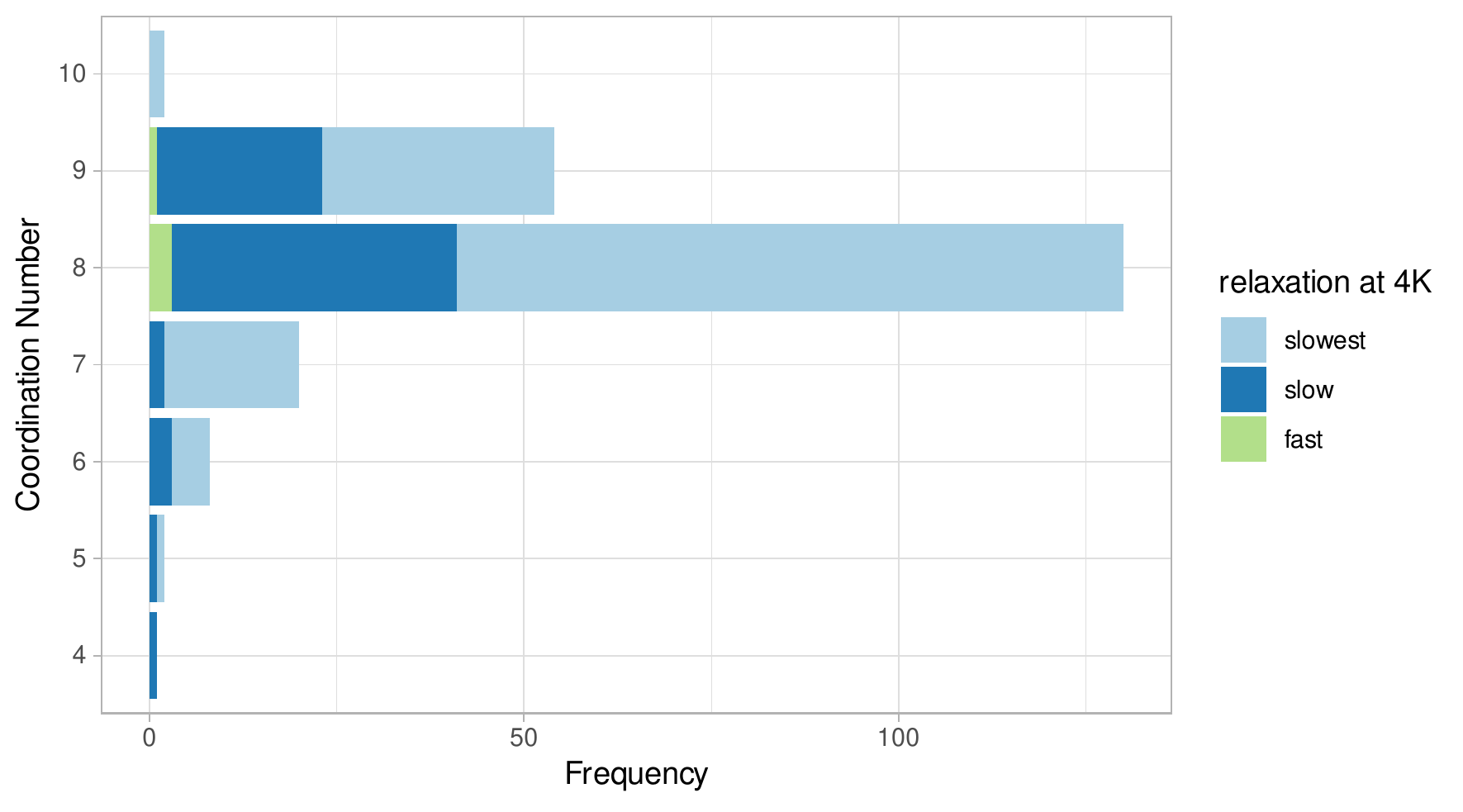}
  \includegraphics[width=0.48\textwidth]{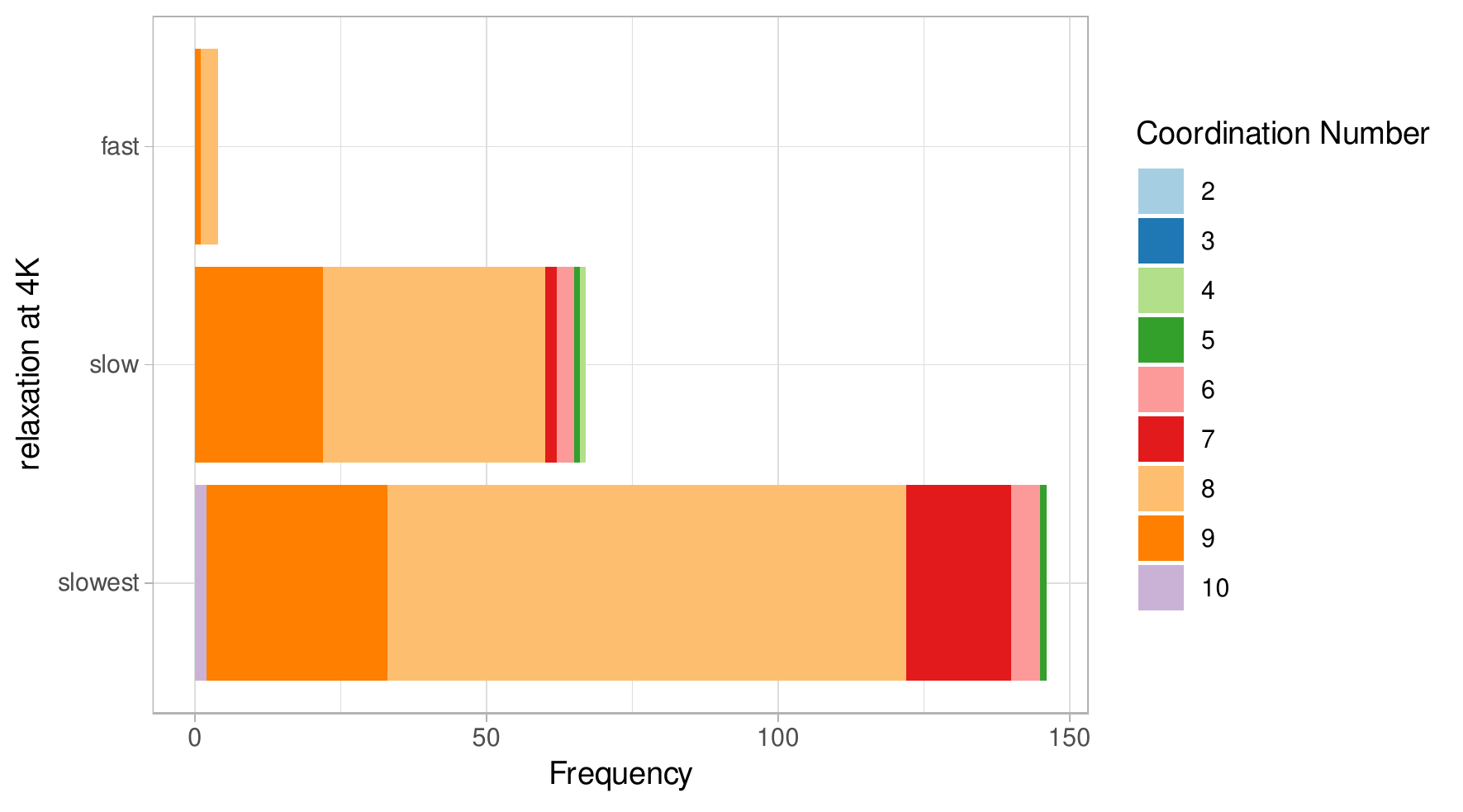}
  \caption{\textcolor{black}{Relationship between the coordination number and $\tau_{4\mathrm{K}}$ within the SIMDAVIS dataset}.}
  \label{figSI:tauvscoordnum4k}
\end{figure*}

\begin{figure*}[h]
\centering
  \includegraphics[width=0.67\textwidth]{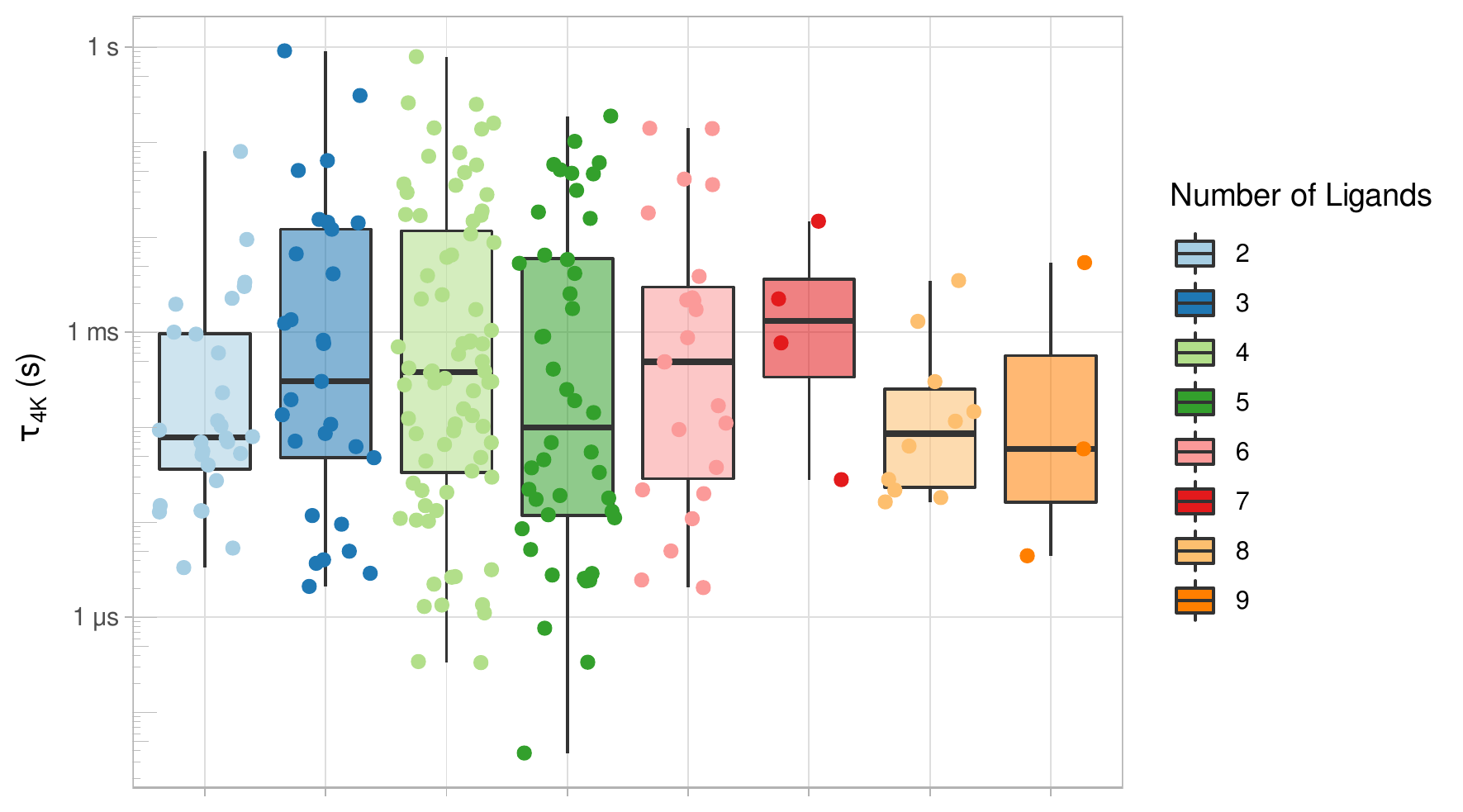}
  \includegraphics[width=0.48\textwidth]{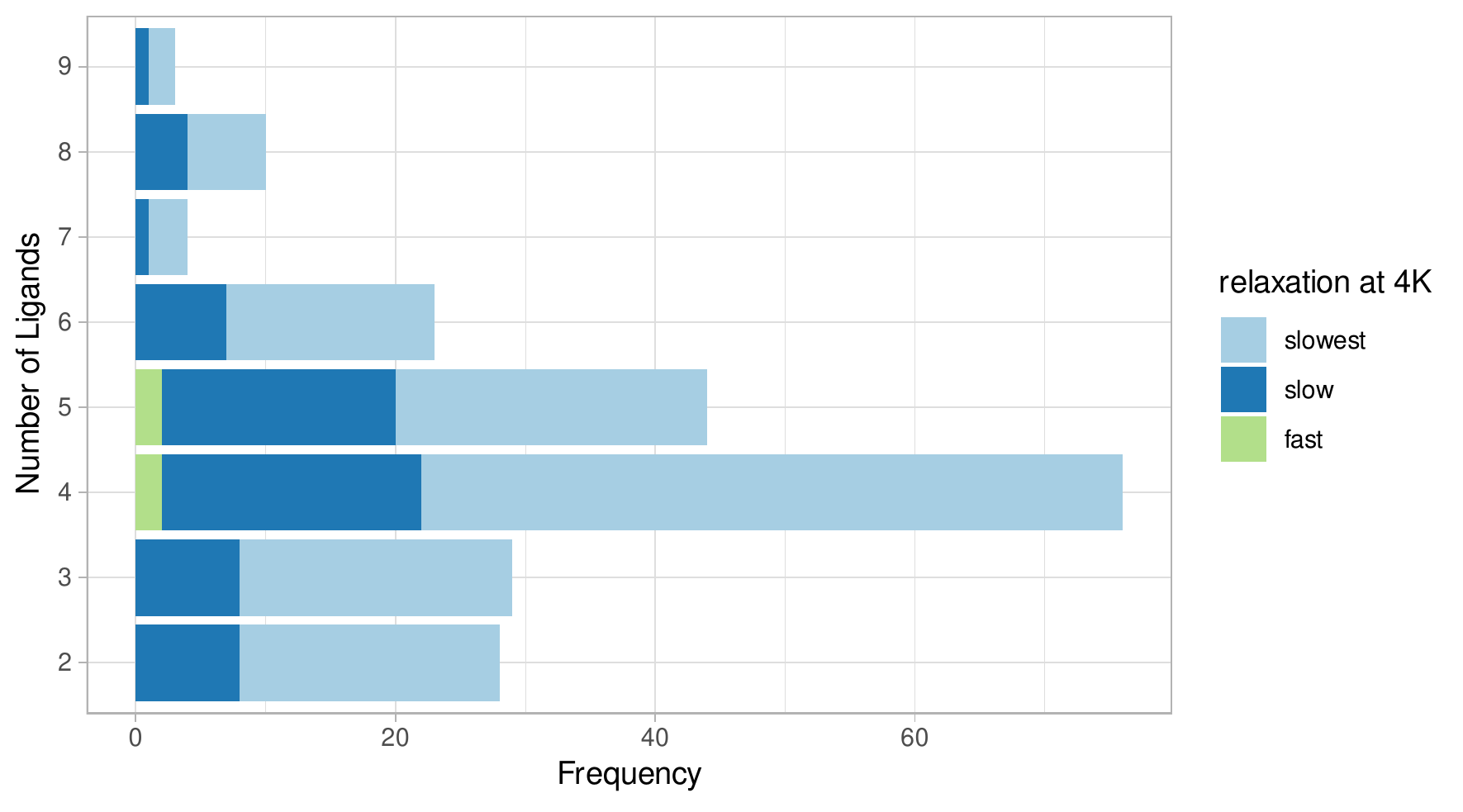}
  \includegraphics[width=0.48\textwidth]{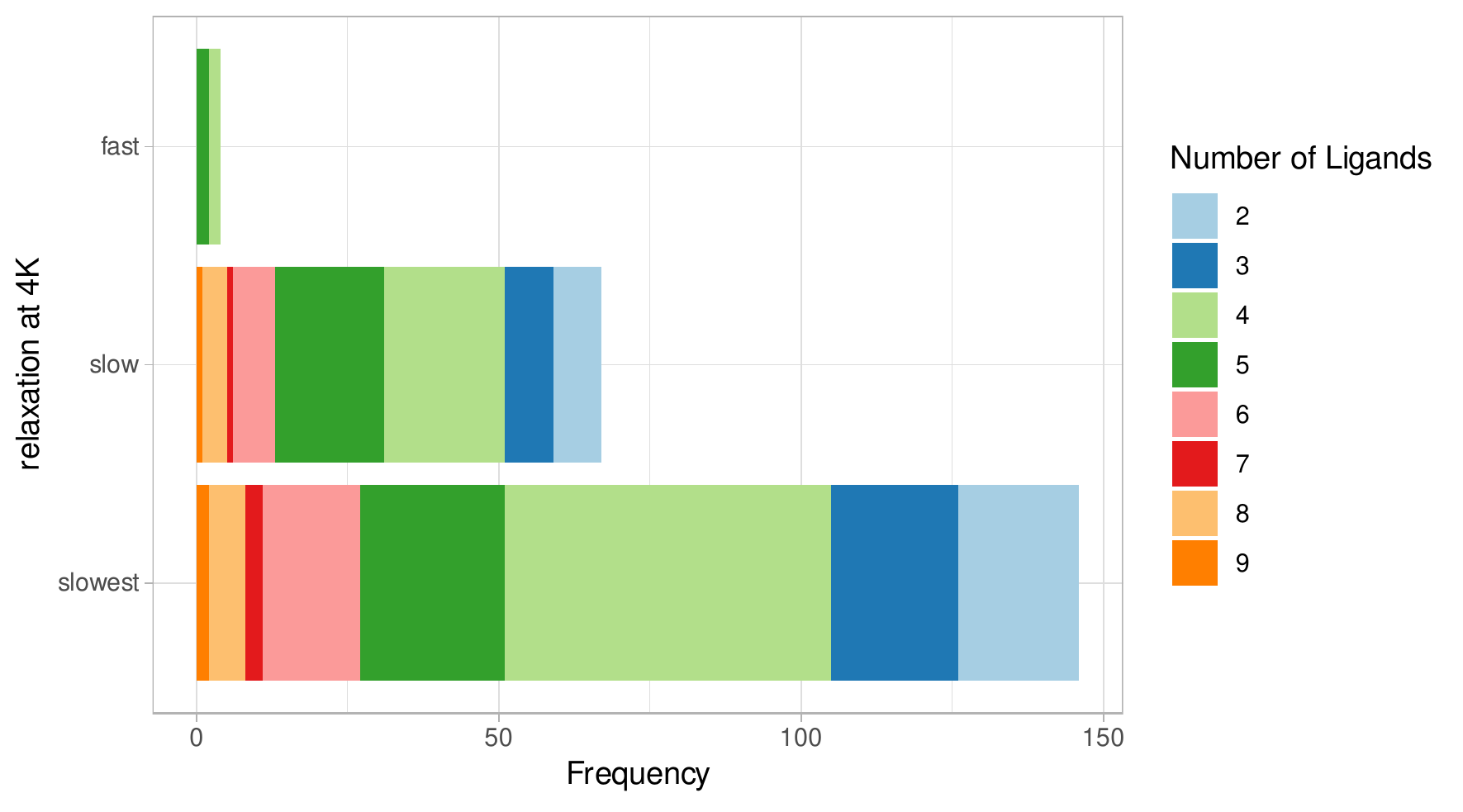}
  \caption{\textcolor{black}{Relationship between the number of ligands and $\tau_{4\mathrm{K}}$ within the SIMDAVIS dataset}.}
  \label{figSI:tauvsnumligands4k}
\end{figure*}

\begin{figure*}[h]
\centering
  \includegraphics[width=0.67\textwidth]{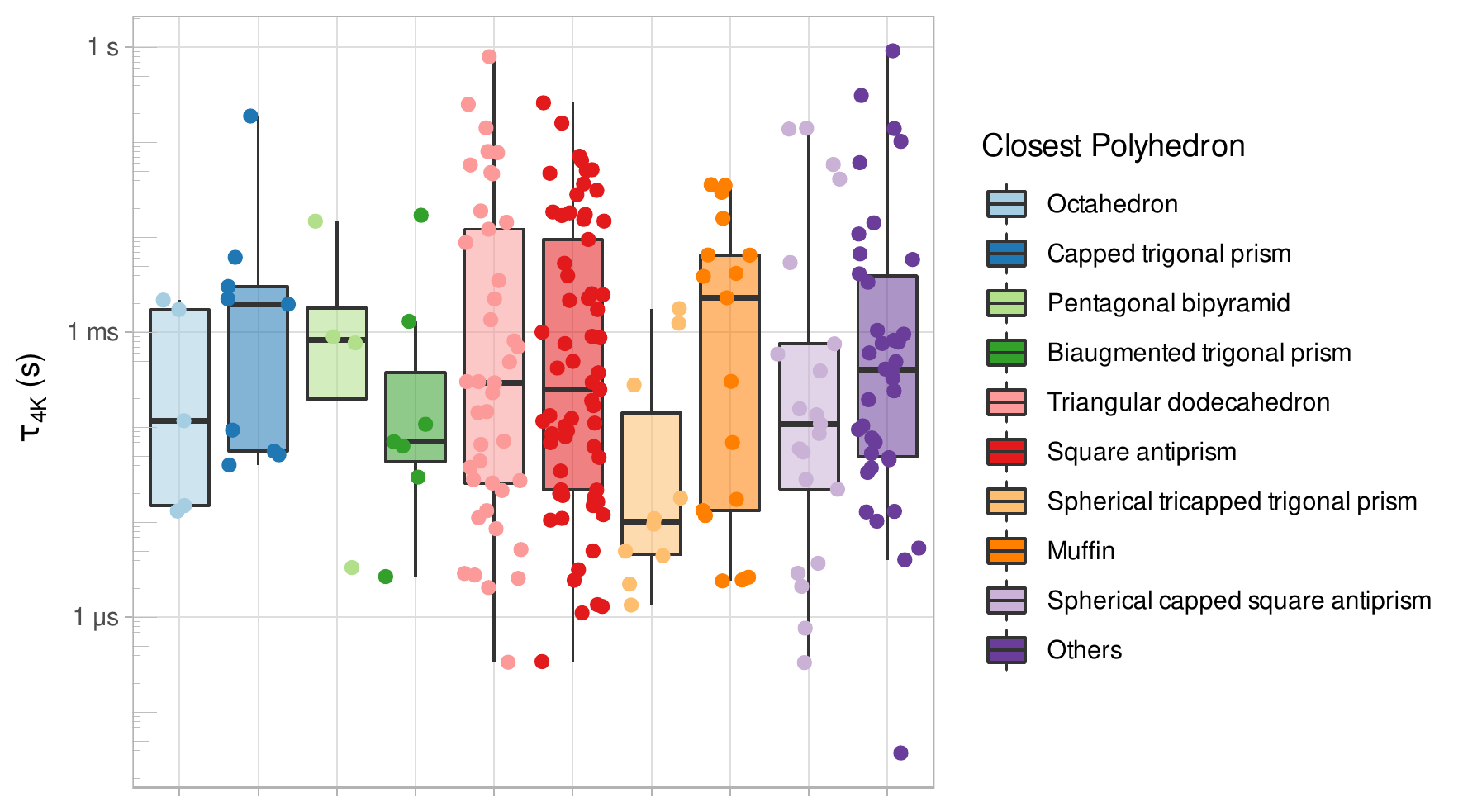}
  \includegraphics[width=0.48\textwidth]{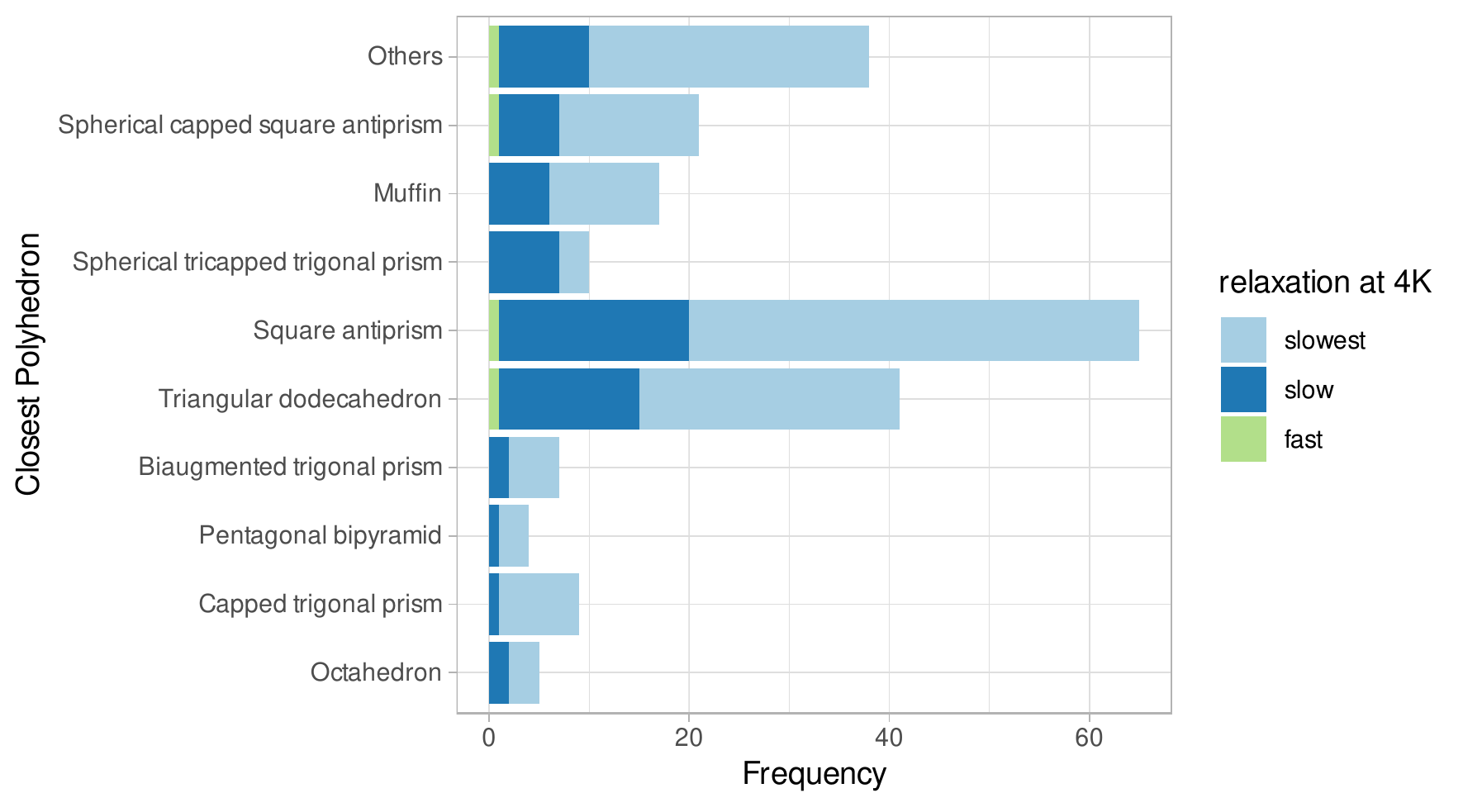}
    \includegraphics[width=0.48\textwidth]{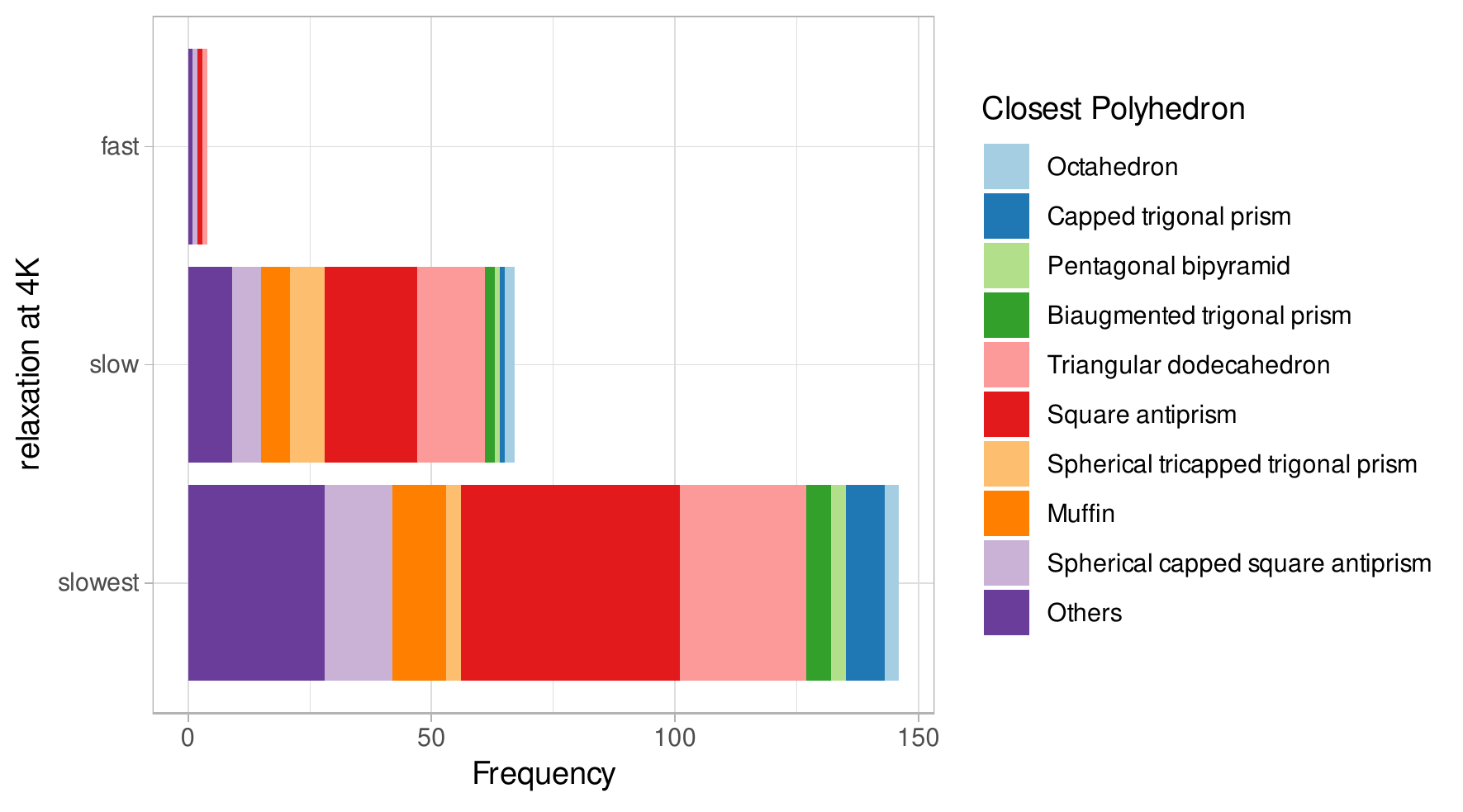}
  \caption{\textcolor{black}{Relationship between the coordination polyhedron and $\tau_{4\mathrm{K}}$ within the SIMDAVIS dataset}.}
  \label{figSI:tauvspoly4k}
\end{figure*}

\textcolor{black}{Overall, in terms of coordination chemistry, the recipe for p-bits that one can reliably operate relatively fast, even at $T=4$ K, is of course "low $U_\mathrm{eff}$". What we see here is that "relatively fast" in this case is mostly the ms-$\mu$s regime, and that typical coordination shapes are the ones that generally result in poorly behaved SIMs, such as octahedra, biaugmented trigonal prisms or spherical tricapped trigonal prisms.}

\clearpage
\subsection{\textcolor{black}{Lanthanide ion}}

\textcolor{black}{In terms of the Kramers vs non-Kramers character, the behavior at $T=4$~K coincides with the behavior at $T=300$~K, in the sense that oblate ions present slower relaxation, in this case significantly slower. Again for the Kramers vs non-Kramers the difference here is much more marked at $T=4$~K, with Kramers ions relaxing much slower than non-Kramers. In terms of the particular lanthanide, the trends are similar at both temperature limits, with Nd, Dy and Yb being slightly slower, and Tb, Er being slighly faster.}

\begin{figure*}[h]
\centering
  \includegraphics[width=0.67\textwidth]{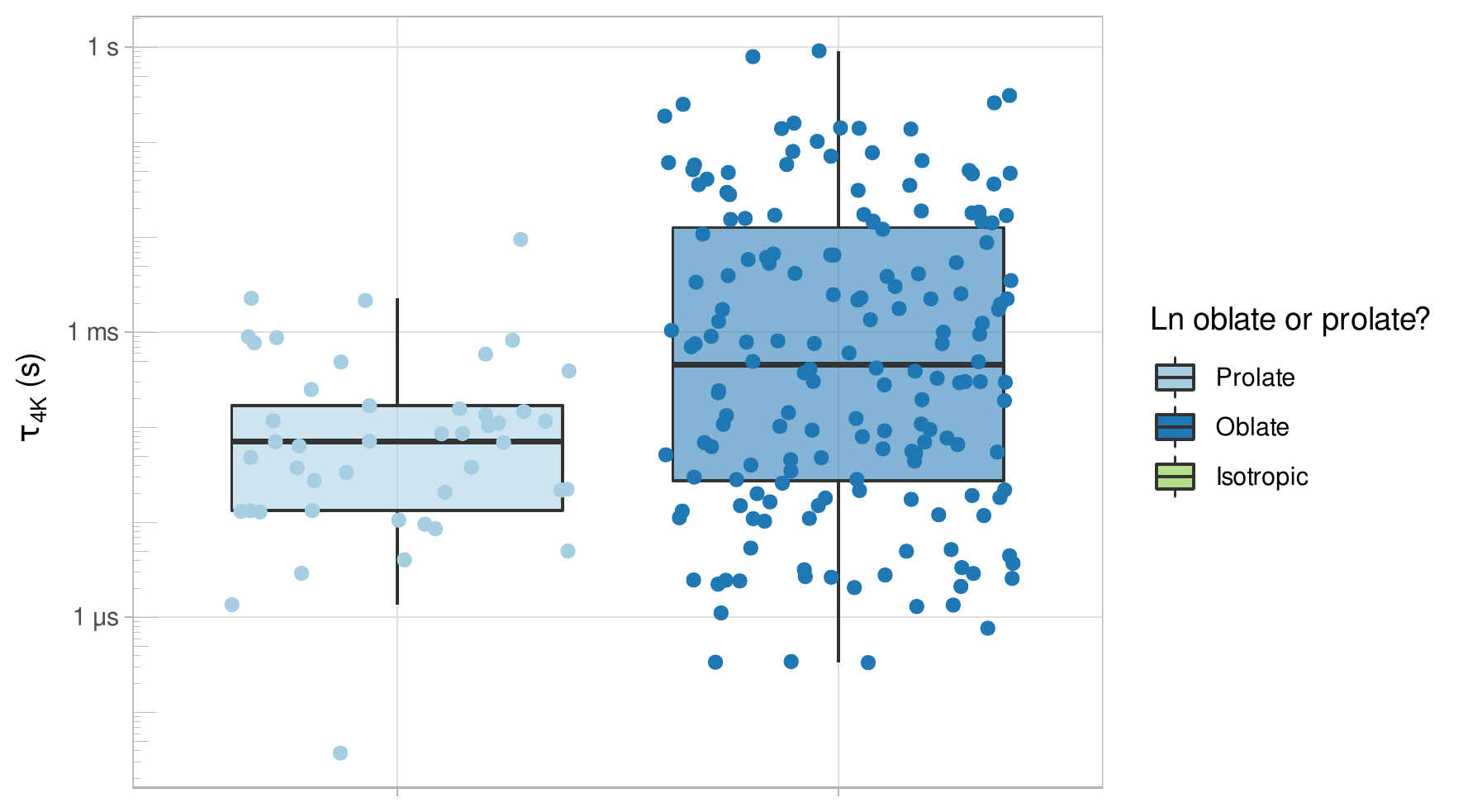}
  \includegraphics[width=0.48\textwidth]{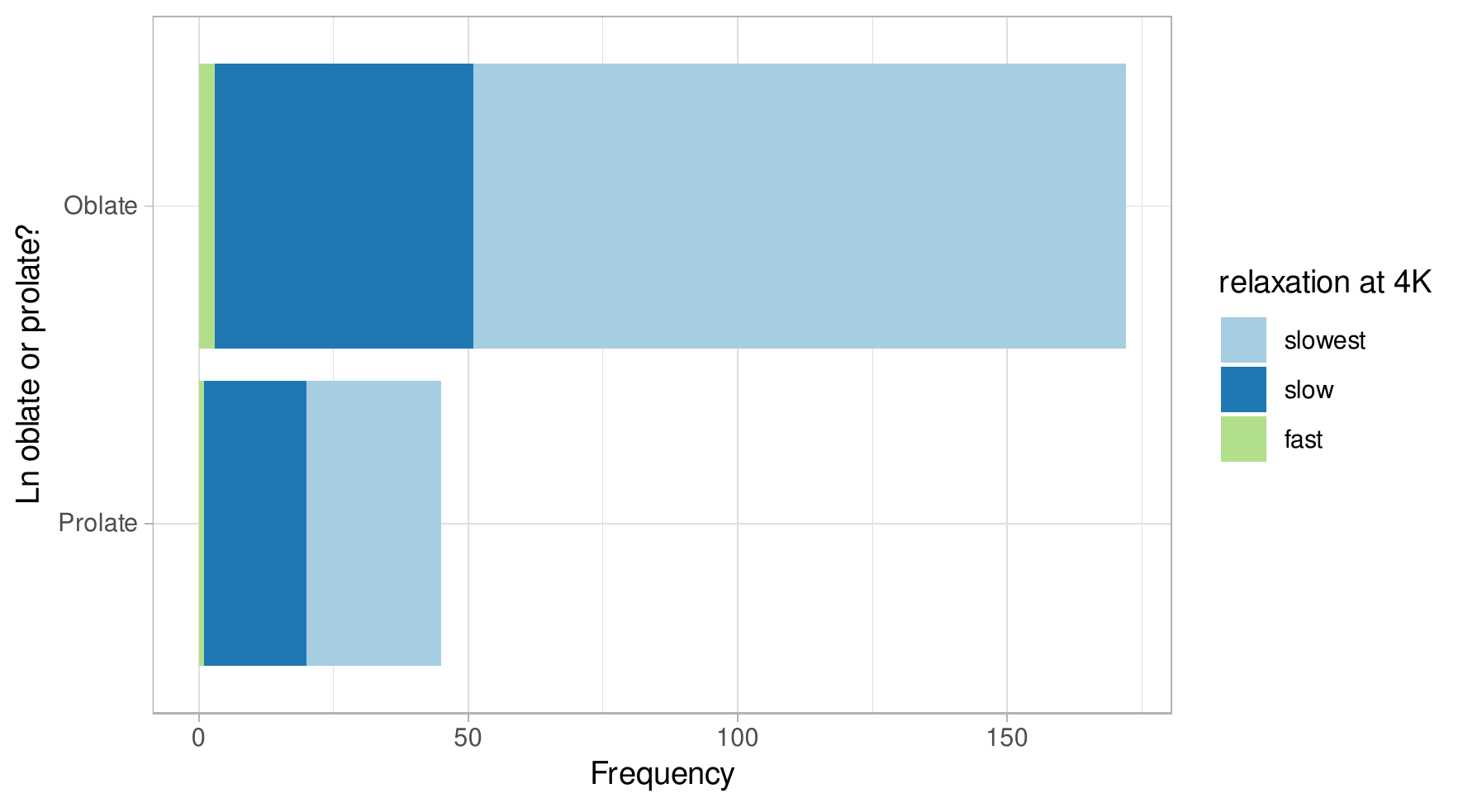}
  \includegraphics[width=0.48\textwidth]{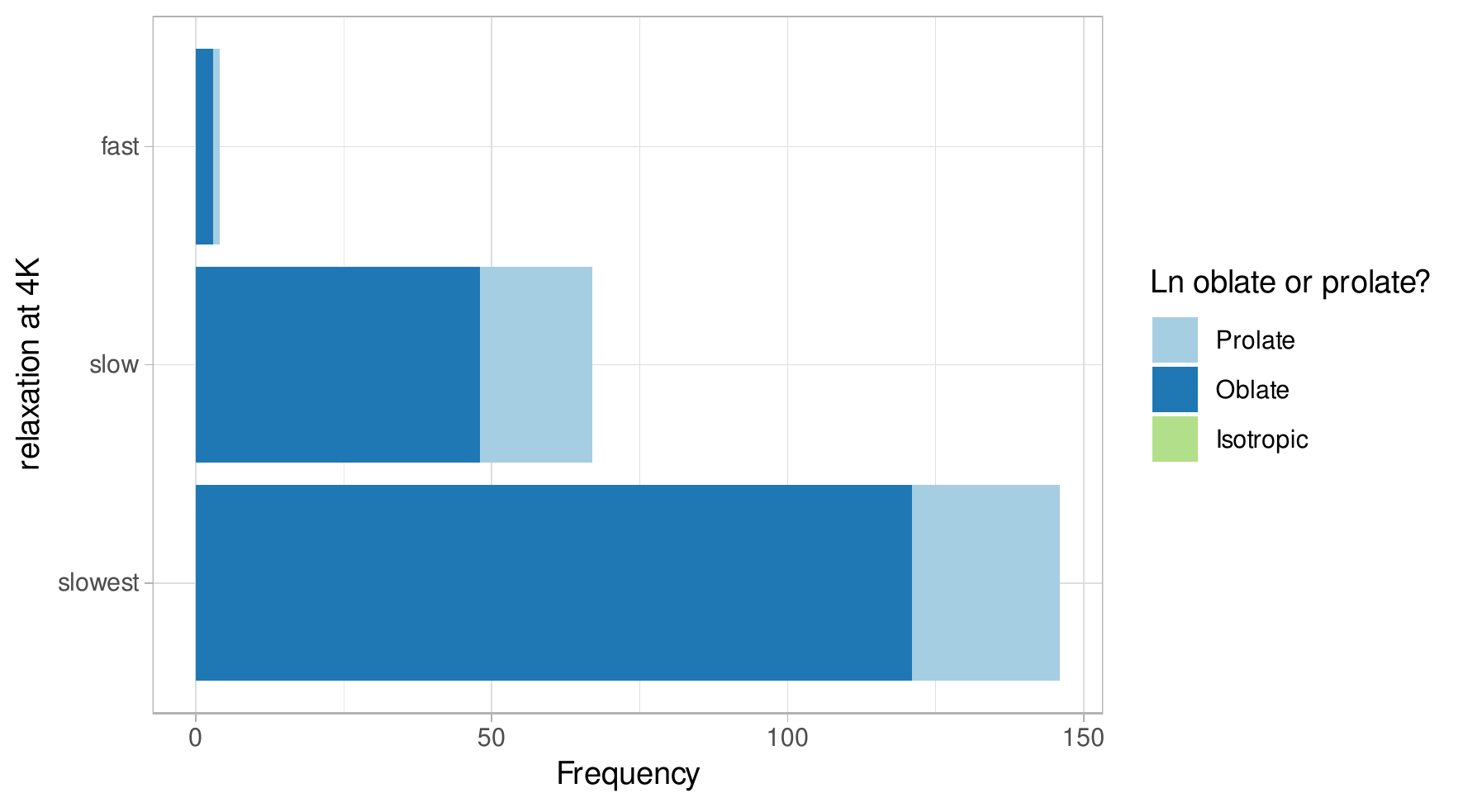}
  \caption{\textcolor{black}{Relationship between the lanthanide ion's anisotropy and $\tau_{4\mathrm{K}}$ within the SIMDAVIS dataset}.}
  \label{figSI:tauvsLnani4k}
\end{figure*}

\begin{figure*}[h]
\centering
  \includegraphics[width=0.67\textwidth]{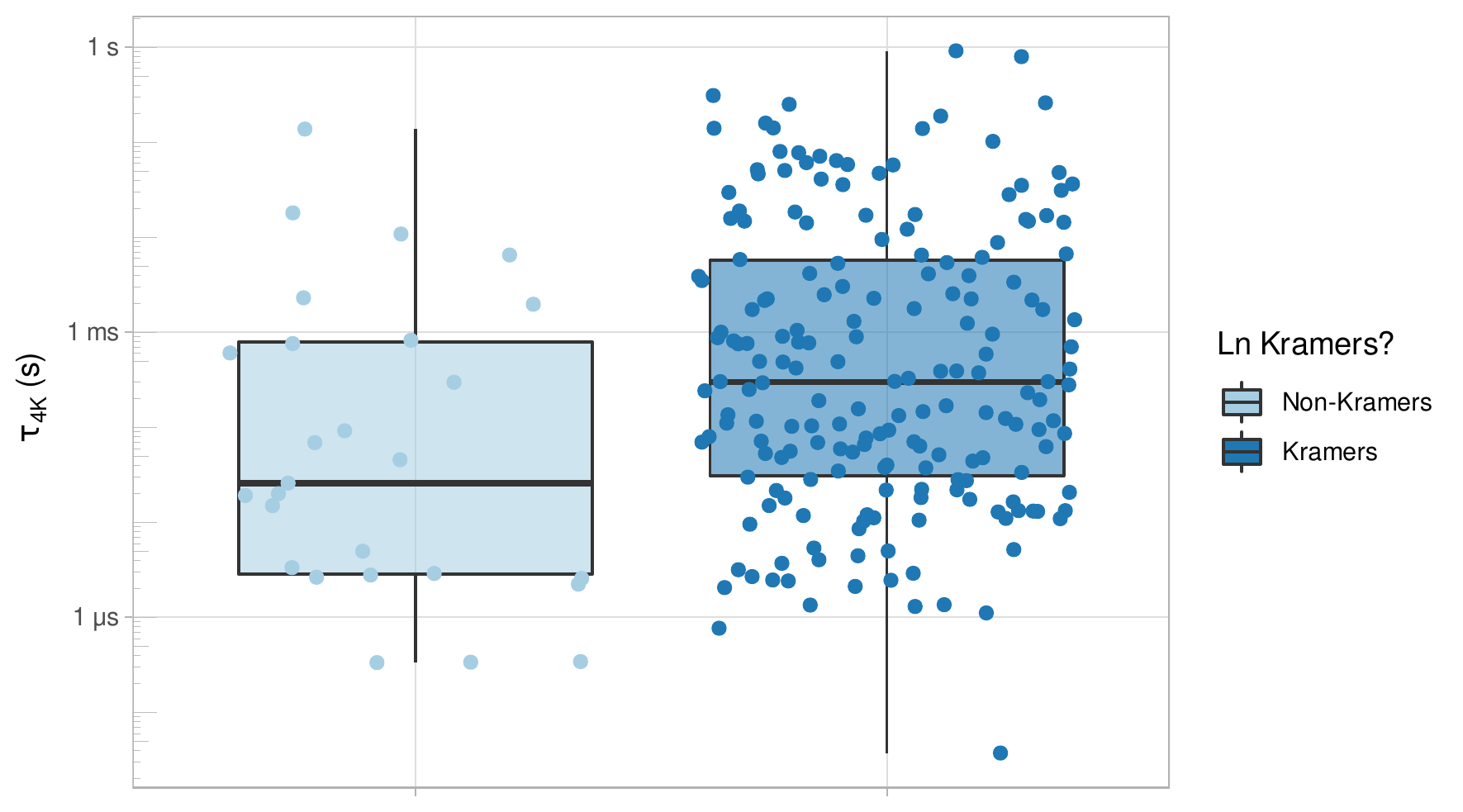}
  \includegraphics[width=0.48\textwidth]{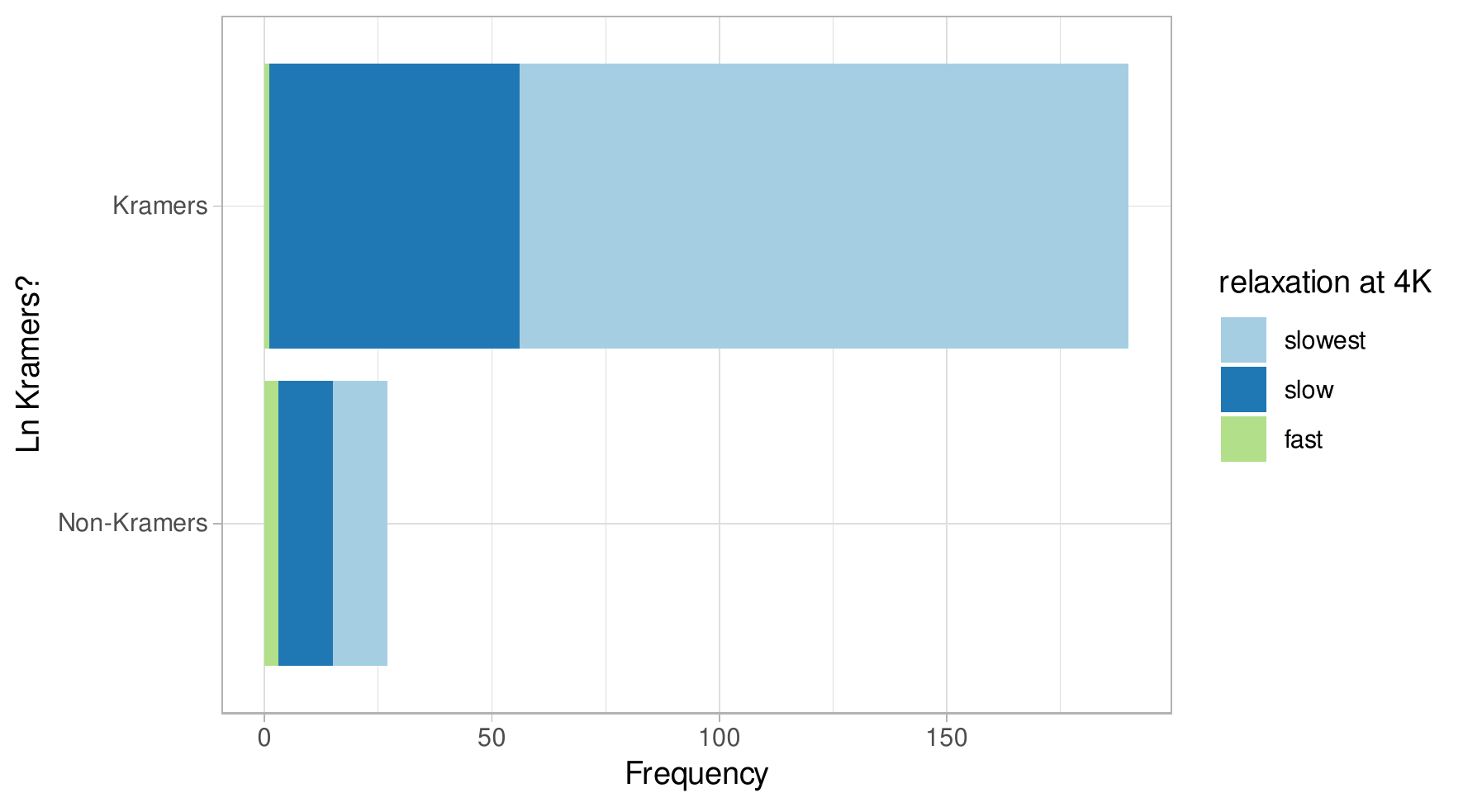}
  \includegraphics[width=0.48\textwidth]{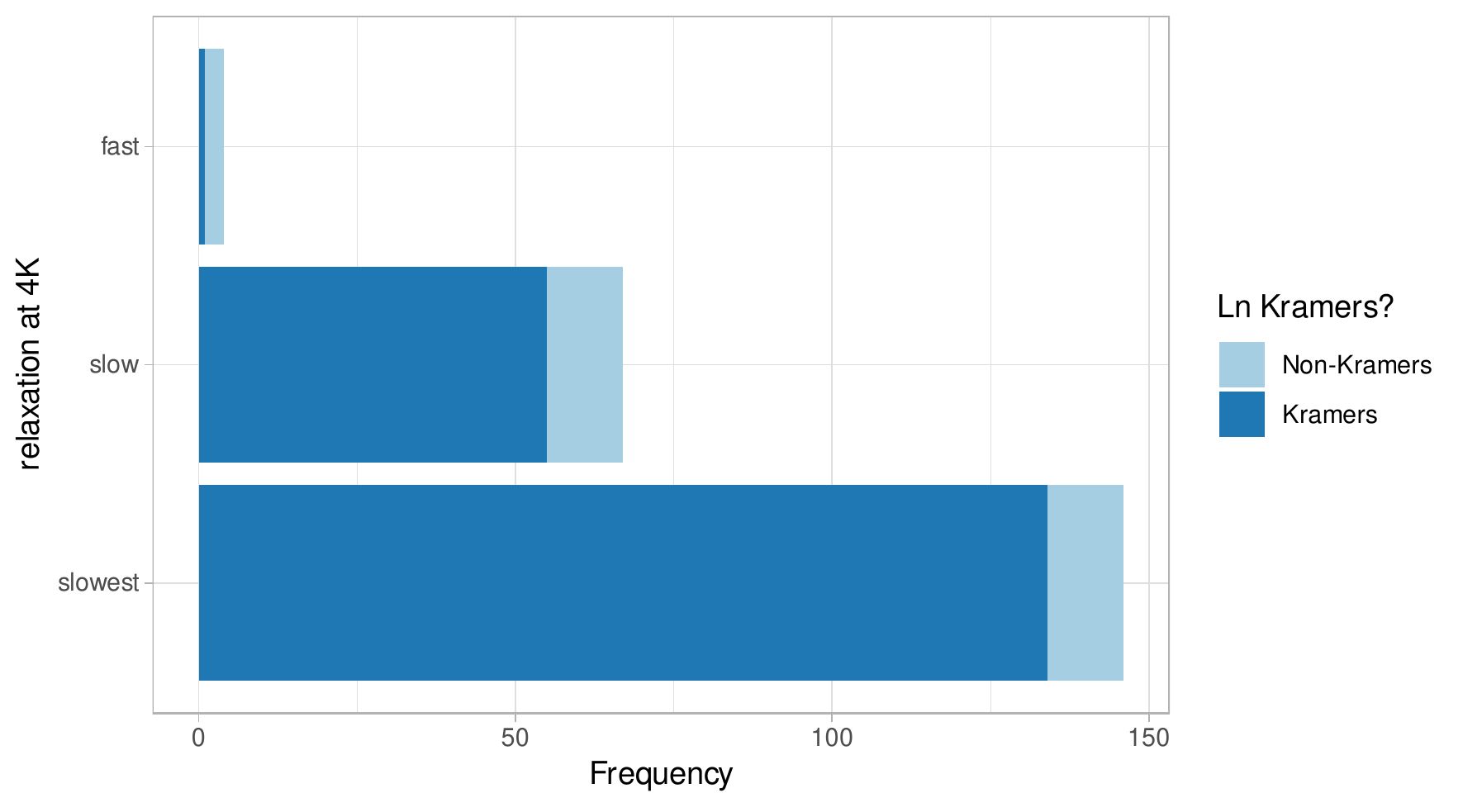}
  \caption{\textcolor{black}{Relationship between the lanthanide ion Kramer's or non-Kramer's character and $\tau_{4\mathrm{K}}$ within the SIMDAVIS dataset}.}
  \label{figSI:tauvsLnKra4k}
\end{figure*}

\clearpage

\begin{figure*}[h]
\centering
  \includegraphics[width=0.67\textwidth]{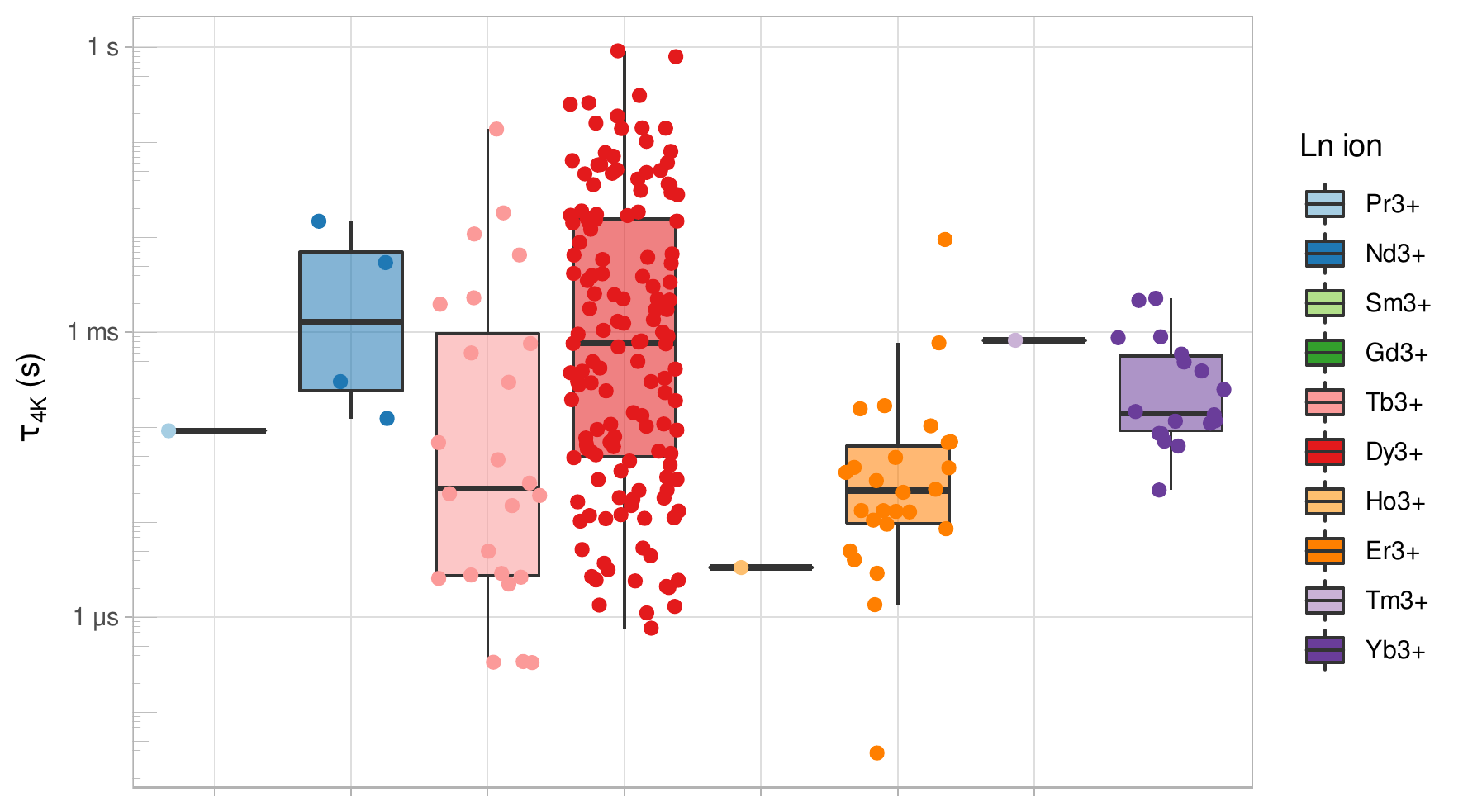}
  \includegraphics[width=0.48\textwidth]{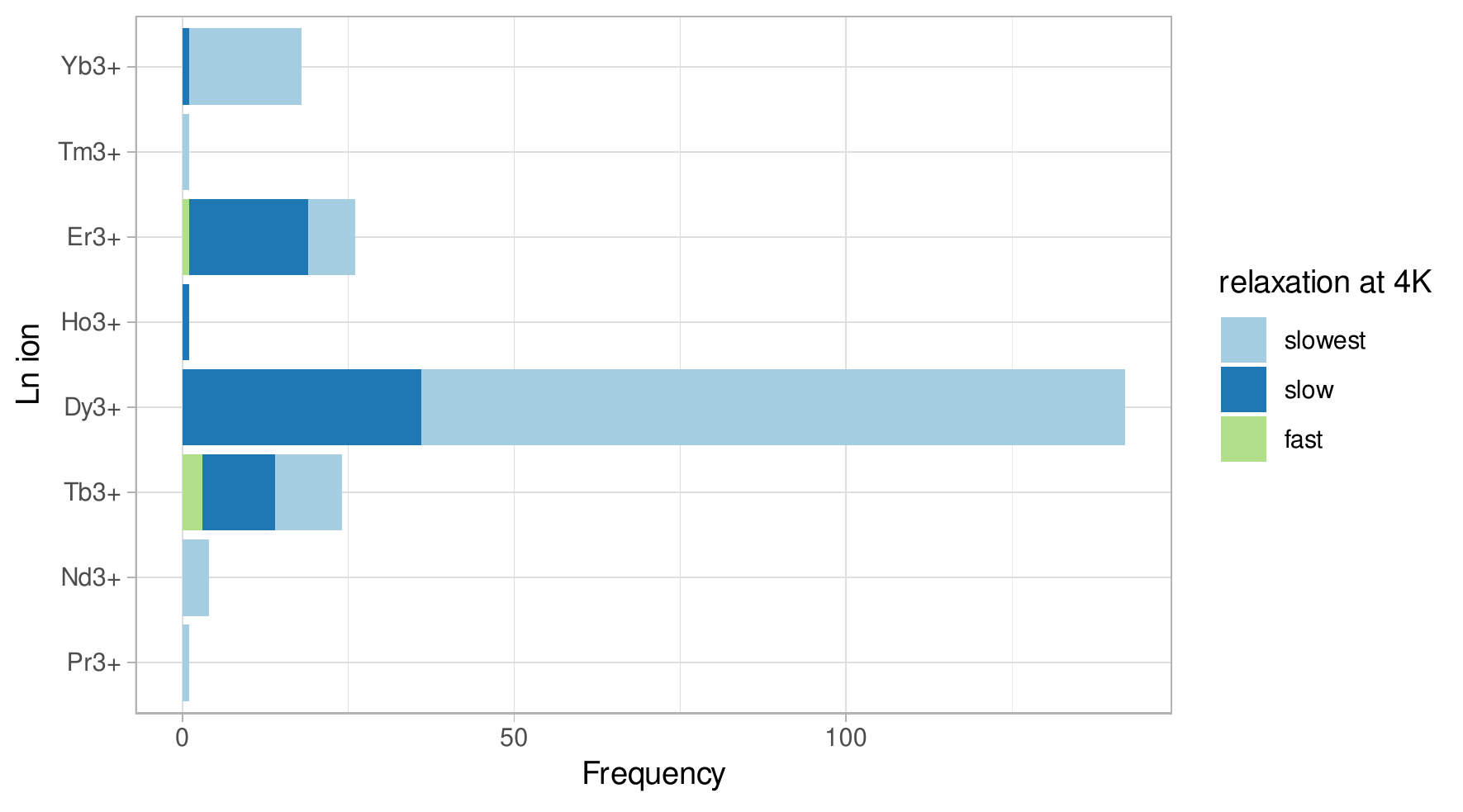}
  \includegraphics[width=0.48\textwidth]{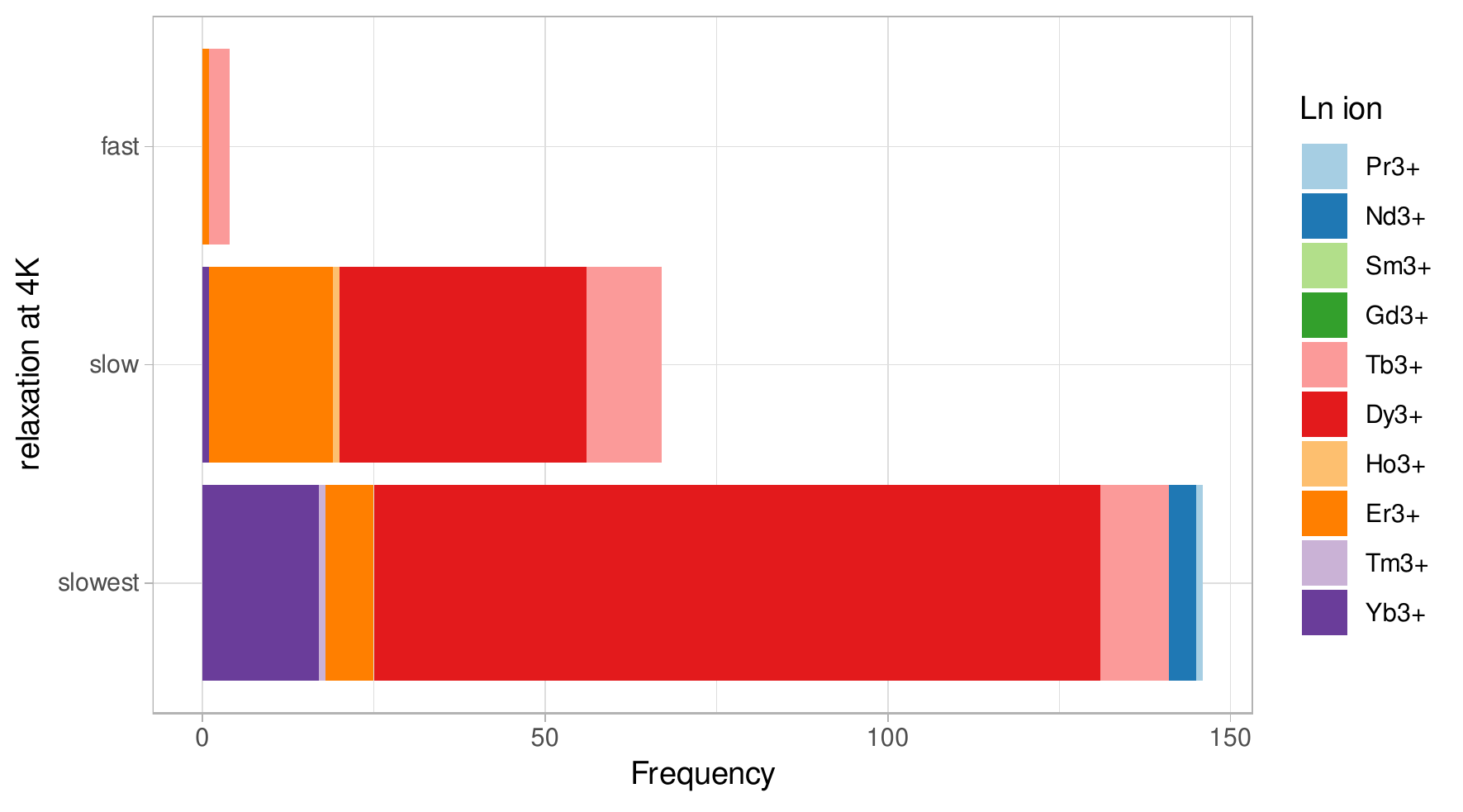}
  \caption{\textcolor{black}{Relationship between the lanthanide ion and $\tau_{4\mathrm{K}}$ within the SIMDAVIS dataset}.}
  \label{figSI:tauvsLnion4k}
\end{figure*}

\clearpage
\subsection{\textcolor{black}{Relaxation behavior}}
\textcolor{black}{Finally, we considered the relation between estimated relaxation time at $T=4$~K and observed relaxation behavior in terms of hysteresis and ac susceptometry. In the case of the hysteresis behavior, we see a different behavior for $\tau_{4\mathrm{K}}$ compared with $\tau_{300\mathrm{K}}$, e.g. the systems presenting pinched hysteresis present lower $\tau_{4\mathrm{K}}$ than the ones with no hysteresis. In case of the ac susceptometry, the behavior is similar at both temperatures, e.g. the samples that present frequency-dependent $\chi''$ relax slower than the ones which present no frequency dependence.}

\begin{figure*}[h]
\centering
  \includegraphics[width=0.67\textwidth]{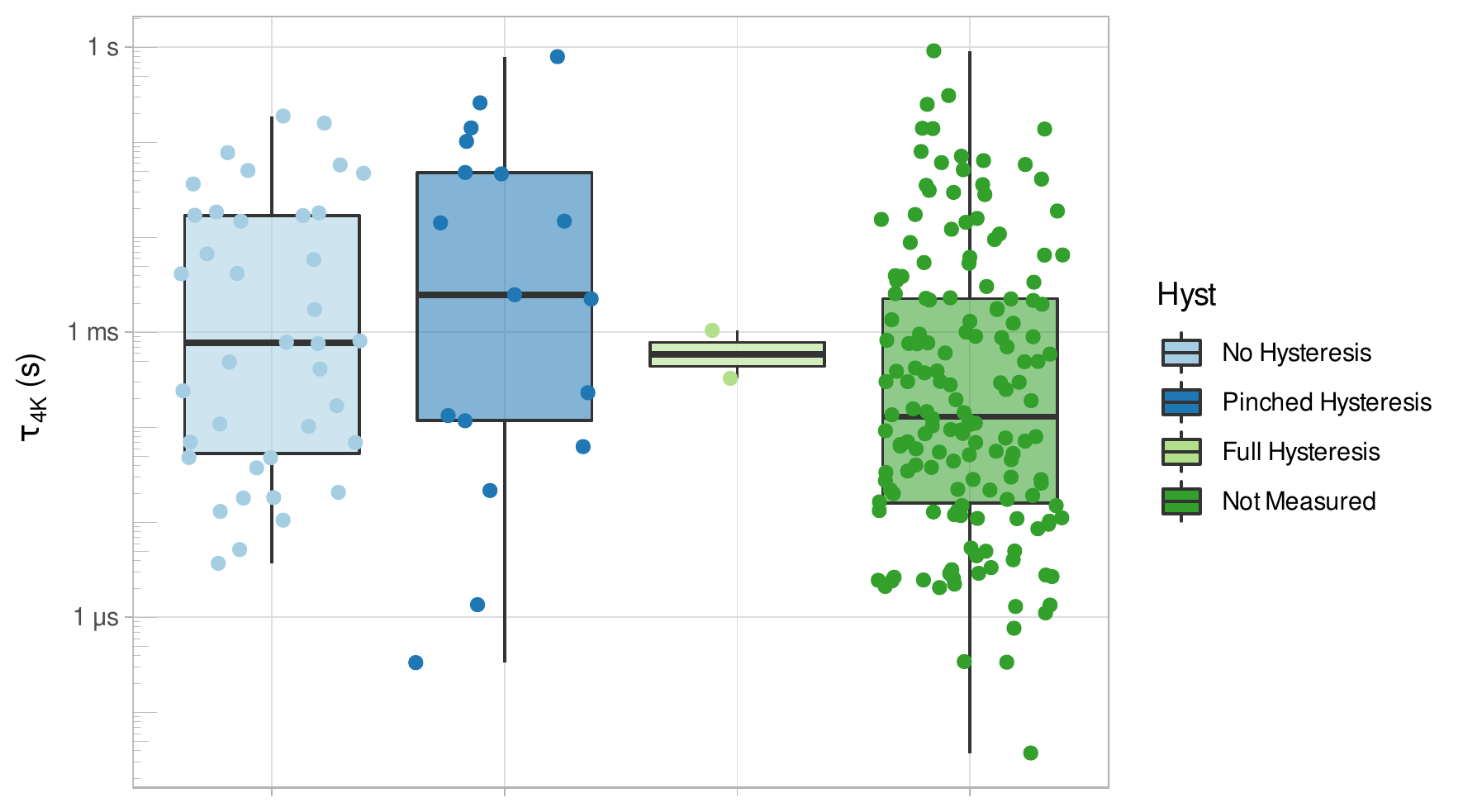}
  \includegraphics[width=0.48\textwidth]{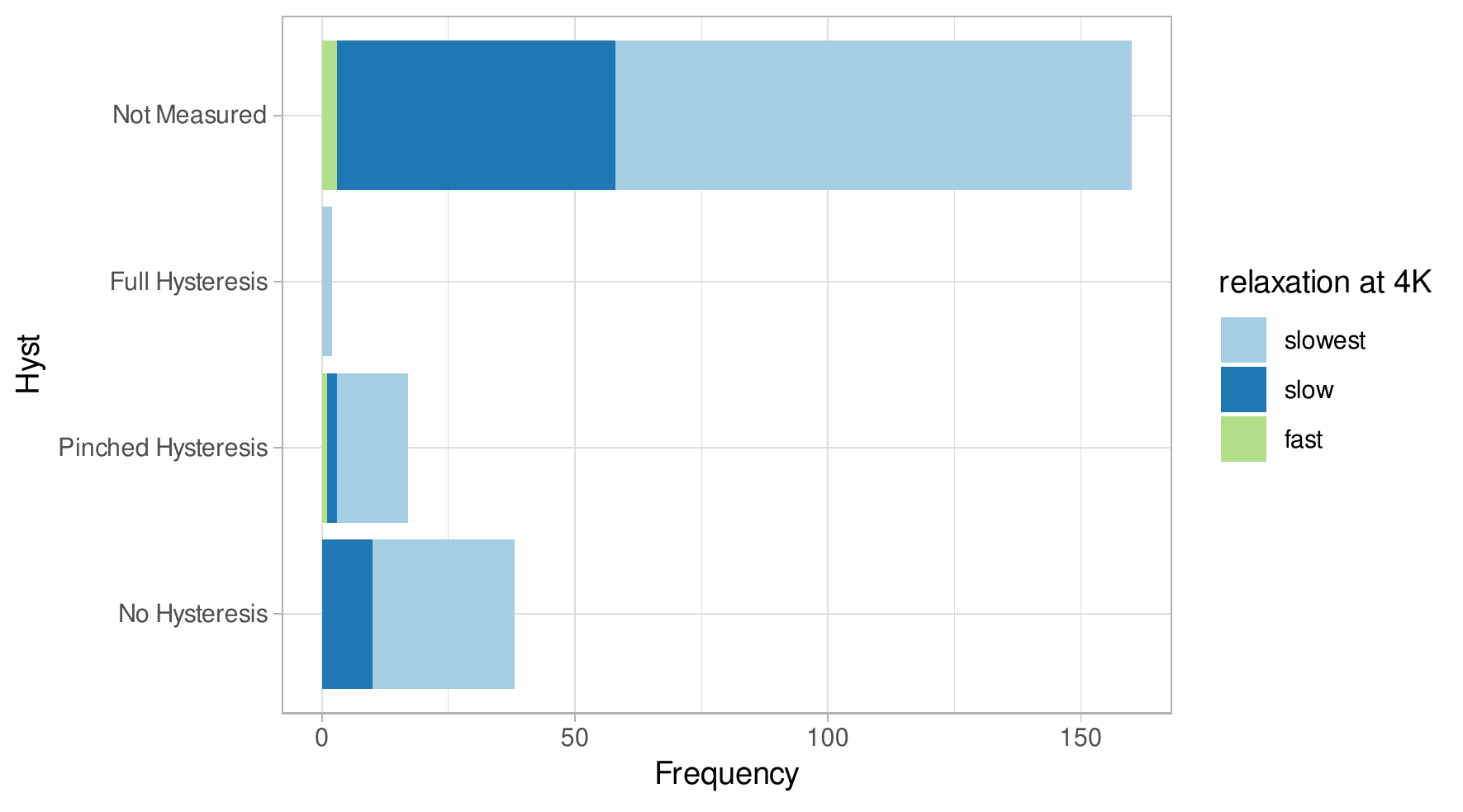}
  \includegraphics[width=0.48\textwidth]{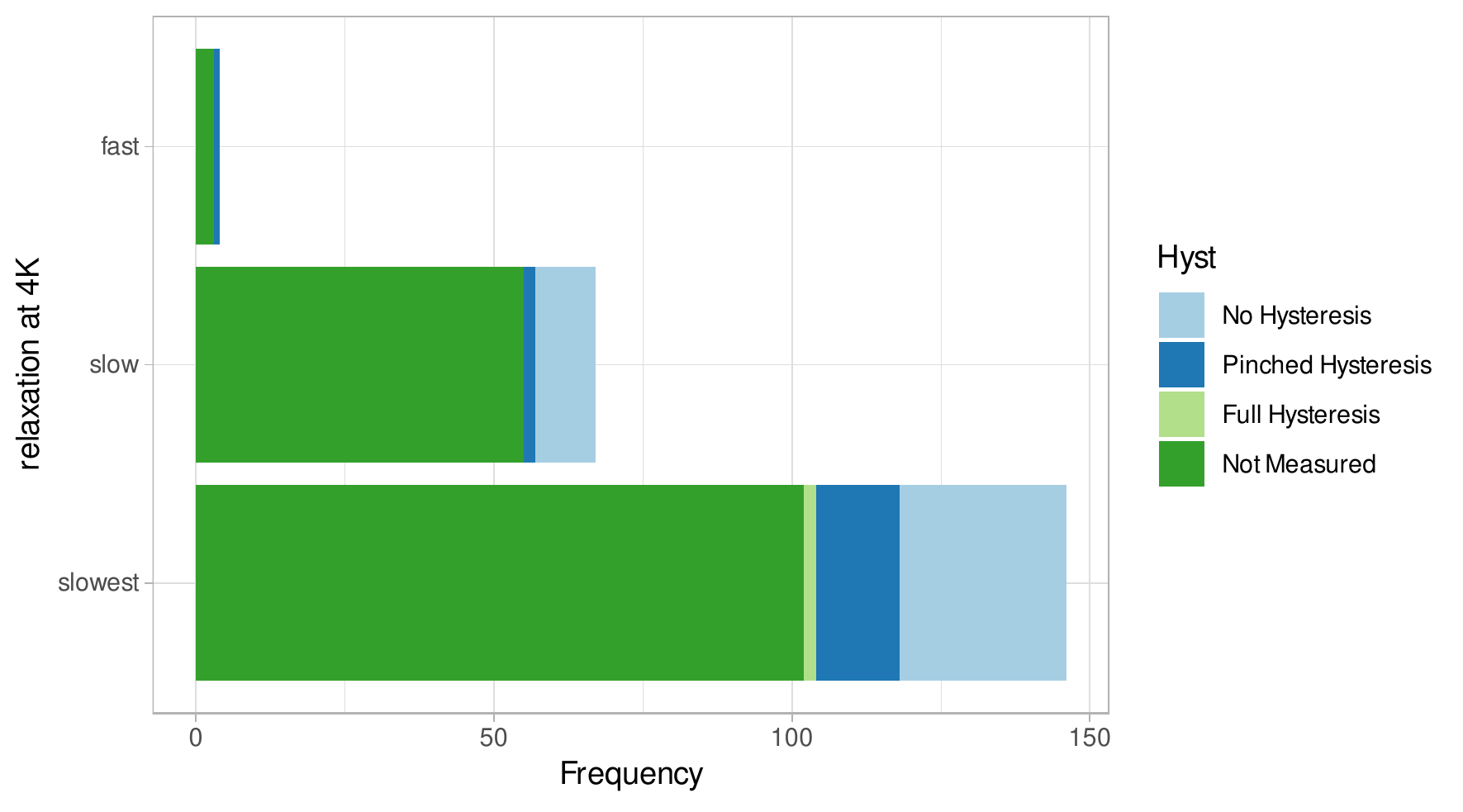}
  \caption{\textcolor{black}{Relationship between the hysteresis behavior and $\tau_{4\mathrm{K}}$ within the SIMDAVIS dataset}.}
  \label{figSI:tauvsHyst4k}
\end{figure*}

\begin{figure*}[h]
\centering
  \includegraphics[width=0.67\textwidth]{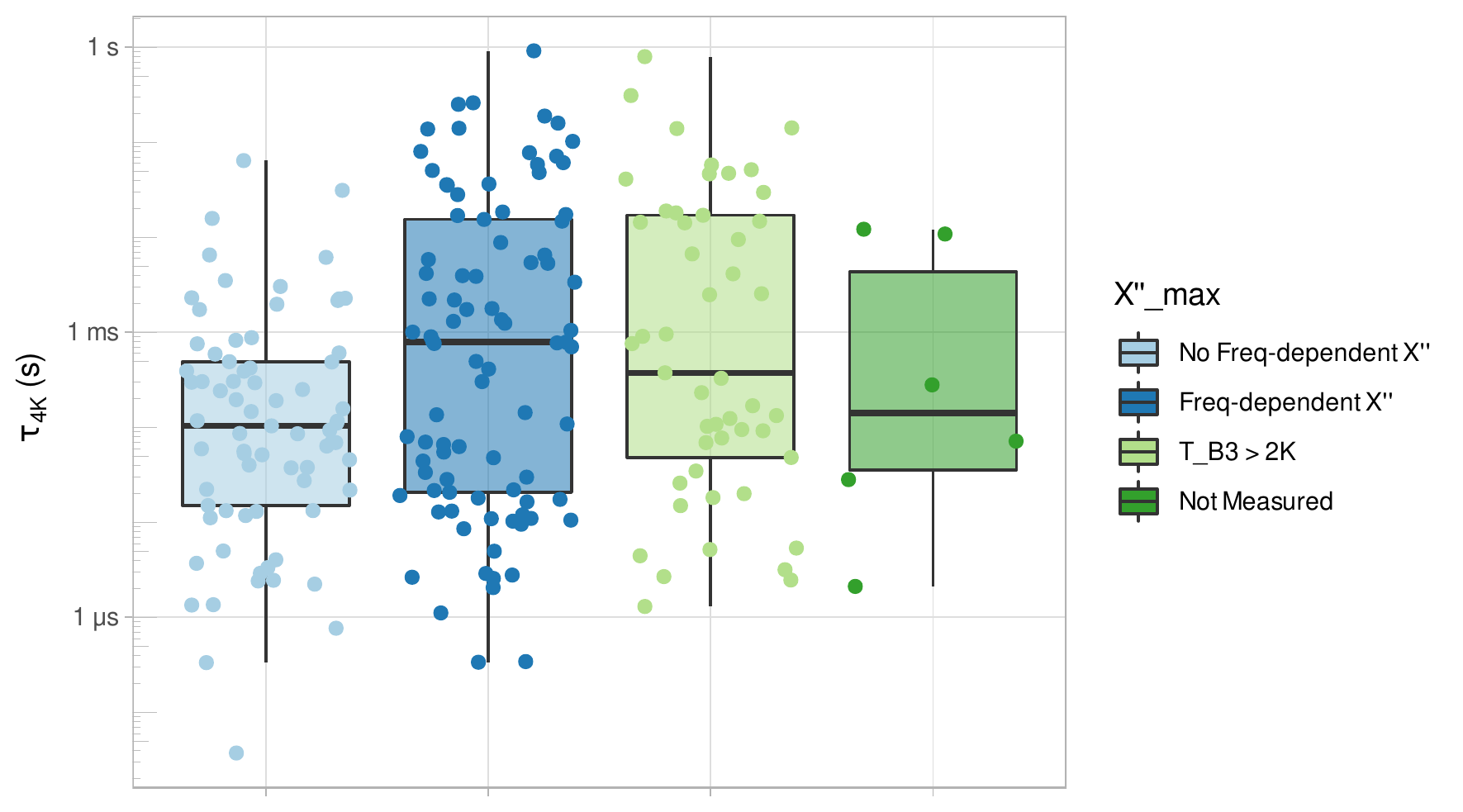}
  \includegraphics[width=0.48\textwidth]{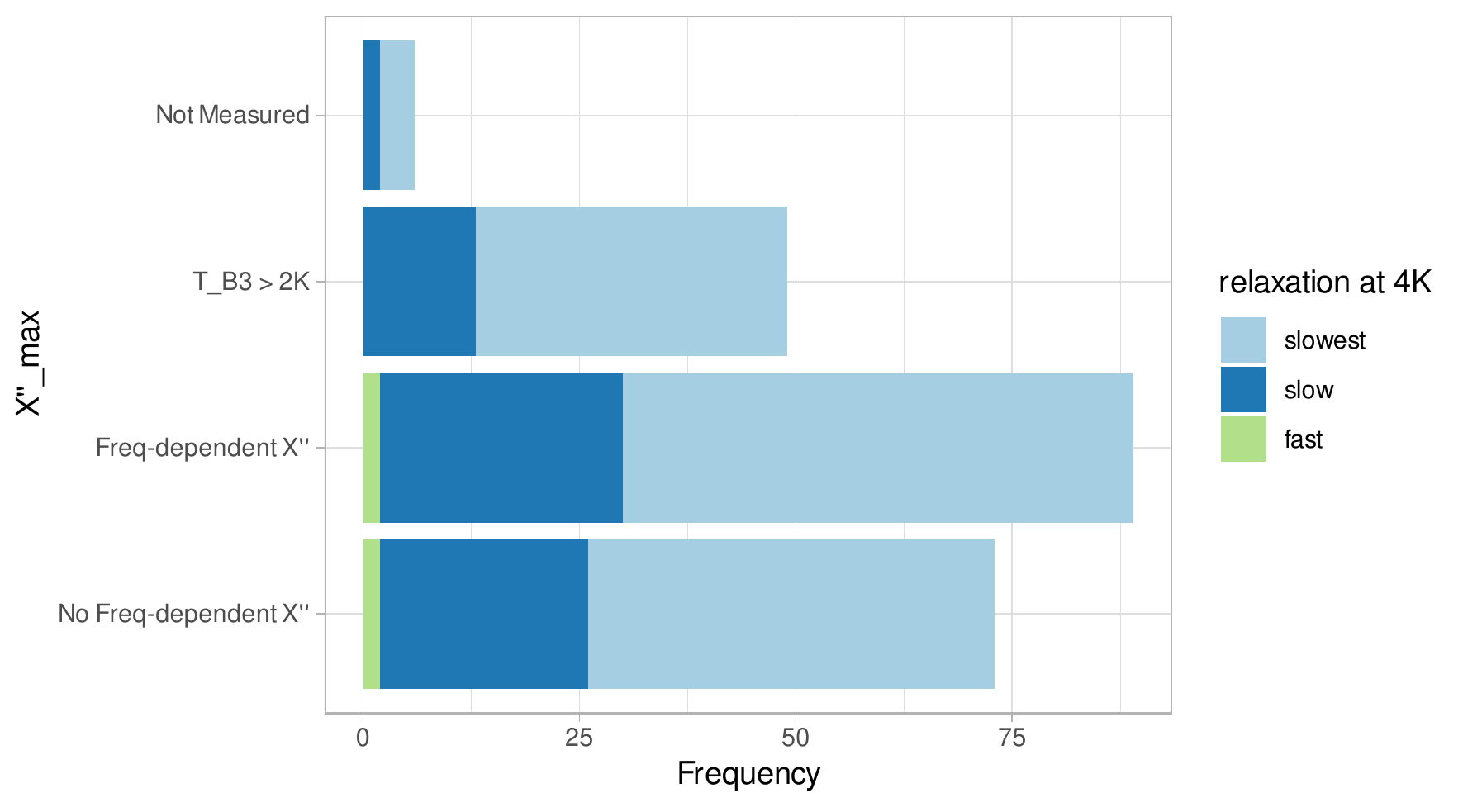}
  \includegraphics[width=0.48\textwidth]{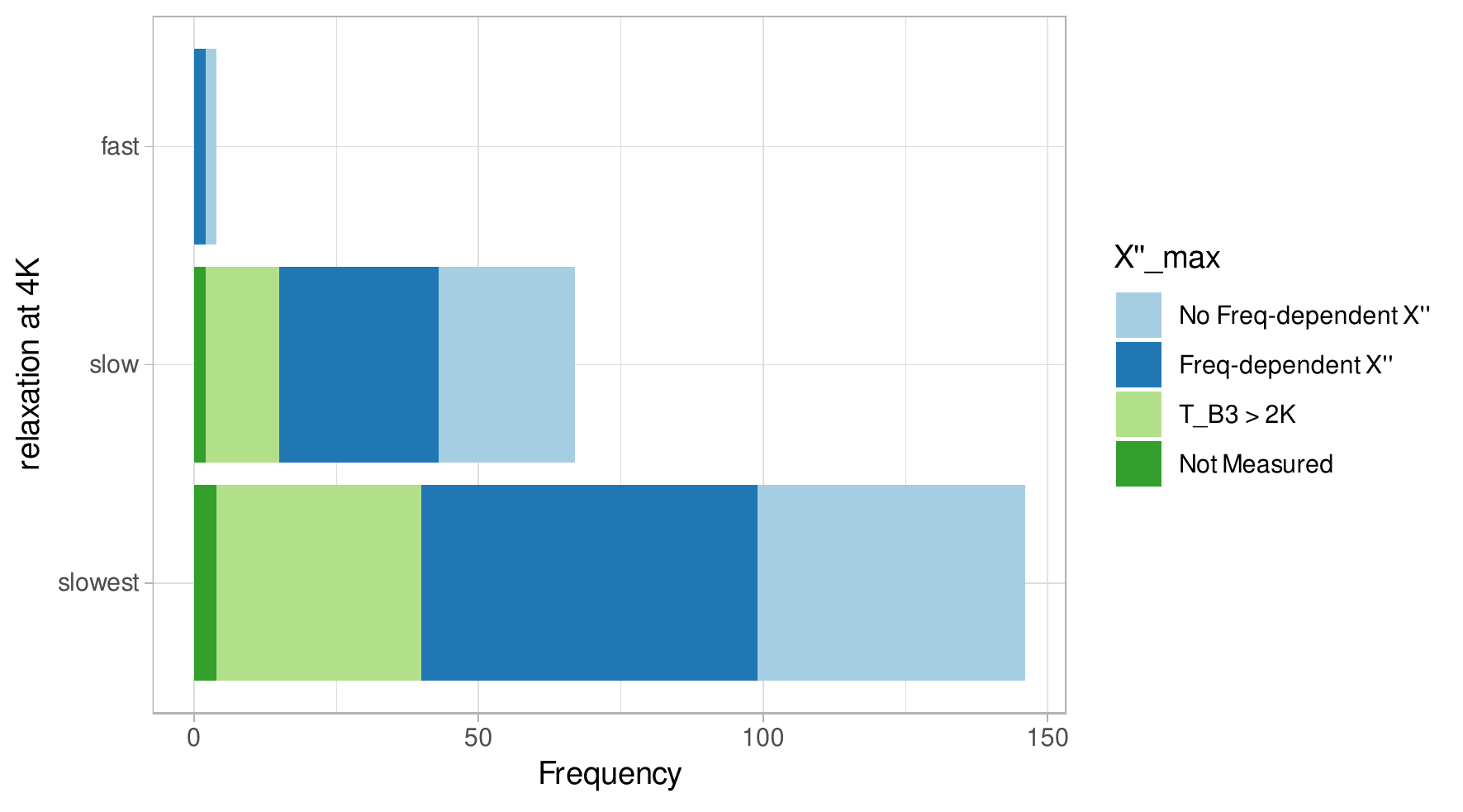}
  \caption{\textcolor{black}{Relationship between the ac out of phase behavior and $\tau_{4\mathrm{K}}$ within the SIMDAVIS dataset}.}
  \label{figSI:tauvsXim4k}
\end{figure*}
\clearpage

\textcolor{black}{\section{Detailed instructions for reproducibility}}

\textcolor{black}{This part offers a help for reproducing the graphics. As we comment in the \textit{readme.txt}, the following modules have to be installed:
\begin{enumerate}
\item numpy
\item matplotlib.pyplot
\item pandas
\item scipy.optimize
\item Collections (Counter)
\item random
\item math
\item time
\end{enumerate}}

\textcolor{black}{Moreover, the following files must be in the same folder containing the \textit{main.py} file:
\begin{enumerate}
\item main.py (Part of the code where the user specifies the parameters).
\item read\_data.py (Description of the system).
\item mag\_relaxation.py (Relaxation Mechanisms, Total probability for spin flipping).
\item Bolztmann\_distribution.py (Zeeman effect, Boltzmann distribution, single probabilities to pass from 0 to 1, and vice versa).
\item mean\_matrix\_state.py (For a two p-bit network, where the collective state is studied in this function).
\item association.py (For a two p-bit network, the association factor is calculated).
\item plotting.py (Graphical Representation of the results).
\item full\_data\_file.csv (File which contains all the information for few systems, from the SIMDAVIS dataset)
\end{enumerate}}

\textcolor{black}{It is important to emphasize that the file named \textit{main.py} contains the body of the simulator; the user could only change the configuration section at the beginning of the script (lines 25-51). In this version of STOSS we are capable to simulate three main scenarios:
\begin{enumerate}
\item Magnetization decays at different temperatures at constant magnetic field.
\item Magnetization decays at different temperatures at changeable magnetic field.
\item Magnetization decays at different temperatures of two p-bit networks.
\end{enumerate}
}

\begin{figure*}[h]
\centering
  \includegraphics[width=\columnwidth]{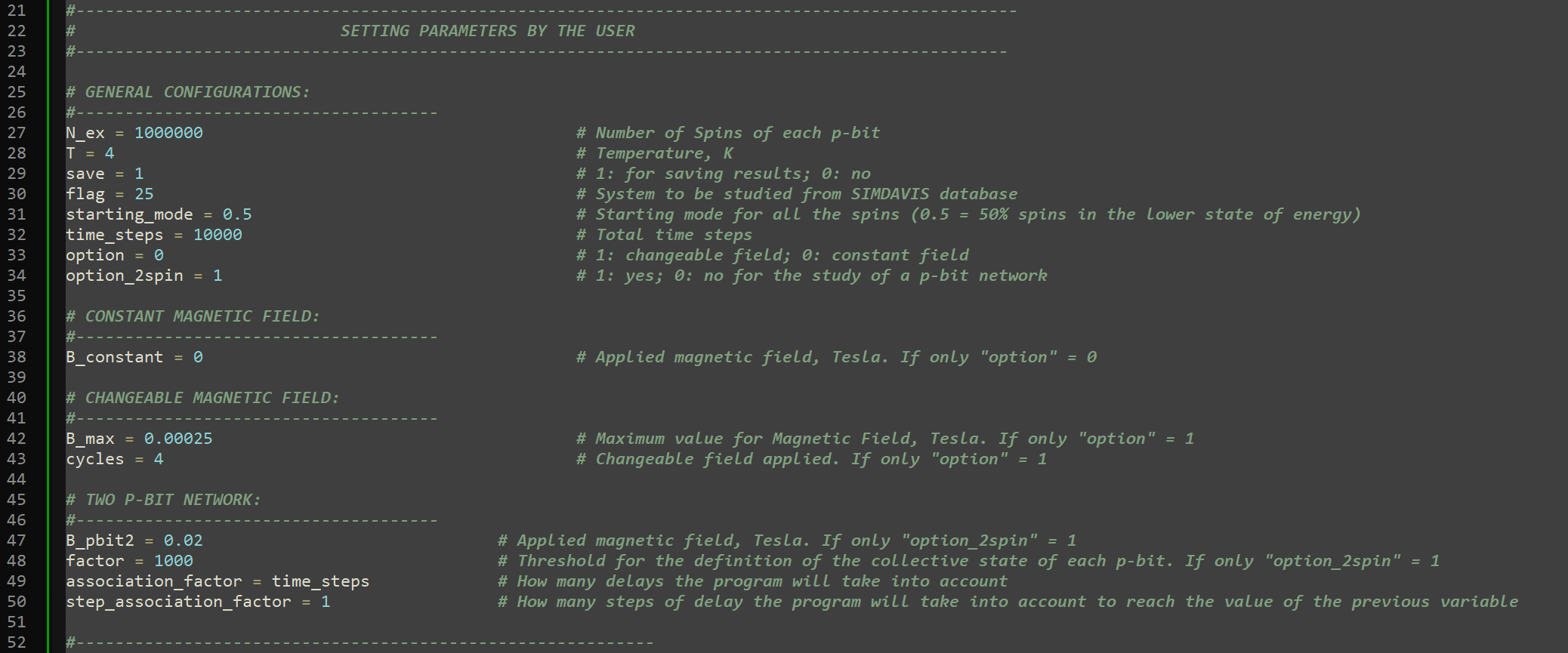}
  \caption {\textcolor{black}{Parameters that can be modified by the user.}}
  \label{figSI:code}
\end{figure*}

\textcolor{black}{Considering this idea, the user can select the type of simulation just writing the values in each variable. Figure~\ref{figSI:code} shows this part of the simulator.} \\

\clearpage

\textcolor{black}{\subsection{Lanthanide-based, molecular, isolated spin p-bits at constant field}}

\begin{figure*}[h]
\centering
  \includegraphics[width=0.67\textwidth]{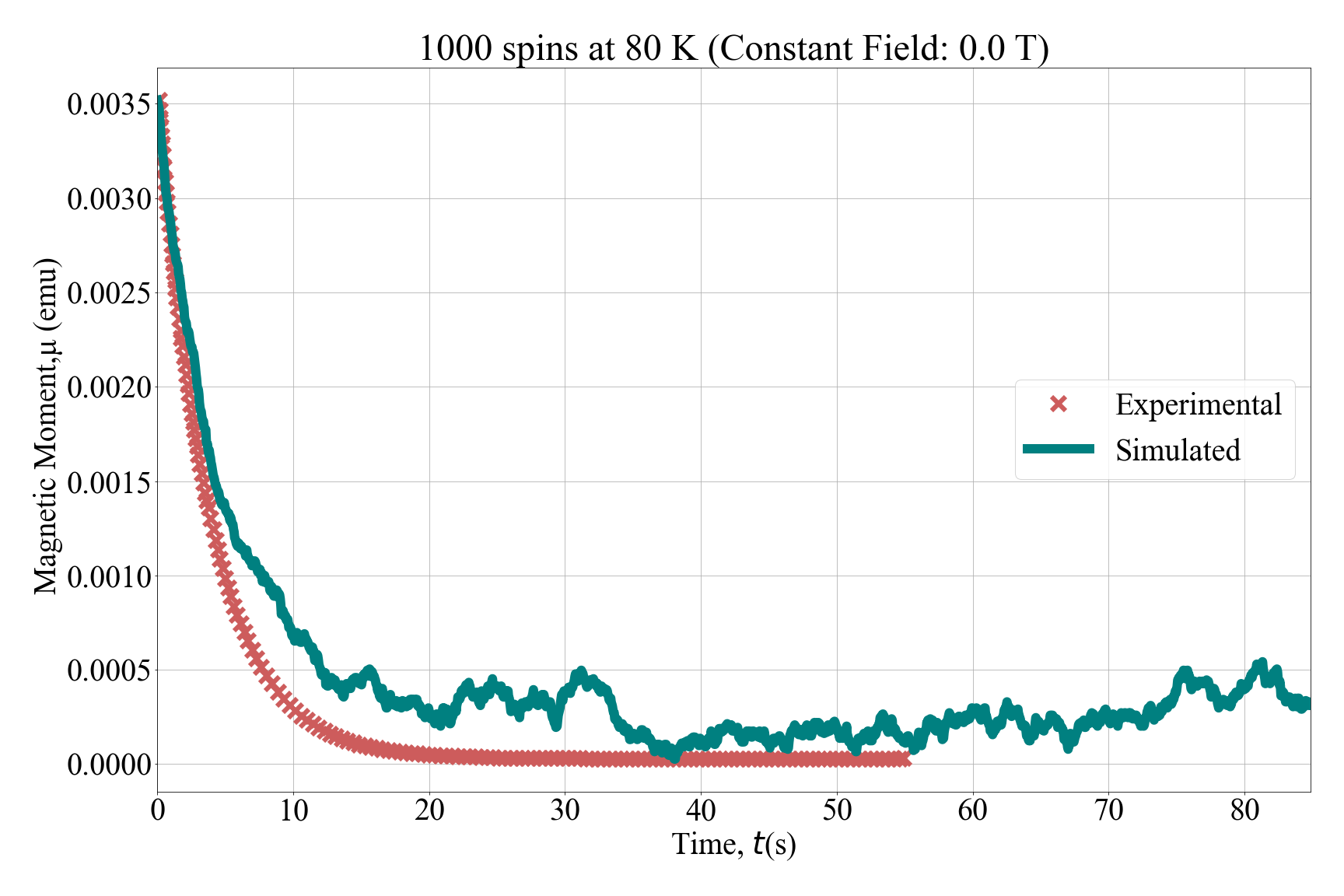}
  \caption {\textcolor{black}{Relaxation plot employing $N = 10^3$ spins at 80K for [(Cp$^\mathrm{iPr5}$)Dy(Cp$^\mathrm{*}$)]$^+$ (Cp$^\mathrm{iPr5}$, penta-iso-propylcyclopentadienyl; Cp$^\mathrm{*}$, pentamethylcyclopentadienyl).}}
  \label{figSI:case_a_1}
\end{figure*}

\begin{figure*}[h]
\centering
  \includegraphics[width=0.67\textwidth]{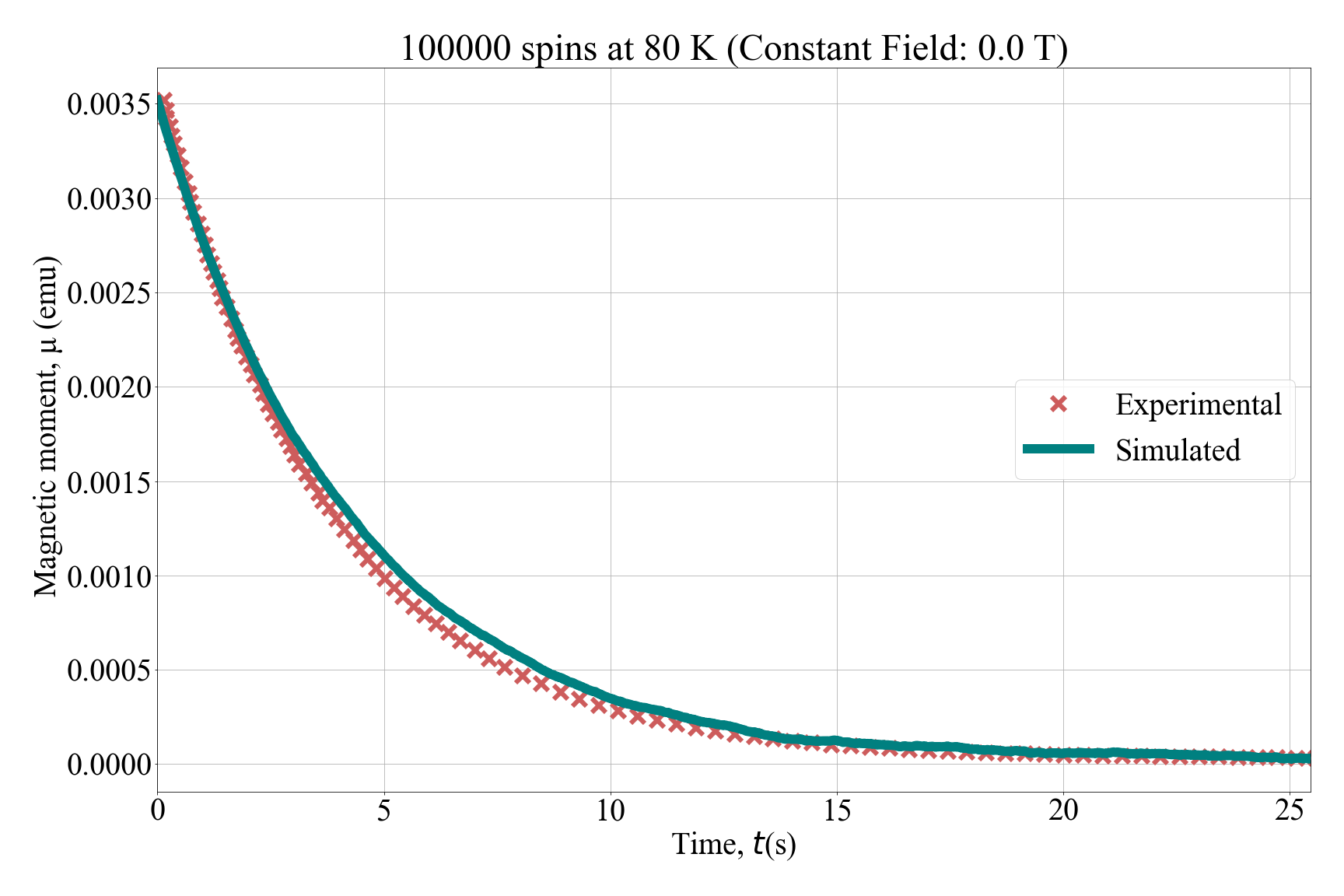}
  \caption {\textcolor{black}{Relaxation plot employing $N = 10^4$ spins at 80K for [(Cp$^\mathrm{iPr5}$)Dy(Cp$^\mathrm{*}$)]$^+$ (Cp$^\mathrm{iPr5}$, penta-iso-propylcyclopentadienyl; Cp$^\mathrm{*}$, pentamethylcyclopentadienyl).}}
  \label{figSI:case_a_2}
\end{figure*}

\textcolor{black}{ As an example, we present the experiment at 80K employing $N = 10^3$ (Figure~\ref{figSI:case_a_1}) and $N = 10^4$ (Figure~\ref{figSI:case_a_2}) spins, respectively. In the case of 80, 40, and 2K, the program compares the results of the magnetic moment with the experimental data extracted from the work by Guo et al.~\cite{Guo2018}}

\begin{figure*}[h]
\centering
  \includegraphics[width=0.67\textwidth]{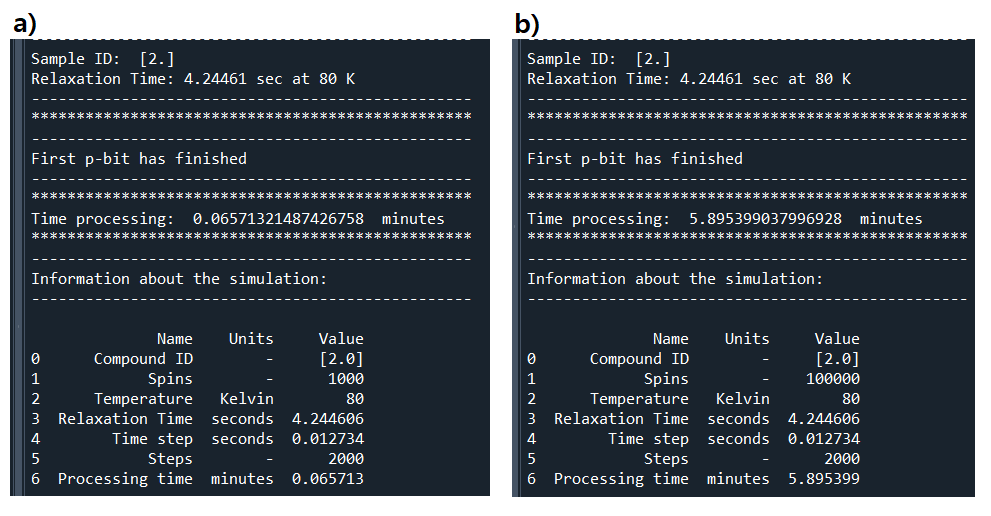}
  \caption {\textcolor{black}{Summary file of the simulation employing $N = 10^3$ (a) and $N = 10^4$ (b) spins at 80K for [(Cp$^\mathrm{iPr5}$)Dy(Cp$^\mathrm{*}$)]$^+$ (Cp$^\mathrm{iPr5}$, penta-iso-propylcyclopentadienyl; Cp$^\mathrm{*}$, pentamethylcyclopentadienyl).}}
  \label{figSI:time1}
\end{figure*}

\textcolor{black}{Figure~\ref{figSI:time1} shows the summary from the simulation process for both cases: $N = 10^3$ (a) and $N = 10^4$ (b) spins. We also present the parameters (Table~\ref{tab:t1}) which must be written by the user to reproduce those results. Given the stochasticity nature of the STOSS, the results could vary slightly.}

\begin{table}[H]
\centering 
\caption{\textcolor{black}{Parameters at 80K for  [(Cp$^\mathrm{iPr5}$)Dy(Cp$^\mathrm{*}$)]$^+$, where DNA = Does Not Apply.}}

\begin{tabular}{@{}cccc@{}}
\toprule
Variable & Figure~\ref{figSI:case_a_1}  & Figure~\ref{figSI:case_a_2}\\  \midrule
 N\_ex          &1000   &10000 \\ 
 T              & 80    & 80  \\
save            & 1     & 1 \\
 flag           & 22    & 22 \\
 starting\_mode & 0     & 0 \\ 
 time\_steps    & 2000  & 2000\\
 option         & 0     & 0  \\
option\_2spin   & 0     & 0 \\
 B\_constant    & 0     & 0\\ 
B\_max          & DNA   & DNA \\
cycles          & DNA   & DNA \\
 B\_pbit2       & DNA   & DNA \\
  factor        & DNA   & DNA \\
 association\_factor & DNA & DNA \\ 
step\_association\_factor & DNA & DNA \\  \bottomrule

\label{tab:t1}
\end{tabular}
\end{table}

\clearpage

\textcolor{black}{\subsection{Lanthanide-based, molecular, dynamically driven spin p-bits}}

\begin{figure*}[h]
\centering
  \includegraphics[width=0.67\textwidth]{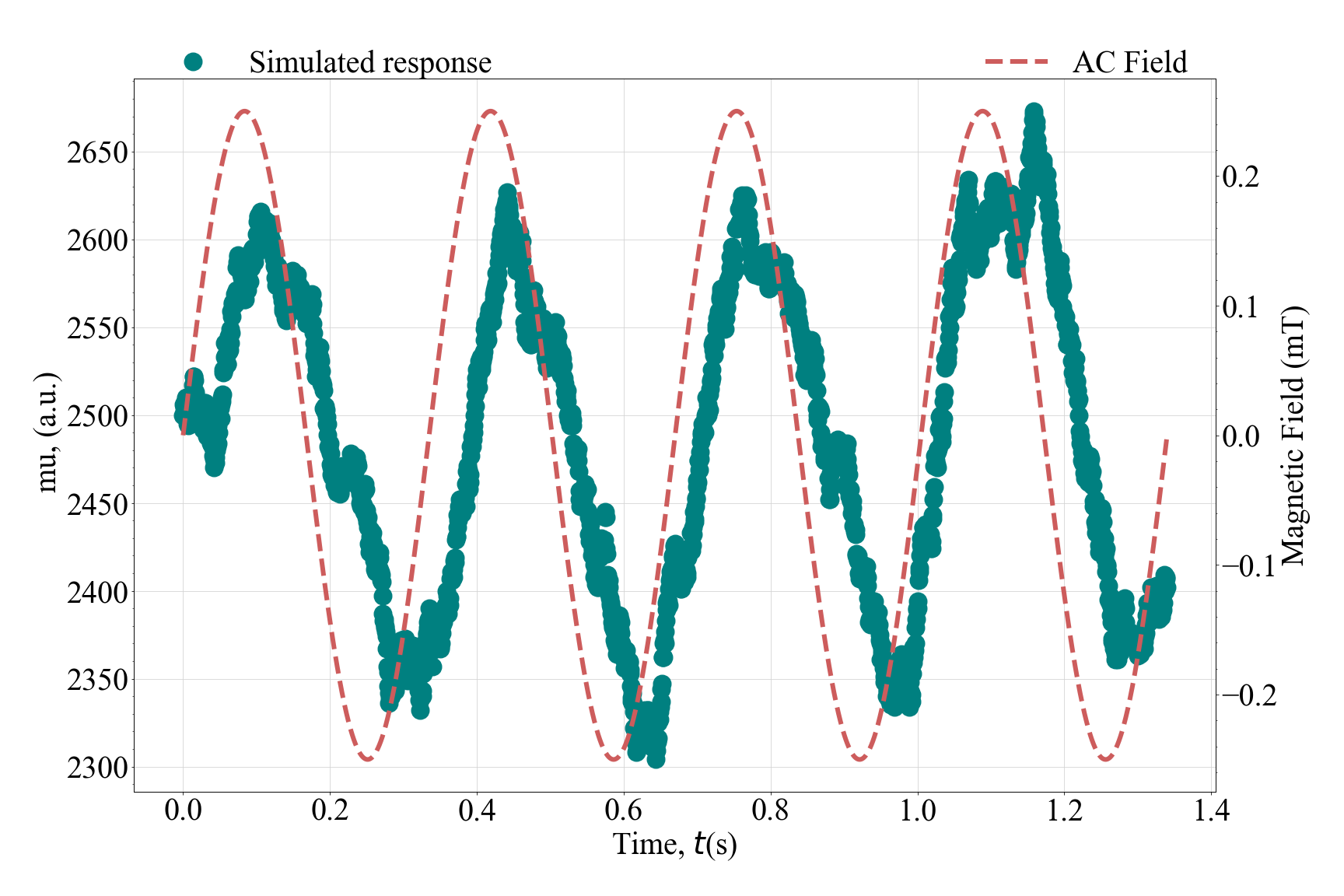}
  \caption {\textcolor{black}{Relaxation plot employing $N = 5\cdot10^3$ spins at 20mK for [Dy(bath)(tcpb)\textsubscript{3}] and frequency 3 Hz.}}
  \label{figSI:case_b_1}
\end{figure*}

\begin{figure*}[h]
\centering
  \vspace*{-10mm}
  \includegraphics[width=0.67\textwidth]{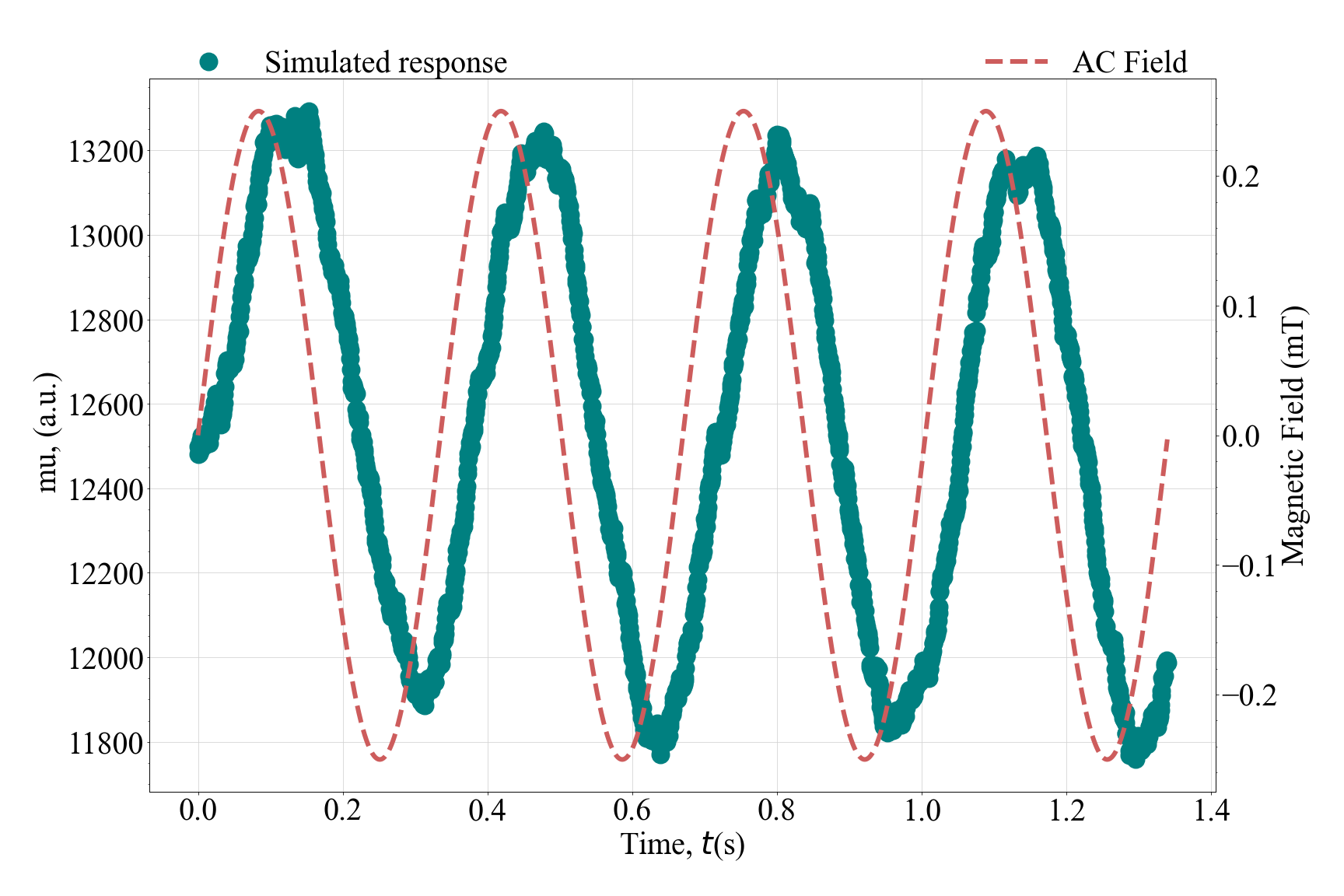}
  \caption {\textcolor{black}{Relaxation plot employing $N = 2.5\cdot10^4$ spins at 20mK for [Dy(bath)(tcpb)\textsubscript{3}] and frequency 3 Hz.}}
  \label{figSI:case_b_2}
\end{figure*}

\begin{figure*}[h]
\centering
  \includegraphics[width=0.67\textwidth]{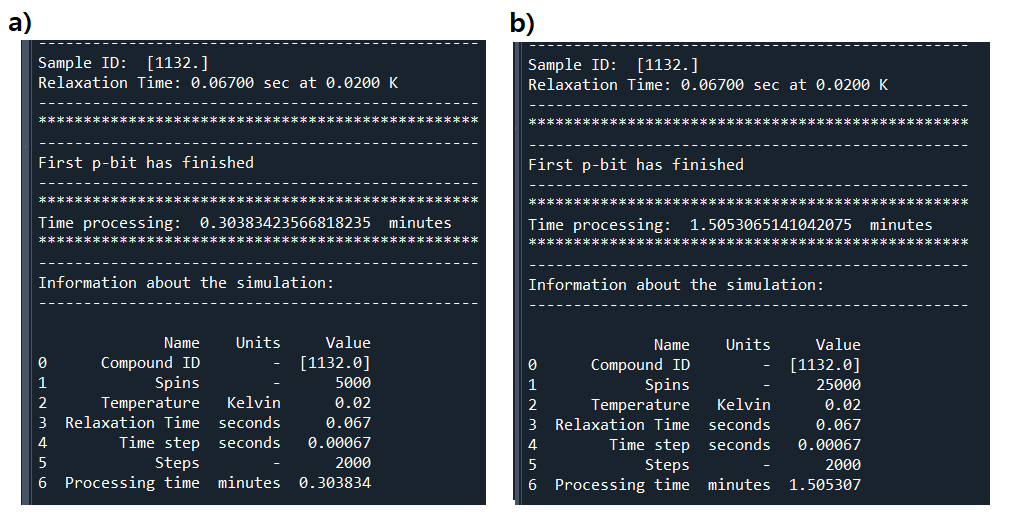}
  \caption {\textcolor{black}{Summary file of the simulation employing $N = 5\cdot10^3$ (a) and $N = 2.5\cdot10^4$ (b) spins at 20mK for [Dy(bath)(tcpb)\textsubscript{3}].}}
  \label{figSI:t2}
\end{figure*}

\textcolor{black}{As we could see in the previous part, all the information to simulate this experiment is shown in the (Table~\ref{tab:t2}). The result is presented in the Figure~\ref{figSI:case_b_1} with $N = 5\cdot10^3$ spins and $N = 2.5\cdot10^4$ spins in the Figure~\ref{figSI:case_b_2}. Moreover, the summary of each simulation is shown in Figure~\ref{figSI:t2}.}

\begin{table}[H]
\centering 
\caption {\textcolor{black}{Parameters at 20mK for [Dy(bath)(tcpb)\textsubscript{3}] and frequency 3 Hz, where DNA = Does Not Apply.}}
\begin{tabular}{@{}cccc@{}}
\toprule
 Variable       & Figure~\ref{figSI:case_b_1}     & Figure~\ref{figSI:case_b_2}\\  \midrule
 N\_ex          & 5000                          & 25000 \\ 
 T              & 0.02                          & 80 0.02  \\
save            & 1                             & 1 \\
 flag           & 26                            & 26 \\
 starting\_mode & 0.5                           & 0.5 \\ 
 time\_steps    & 2000                          & 2000\\
 option         & 1                             & 1  \\
option\_2spin   & 0                             & 0 \\
 B\_constant    & 0                             & 0\\ 
B\_max          & 0.00025                       & 0.00025 \\
cycles          & 4                             & 4 \\
 B\_pbit2       & DNA                           & DNA \\
  factor        & DNA                           & DNA \\
 association\_factor & DNA                      & DNA \\ 
step\_association\_factor & DNA                 & DNA \\  \bottomrule

\label{tab:t2}
\end{tabular}

\end{table}
\clearpage

\textcolor{black}{\subsection{Lanthanide-based, molecular spin p-bit network}}

\begin{figure*}[h]
\centering
  \includegraphics[width=0.67\textwidth]{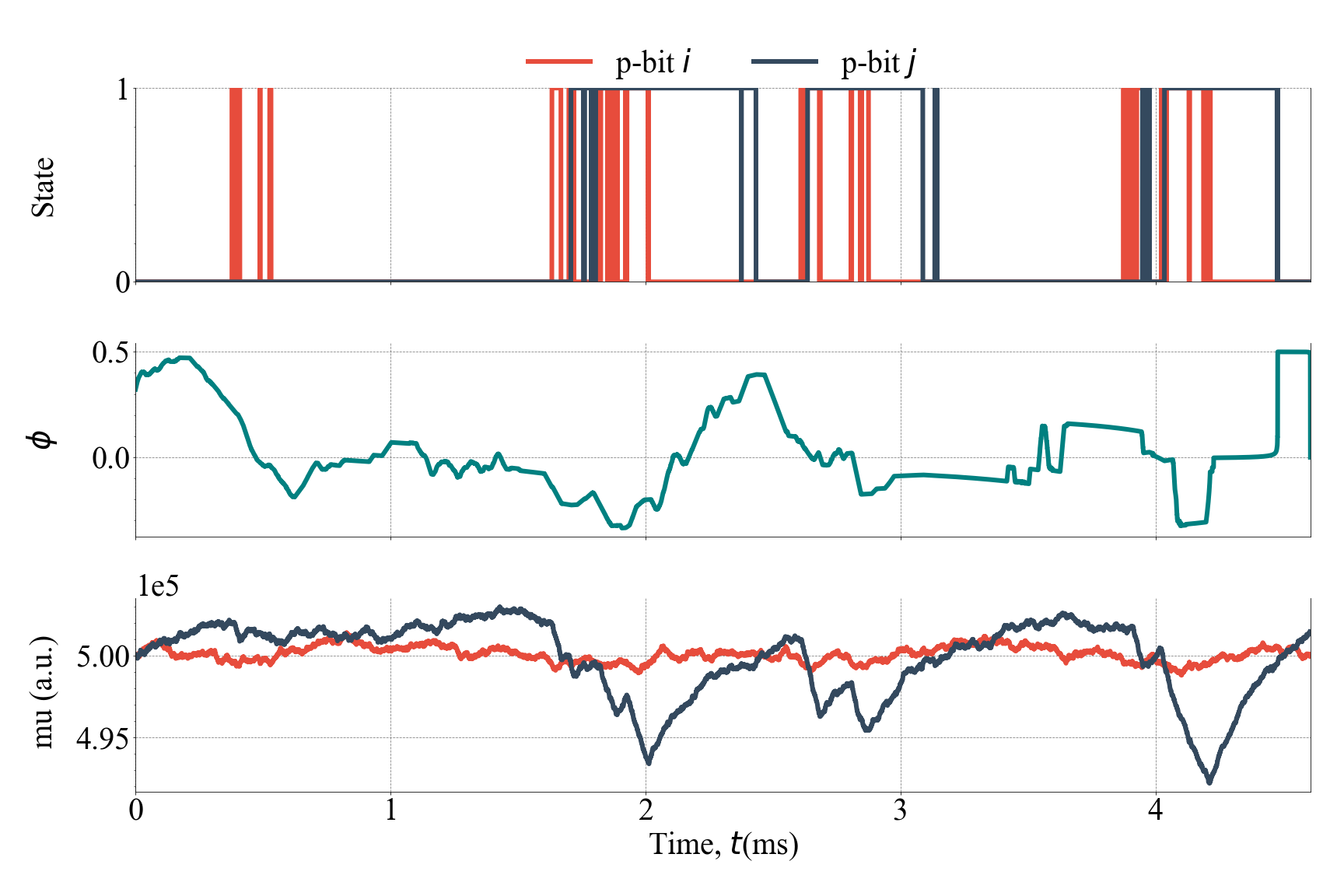}
  \caption {\textcolor{black}{Simulation employing $N = 10^6$ spins per p-bit at 4 K  for [Dy(obPc)\textsubscript{2}] Cd[Dy(obPc)\textsubscript{2}]. Where part a) represents the state of each p-bit, b) the association factor, and c) the relaxation behaviour.}}
  \label{figSI:case3}
\end{figure*}

\begin{figure*}[h]
\centering
  \includegraphics[width=0.35\textwidth]{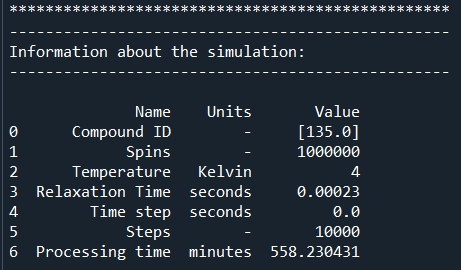}
  \caption {\textcolor{black}{Simulation employing $N = 10^6$ spins per p-bit at 4 K  for [Dy(obPc)\textsubscript{2}] Cd[Dy(obPc)\textsubscript{2}].}}
  \label{figSI:time3}
\end{figure*}

\textcolor{black}{This part of the simulator follows the calculations of two p-bit network. To reproduce the result, we present the Table~\ref{tab:t3} with all the necessary values and its simulation (Figure~\ref{figSI:case3} ). In contrast,  Figure~\ref{figSI:time3} shows the summary part where we could see the computational cost when we increase the number of spins of each p-bit to 1 million with 10,000 total time steps.}

\begin{table}[H]
\centering 
\caption {\textcolor{black}{Parameters at 4 K to analyze the effect of a p-bit to another, for [Dy(obPc)\textsubscript{2}] Cd[Dy(obPc)\textsubscript{2}].}}
\begin{tabular}{@{}cccc@{}}
\toprule
 Variable                   & Figure~\ref{figSI:case3} \\  \midrule
 N\_ex                      & 1000000 \\ 
 T                          & 4 \\
save                        & 1  \\
 flag                       & 25 \\
 starting\_mode             & 0.5 \\ 
 time\_steps                & 10000\\
 option                     & 0  \\
option\_2spin               & 1 \\
 B\_constant                & 0 \\ 
B\_max                      & DNA  \\
cycles                      & DNA \\
 B\_pbit2                   & 0.02 \\
  factor                    & 1000 \\
 association\_factor        & time\_steps \\ 
step\_association\_factor   & 1 \\ \bottomrule

\label{tab:t3}
\end{tabular}

\end{table}

\cleardoublepage

\bibliographystyle{apsrev4-2}
\bibliography{pbits}

\clearpage